\documentclass{aa}
\usepackage{natbib}
\usepackage{graphicx}
\usepackage{subfigure}
\usepackage{longtable}
\usepackage{txfonts}
\bibpunct[;]{(}{)}{;}{a}{}{;}

\def\lesssim{\mathrel{\hbox{\rlap{\hbox{\lower4pt\hbox{$\sim$}}}\hbox{$<$}}}}
\def\gtrsim{\mathrel{\hbox{\rlap{\hbox{\lower4pt\hbox{$\sim$}}}\hbox{$>$}}}}

\newcommand{\beq}{\begin{equation}}
\newcommand{\eeq}{\end{equation}}
\newcommand{\lab}{\label} 
\newcommand{\degree}{\mbox{$^\circ$}}
\newcommand{\ha}{\mbox{H$\alpha$}}
\newcommand{\hb}{\mbox{H$\beta$}}
\newcommand{\lha}{\mbox{$L_\mathrm{H\alpha}$}}
\newcommand{\eha}{\mbox{EW($\mathrm{H\alpha}$)}}
\newcommand{\lb}{\mbox{$L_{B}$}}
\newcommand{\lk}{\mbox{$L_{K}$}}

\newcommand{\lfuv}{\mbox{$L_\mathrm{FUV}$}}
\newcommand{\ltir}{\mbox{$L_\mathrm{TIR}$}}

\newcommand{\msun}{\mbox{M$_{\odot}$}}
\newcommand{\lbsun}{\mbox{$L_{B,\odot}$}}
\newcommand{\lksun}{\mbox{$L_{K,\odot}$}}
\newcommand{\kcc}{\mbox{K$_\mathrm{CC}$}}
\newcommand{\rcc}{\mbox{$R_\mathrm{CC}$}}

\newcommand{\acc}{\mbox{$A_\mathrm{CC}$}}
\newcommand{\pcc}{\mbox{$P_\mathrm{CC}$}}
\newcommand{\fcc}{\mbox{$f_\mathrm{CC}$}}
\newcommand{\ncc}{\mbox{$N_\mathrm{CC}$}}
\newcommand{\mlcc}{\mbox{$m_\mathrm{l}^\mathrm{CC}$}}
\newcommand{\mucc}{\mbox{$m_\mathrm{u}^\mathrm{CC}$}}
\newcommand{\ml}{\mbox{$m_l$}}
\newcommand{\mup}{\mbox{$m_u$}}
\newcommand{\NCCExp}{\langle N_\mathrm{CC} \rangle}

\begin{document}

\title{A comparison between star formation rate diagnostics and rate of  core collapse supernovae within 11 Mpc}

\author{M.T.~Botticella\inst{1,2}, S.J.~Smartt\inst{2}, R.C.~Kennicutt, Jr.\inst{3},  E.~Cappellaro\inst{1},  M.~Sereno\inst{4,5}, J.C.~ Lee\inst{6}}
\offprints{M.T. Botticella, \email{mariateresa.botticella@oapd.inaf.it}}

\institute{INAF- Osservatorio Astronomico di Padova, Vicolo dell'Osservatorio 5, 35122 Padova, Italy
\and Astrophysics Research Centre, School of Mathematics and Physics,  Queen's University Belfast, Belfast BT7 1NN,  UK 
\and  Institute of Astronomy, University of Cambridge, Madingley Road, Cambridge CB3 0HA, UK
\and  Dipartimento di Fisica, Politecnico di Torino, Corso Duca degli Abruzzi 24, 10129, Torino, Italy
\and INFN, Sezione di Torino, Via Pietro Giuria 1, 10125, Torino, Italy
\and Carnegie Fellow, Carnegie Observatories, 813 Santa Barbara Street, Pasadena, CA 91101
}

\date{Received .../ Accepted ...}

\abstract{}
{The core collapse supernova rate provides a strong lower limit for the star formation rate.  
Progress in using it as a cosmic star formation rate tracer requires some confidence that
it is consistent with more conventional star formation rate diagnostics in the nearby Universe.  
This paper compares standard star formation rate measurements based on H$\alpha$, Far Ultraviolet and Total Infrared galaxy luminosities  with the observed core collapse supernova rate in the same galaxy sample. 
The comparison can be viewed from two perspectives. Firstly, by adopting an estimate 
of the minimum stellar mass to produce a core collapse supernova one can determine a star formation
rate from supernova numbers. Secondly, the radiative star formation rates can
be assumed to be robust and then the supernova statistics provide a constrain on 
the minimum stellar mass for core collapse supernova progenitors. }
{The novel aspect of this study is that
 H$\alpha$, Far Ultraviolet and Total Infrared luminosities are now available 
for a complete galaxy sample within the local 11Mpc volume  and the number of discovered supernovae in this sample within the last 13 years is high enough to perform a meaningful statistical comparison. 
We exploit the multi-wavelength dataset  from 11HUGS,  a  volume-limited survey  designed to provide a census of  star formation rate in the local Volume. There are 14 supernovae discovered in this
sample of galaxies within the last 13 years. Although one could argue that this may not be
complete, it is certainly a robust lower limit. }
{Assuming a lower limit for core collapse of 8\,\msun\, (as proposed by direct detections of SN progenitor stars and white dwarf progenitors), the core-collapse supernova rate 
matches the star formation rate from the Far Ultraviolet luminosity. However the star formation rate 
based on \ha\, luminosity is lower than these two estimates by a factor of  nearly  2. 
If we assume that the Far Ultraviolet or  \ha\,  based luminosities are a true reflection 
of the star formation rate, we find that the minimum mass for core collapse supernova  progenitors is $8 \pm 1$\,\msun\, and $ 6\pm 1$\,\msun, respectively. } 
{The estimate of the minimum mass for core collapse supernova  progenitors obtained exploiting  Far Ultraviolet data  is in good agreement with that from the direct detection of  supernova progenitors.
 The concordant results by  these independent methods  point toward a constraint of $8\pm 1$\,\msun\, on the lower mass limit for progenitor stars of core collapse  supernovae.}


\titlerunning{ Mass range of  CC SN progenitors}
\authorrunning{Botticella et al.}
\maketitle

\section{Introduction}
The progenitors of core collapse supernovae (CC SNe)
are massive stars, either single or in binary systems, that complete exothermic nuclear burning, up to the development
of an iron core that cannot be supported by any further nuclear fusion reactions or by electron degeneracy pressure. The subsequent collapse of the iron core results in the formation of a compact object, a neutron star or a black hole, accompanied by the high-velocity ejection of a large fraction of the progenitor mass. 
The SNe ejecta  sweep, compress and heat the interstellar medium,  and
release the heavy elements  which are produced during the progenitor
evolution and in the explosion itself . This can further trigger
subsequent star formation process  \cite[e.g.][]{McKee1977}, hence
having a profound effect on galaxy evolution. 

Due to the short lifetime of their progenitor stars,  the rate of
occurrence of CC SNe closely follows the current star formation rate (SFR) in a stellar system.
 The evolution of the CC SN rate  with redshift  is a probe of the SF
 history (SFH)  and  allows us  to constrain  the chemical enrichment
 of the galaxies and the effect of  energy/momentum feedback.  
Poor statistics is a major limiting factor for using the CC SN rate as
a tracer of the SFR. At low redshift the difficulty is in sampling
large enough volumes of the local Universe to ensure significant
statistics \cite[e.g.][]{Kennicutt1984}.  
While  at high redshift  the difficulty  lies in detecting and typing complete samples of intrinsically faint SNe \citep{Botticella2008,Bazin2009,Li2010}.  Moreover some fraction of CC SNe are missed by optical searches, since they are embedded in dusty spiral arms or galaxy nuclei. This fraction may change with redshift, if the amount of dust in galaxies evolves with time.  
Progress in using CC SN rates as SFR tracers requires accurate measurements of rates at various cosmic epochs and in different environments. 
Furthermore it requires a meaningful comparison  with other SFR diagnostics  to verify its reliability and to analyse its main limitations. 

The CC SN rate is also a  powerful tool to investigate the nature of SN progenitor stars and to test stellar evolutionary models.
Different sub-types of CC SNe have been identified on the basis of their spectroscopic and photometric properties and a possible 
sequence has been proposed on the basis of the progenitor mass loss history   with the most massive stars losing the largest fraction of their initial mass \citep{Heger2003}.
However, this simple  scheme where only the mass loss drives the
evolution of massive stars cannot easily explain the variety of  observational properties showed by CC SNe of the same sub-type and the relative numbers of  different  sub-types \citep{2009ARA&A..47...63S}.

In particular, two important outstanding issues are : the minimum mass
of a star that leads to a CC SN (in a single or binary system) ; and
what is the mass range of progenitor stars of different CC SN
sub-types. It is possible to constrain the mass
range of stars that produce CC SNe by comparing the CC SN rate
expected for a given SFR and the observed one in the same galaxy
sample or in the same volume.
  
  In this paper  we exploit  a complete, multi-wavelength dataset collected for a
  volume-limited sample of nearby galaxies to compare different SFR
  diagnostics with the CC SN rate.
  This  provides a method to constrain the cutoff mass for
  CC SN progenitors by exploiting the SFR as traced by Ultraviolet
  (UV) and \ha\, emission. The novelty of  this work consists  in
  studying both  the 
CC SN rate and SFR in the same  well defined galaxy sample.
Thorough the paper we adopt a Hubble constant ($\rm{H}_{0}$) of 75 \,km s$^{-1}$ Mpc$^{-1} $ and  the Vega System for the magnitudes.

\section{The link between CC SN rate and SFR}\label{link}
The instantaneous SFR in a galaxy  is directly traced by the number of
currently existing massive stars  since  these stars have short life
times. Usually the total SFR in a galaxy is obtained  by extrapolating
the massive star SFR  to lower stellar masses given an initial mass function (IMF) describing the relative probability of stars
of different masses forming. 
The  luminosity of a galaxy is a direct and sensitive tracer of its stellar population so it is possible
to directly connect a luminosity to the instantaneous SFR when the  observed emission comes from stars which are short lived or from short lived phases of stellar evolution.
The instantaneous SFR can be calculated from the observed luminosity in a wavelength band F   which satisfies the above requirement from the relation:
\begin{equation}
L_{\rm F}= \frac{\int_{\ml}^{\mup} t_{\rm F}(m)l_{\rm F}(m)\phi(m) dm}{\int_{\ml}^{\mup} m \phi(m) dm} \psi
\end{equation}
where $L_{\rm F}$ is the total galaxy luminosity, $\psi$ is the SFR, $l_{\rm F}(m)$ is the luminosity of a single star of mass $m$,  $t_{\rm F}$ is the characteristic timescale over which a star of mass $m$ emits radiation in the wavelength band F and $\phi(m)$ the IMF. The limits of integration extend over the range of masses of the stars which are expected to emit  radiation in the band F.
The  IMF  generally  is parametrized as
a power law: 
\begin{equation}
\mathrm{dN}=\phi(m)dm \propto m^{\gamma}dm 
\end{equation}
where dN is the number of single stars in the mass range $m$, $m+dm$. 
We adopted  a Salpeter IMF  defined  in the mass range 0.1--100\,\msun\   with $\gamma = -2.35$ \citep{Salpeter1955}.

\begin{figure}
\begin{center}
\includegraphics[width=5cm,angle=-90]{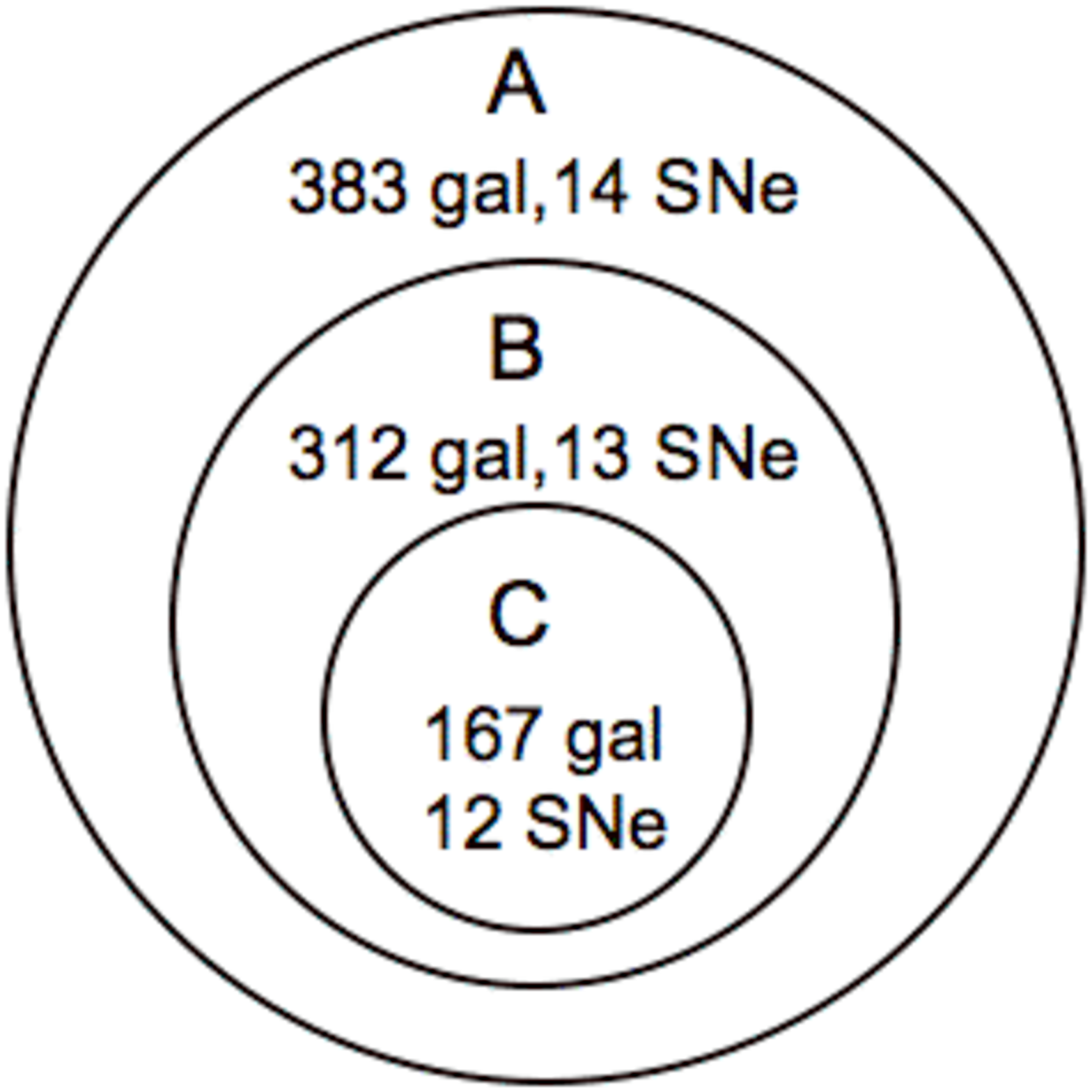}
\end{center}
 \caption{The three galaxy samples selected for our analysis with the numbers of galaxies and  discovered  CC SNe in the last 13 years.} \label{diagr}
\end{figure}

The constant of proportionality  between SFR and luminosity can be derived by assuming an IMF and a stellar evolution model which provides lifetimes of stars as a function of their masses. 
The emission from longer lived stars that encodes part or all of the past galaxy SFH  dominates  in many wavelength bands and only  the UV stellar  continuum,  the emission of optical nebular recombination and forbidden lines  (in particular \ha\,  and [OII]) and  far infrared  (FIR)  emission can be used as probes of the young massive star population.
 Observations made at these wavelengths sample different aspects of the SF activity and are sensitive to different times scales over which SFR is averaged, i. e.  the time interval over which radiation is emitted by a  massive star in the wavelength band of interest (the continuous SF approximation,   \cite{Kennicutt1998}). 
 
UV luminosity directly probes the bulk of the emission from young massive stars with a timescale of order of 100 Myr but it is affected by  contamination from 
evolved stars (i.e., the ``UV upturn''  \cite{OConnell1999} and references therein) and  it is highly sensitive  to dust extinction.

 However, there is an indirect way to utilise the UV emission as a SF tracer:  the UV photons emitted by hot, short-lived stars  ionize the surrounding gas to form an HII region, where recombination produces spectral emission lines so we can assume  that the massive SF is traced by the ionized gas. 
 The  UV continuum and \ha\,  luminosity probe different mass ranges of the massive stellar population: the early and mid B-type
stars (5-15\,\msun) can produce much of a galaxy UV continuum, but contribute little to the photo-ionisation of HII regions \citep{Kennicutt1998}. 
 Moreover, the massive stars that can produce measurable amounts of ionising photons (stars with M $>$ 10\,\msun) have considerably shorter lifetimes (about 10 Myr)  than massive stars that produce the UV continuum.
Of the Balmer lines, \ha\, is the most directly proportional to the ionising UV stellar spectra, because the weaker lines are much more affected by the equivalent absorption lines produced in stellar atmospheres. 
 This SFR indicator is sensitive to the high end of the IMF,  much more than the UV continuum, to dust extinction and to the possible leakage of 
ionising photons. Moreover, it is also susceptible to the stochastic  formation of high mass stars and may not reliably measure the SFR when the activity is low \citep{Lee2009}.

FIR luminosity is a SFR tracer  complementary to the UV and optical ones if we assume that much of the stellar light from new-born stars is absorbed, reprocessed by dust (since the cross section of the dust peaks in UV) and  emerges in the FIR wavelength region. 
The efficacy of  this SFR diagnostic  depends on the fraction of obscured SF and on the optical depth of the dust in star forming regions. The timescale for FIR emission is set by the time it takes massive stars to remove their surrounding material by radiation pressure, expansion of giant HII regions or SN explosions (about 2 Myr).  However, this SFR  indicator is affected by the contribution to dust  heating by  older stars and AGNs.  
Although the FIR luminosity can provide a reliable measure of the SFR only  in the most obscured circumnuclear starburst, the combination of  the dust-attenuated fluxes  in the UV and \ha\,         with measurements of the dust emission in the FIR in the same galaxy sample  can  provide consistent extinction-corrected SFRs \citep{Kennicutt2009}. 

An alternative and complementary approach to trace the SFR is based on the direct observation of the numbers of CC SNe occurring in a sample of galaxies or in a given volume.  
The CC SN rate  (\rcc) is  given, following the formalism by \citet{Blanc2008},  by :
\begin{equation}
\rcc(t) =  \int_{\tau_{min}}^{min(t,\tau_{max})} k(t- \tau) \acc(t- \tau) \fcc(\tau) \psi(t- \tau) d{\tau}
\end{equation}
where t is the time elapsed since the beginning of  SF in the galaxy
under analysis, $\psi$ is the SFR, $k(t-\tau)$ is the number of stars
per unit mass of the stellar generation born at epoch $(t-\tau)$,
\acc$(t-\tau)$ is the number fraction of stars from this stellar
generation that end up as CC SNe,  \fcc\,  is the distribution function of the time intervals between the formation of the progenitor and the SN explosion (delay times) and $\tau_{min}$ and $\tau_{max}$ are respectively the minimum and maximum possible delay times. 
The factor $ k$  is given by:
\begin{equation}
k(\tau)= \frac{\int_{\ml}^{\mup} \phi(m,\tau) dm}{\int_{\ml}^{\mup} m\phi(m,\tau) dm}
\end{equation}
where $\phi(m,\tau)$ is the IMF and \ml--\mup\, is the mass range of the IMF. 
This factor can change if the  IMF evolves with time.

The factor \acc\, can be expressed by:
\begin{equation}
\acc(\tau)=\pcc(\tau) \frac{\int_{\mlcc}^{\mucc} \phi(m,\tau) dm}{\int_{\ml}^{\mup} \phi(m,\tau) dm}
\end{equation}
where \pcc\,    is the probability that a star with suitable mass (i.e., in the range \mucc--\mlcc) to become a  CC SN actually does it.  This probability depends on SN progenitor models and on stellar evolution assumptions.  
The factor \acc\, can vary with galaxy evolution, for example due to the effects of higher metallicities and$/$or the possible evolution of IMF.  
We assume that all stars with suitable mass (\mlcc-- \mucc) become CC SNe and  \pcc($\tau$)=1.
In the following we also assume that $k$ and \acc\, do not vary with time,
that the delay time for CC SN ($\sim 3-20$ Myr) is negligible and  that the SFR has remained constant over this timescale obtaining a direct relation between CC SN rate and SFR:
\begin{equation}
\rcc(t) =  K_{CC} \times  \psi(t)
\end{equation}
where the scaling factor between CC SN rate and SFR is given by  the number  fraction of stars per unit mass that produce CC SNe :  
\begin{equation}
K_{CC}=  \frac{\int_{\mlcc}^{\mucc} \phi(m) dm}{\int_{\ml}^{\mup} m \phi(m) dm} .
\end{equation}
The estimate of the \rcc\, in a galaxy sample needs a systematic SN
search with a known surveillance time for each galaxy, i.e. the control time.
The control time of a single observation of a given galaxy for a given
SN type is defined as the total period of time during which the SN is
bright enough to be detected, while observing that galaxy \citep{Zwicky1938}.
How long a SN is observable in a given galaxy depends on the SN light
curve,  host galaxy distance and extinction and on several
characteristics of the SN search, such as the limiting magnitude. To
determine the total control time of a SN search, we need information
on the distribution in time of the single observations of each galaxy
and to combine appropriately the control time of each single
observation. Another possible approach is to collect as many SNe as
possible and the define the galaxy sample from which they emerged (the
method of the fiducial sample, e.g. \citet{Tammann1977}). In this case
the galaxies without SNe enter the sample only according to some
selection criteria (e.g.  if contained in a given volume).

\begin{figure*}
\begin{center}
$
\begin{array}{c@{\hspace{.1in}}c@{\hspace{.1in}}c}
\includegraphics[width=6cm]{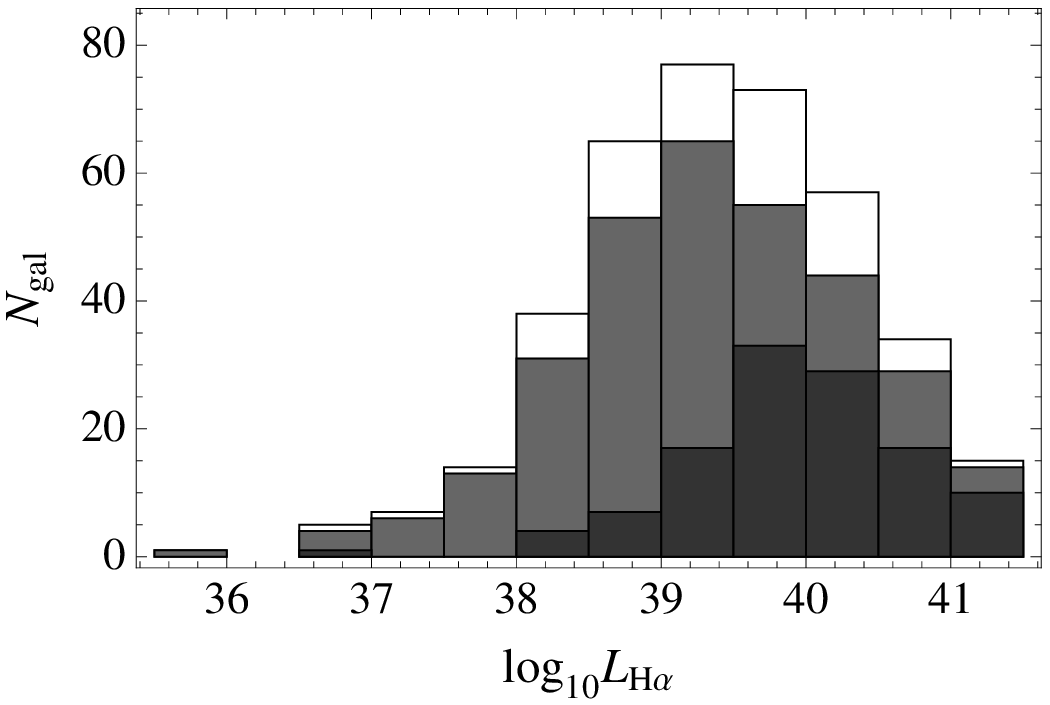} &
\includegraphics[width=6cm]{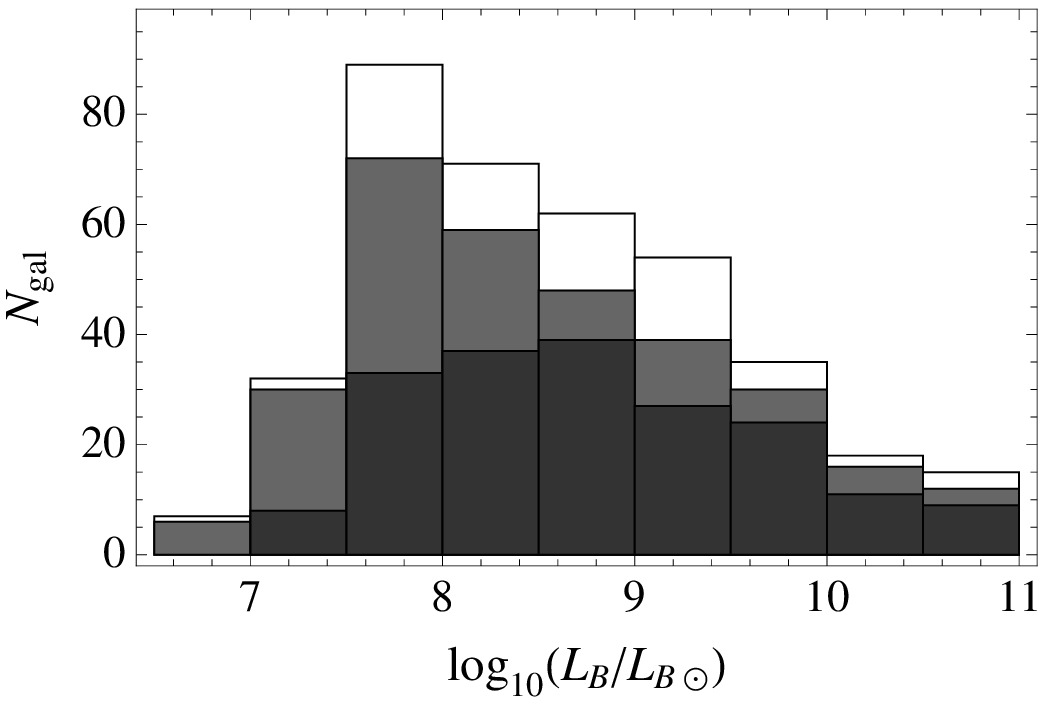}
\end{array}
$
\end{center}
\caption{Distribution of \lha (left panel) and \lb (right panel), corrected for extinction, in the sample A (empty histogram), sample B (gray histogram) and sample C (black histogram).}
\label{distlblha}
\end{figure*}

\section{Galaxy sample}\label{gal}
Our galaxy sample is based upon the catalogue of the ``11 Mpc  \ha\,   and Ultraviolet Galaxy Survey'' (11HUGS). 
11HUGS was designed to provide a census of SFR in the Local Volume, to characterise the population of the star forming  galaxies  and to constrain  the temporal behaviour of the SF in low mass galaxies. 
The design of the 11HUGS survey, its completeness properties, the observations, the data processing and the characteristics of the galaxy sample  are  described  in \citet{Kennicutt2008} and \citet{Lee2010}.

 A distance-limit of  11 Mpc  was adopted to simultaneously obtain a sample that  is statistically significant and nearly complete.
Direct stellar distances are  available  for most galaxies within
$\sim 5$\,Mpc  while distances for other galaxies  are obtained using
the galaxy radial velocity  corrected according to the Local Group
flow model provided by \cite{Karachentsev1996} and the Hubble constant. 
The galaxy selection consists of two steps:  the ``primary" sample (261 galaxies) has  limits on apparent magnitude ($B \le 15$\,mag), Galactic latitude ($|b| \ge 20\degree$)  and Third Reference Catalogue of Bright Galaxies (RC3) type (T$\ge 0$).  The ``secondary" sample  includes additional 175 galaxies which are either below the magnitude and Galactic latitude limits or  lenticular types.  The primary sample aims to be as complete as possible in its inclusion of known nearby star-forming galaxies while the overall sample  is complete to  $M_{B} \le -15$\,mag and $M_{HI} > 2 \times 10^{8}$\,\msun\,   for  $|b| > 20\degree$ at the edge of the 11 Mpc volume \citep{Kennicutt2008}.  Over the 80\% of the sample are dwarf galaxies and  low surface brightness systems with SFRs lower than that of  the Large Magellanic Cloud.

The \ha\,   observations were obtained with the Bok 2.3\,m telescope on the Steward Observatory, the Lennon  1.8\,m Vatican Advanced Technology Telescope, the 0.9\,m telescope at Cerro Tololo Interamerican Observatory \citep{Kennicutt2008}.
The \textsl{GALEX}   UV imaging primarily targeted the $|b| > 30\degree$, $B \le 15.5$\,mag subset of galaxies.  The more restrictive  latitude limit was imposed to avoid excessive Galactic extinction  and fields with bright foreground stars.  Deep, single orbit  imaging  in the far-UV (FUV) and near-UV (NUV) bands was obtained for each galaxy  following the strategy of the  \textsl{GALEX} Nearby Galaxy Survey \citep{GildePaz2007}. 
\cite{Lee2010}  provide  full details on  \textsl{GALEX} observations and photometry  for 390  galaxies: 256 have  $|b| > 30\degree$ and $B \le 15.5$\,mag, 120 have lower latitude and fainter magnitude and have been observed by other programmes  while 27 galaxies  are not included in the 11 HUGS sample.
Fig 1 in  \cite{Lee2010} shows the resultant  \textsl{GALEX} coverage of the overall 11 HUGS sample.

The data from 11HUGS have further been augmented by Spitzer observations through the composite Local Volume Legacy
\footnote{http://www.ast.cam.ac.uk/research/lvls/} 
(LVL) program and  data from the Two Micron All Sky Survey (2MASS) obtained  at 1.25, 1.65,
and 2.17 $\mu$m. 
The sample of LVL  (258 galaxies) consists of two tiers: the inner one includes 69 early and late type galaxies  within 3.5 Mpc that lie outside the Local group for which also Hubble Space Telescope observations exist  from  the ACS Nearby Galaxy Survey Treasury program  and the outer one  include a subset of 11 HUGS primary sample with more stringent  limits on Galactic latitude.  The observational strategy, data processing and photometry measurements are detailed in \citet{Dale2009}. 
\textsl{Spitzer} MIR (IRAC)  and  FIR  (MIPS) data  have been obtained for 180 galaxies and the globally integrated 0.15--160 $\mu$m spectral energy distribution is obtained  from \textsl{GALEX}, 2MASS, IRAS and Spitzer data.
Differences in the selection of \textsl{GALEX} and \textsl{Spitzer} samples are due to different adopted distances for some galaxies (see \citet{Lee2010} for more details).

 For our analysis we considered  three different samples:  383  galaxies   (88\% of  11HUGS sample)  with measured flux in the \ha\,   (sample A\footnote{Few  galaxies  edge-on (with  \ha\, extinction estimates under-estimated), or  with very high Galactic foreground extinction (with uncertain corrections especially in the FUV) have been removed from our galaxy samples (UGC 2847, NGC 5055, NGC 5195, NGC 6744).}
),  312 galaxies (71\%) with both measured flux in the  \ha\,   and FUV (sample B) and  167 galaxies  (38\%) with measured flux in \ha\,,  FUV and TIR  (sample C, see Fig.~\ref{diagr}).

\subsection{Galaxy sample A : H$\alpha$  luminosities}
Integrated \ha\, luminosity (\lha)  are taken from \citet{Kennicutt2008} after applying the following corrections:
\begin{itemize}
\item emission of the [NII] ($\lambda\lambda 6548,6583$) satellite forbidden lines;
\item  underlying stellar absorption by subtracting a scaled $R$ band image  from the narrow band image;
\item Galactic foreground extinction  exploiting the relationship between colour excess and extinction ($A_{\rm{H}\alpha}=2.5 \times E(B-V)$\,mag)  by using  values based on the maps of \citet{Schlegel1998} and  the \citet{Cardelli1989} extinction law with $R_{V}=3.1$.   
\end{itemize}

No correction of \ha\, fluxes for internal extinction  was applied  in the published catalogue.
For about $20\% $ of the sample  it was possible to estimate the internal dust extinction via the Balmer decrement,  since spectroscopic measurements of \ha/\hb\,  from the literature are available. We assumed  a case B recombination ratio and the \citet{Cardelli1989} extinction law with $R_V=3.1$ and used the following relation between  $A_{\rm{H}\alpha}$ and \ha/\hb \, flux ratio:
\begin{equation}
 A_{\rm{H}\alpha}  =5.91  \log  \frac{f_{{\rm{H}\alpha}}}{f_{{\rm{H}\beta}}}  -2.70
\end{equation}

For the galaxies without measurements of the Balmer decrement we adopted an empirical  correction scaling with parent galaxy luminosity  following the algorithm of \citet{Lee2009}:
\begin{eqnarray}\label{eqLee}
 A_{\rm{H}\alpha} &= 0.10                                                              &   if   M_{B} > -14.5  \nonumber \\
A_{\rm{H}\alpha}  & =  1.971 +0.323 M_{B}  + 0.0134 M_{B} ^{2}  & if  M_{B}  \le -14.5 \nonumber
\end{eqnarray}

 The SFRs  have been  estimated by adopting the  conversion factor  by \cite{Kennicutt1998} :
\begin{equation}
SFR(\msun\,  yr^{-1}) =7.9 \times 10^{-42} \lha (erg s^{-1})  
\end{equation}
that assumes  a Salpeter IMF in the mass range 0.1--100\,\msun,  solar metallicity and a constant SFR  for at least the past $\sim 10$\,Myr.

The ratio of  \ha\,  flux to the underlying continuum intensity, expressed as  an integrated  equivalent width (\eha), has also been measured for 243 galaxies.  
\eha\,  is an indicator of the ratio  of the current SFR  to the total stellar mass (i.e.  the specific SFR) that is closely related to the so called   stellar birth rate parameter, defined as the ratio of the current SFR to the past averaged SFR. The typical \eha\,  ranges from zero for early type galaxies up to 20--50 $\AA$ for late type galaxies and have values as high as 150 $\AA$ for some irregular and unusually active galaxies \citep{Lee2007}.

\subsection{Galaxy sample B: H$\alpha$ and FUV luminosities}

The procedure used to perform FUV ($\sim$1500$ \AA$) and NUV ($\sim$2200$\AA$)  photometry  and  to measure  the FUV luminosity (\lfuv) is detailed in \cite{Lee2010}.
To determine the
asymptotic magnitudes, the growth curve in each GALEX  band is computed while the aperture fluxes are measured within the outermost elliptical annulus where both FUV
and NUV surface photometry can be performed. This annulus has been defined as the one beyond
which either the flux error becomes larger than 0.8 mag or where the intensity falls below
that of the sky background in both FUV and NUV bands.
The  fluxes have been corrected for  Galactic reddening by using  the relationship $A_{\rm{FUV}}=7.9 \times E(B-V)$,  adopting  $E(B-V)$ values based on the maps of \citet{Schlegel1998} and  the \citet{Cardelli1989} extinction law with $R_{V}=3.1$.
When TIR data are available the correction  for internal extinction is obtained by using the mapping between $A_{\rm{FUV}}$ and the total infrared to UV (TIR/FUV) flux ratio  given by \citet{Buat2005}:
\begin{equation}
A_{\rm{FUV}}=-0.0333 x^{3} + 0.3522 x^{2} + 1.1960 x +0.4967
\end{equation}
where x=log(TIR/FUV).
\citet{Lee2009} compared the \ha\,  and FUV attenuation  finding a good correlation with a slope $A_{\rm{FUV}}/A_{\rm{H}\alpha} =1.8$ that is the expected value for the Calzetti obscuration curve and differential extinction law \citep{Calzetti2001}.
This agreement provides some assurance that the extinction corrections estimated by \citet{Lee2009}  are reasonable  and generally consistent.
When TIR data are not available or when the  equation  gives a negative correction,  the  $A_{\rm{FUV}}$ was obtained scaling the computed $A_{\rm{H}\alpha}$ by a factor 1.8  \citep{Lee2009}.

The SFRs in the sample B have been  estimated by adopting the  conversion factors  by \cite{Kennicutt1998}  that assumes  a Salpeter IMF in the mass range 0.1--100\,\msun\,, solar metallicity  and a constant SFR for at least the past $\sim100$\,Myr:
\begin{equation}
SFR (\msun\, yr^{-1}) =1.4 \times 10^{-28} \lfuv  (erg s^{-1} Hz^{-1}) .
\end{equation}

\subsection{Galaxy sample C: H$\alpha$,  FUV  and TIR luminosities}\label{galsamplec}
In the sample C,  in addition to  \lha\,  and \lfuv,  the total IR luminosity (\ltir)  is obtained  combining  \textsl{Spitzer} MIR and  FIR  fluxes with with 2MASS NIR data  \citep{Dale2009} with the exception of NGC 628, NGC 1058 and NGC 6949 for which data are from  C. Hao  (private communication).
Elliptical apertures were based on capturing all the galaxy emission visible for all infrared images while for a subset of  about 40 galaxies, the infrared-based apertures were slightly enlarged to capture extended
UV emission. For a given galaxy, in most cases the same aperture was used for extracting all infrared flux densities. 
2MASS  fluxes have been extracted  for the
vast majority of the LVL sample using the same apertures and foreground star removals used
to determine IRAC and MIPS fluxes \citep{Dale2009}.
\ltir\, has been used  to obtain reliable extinction corrected SFRs according to the prescription of \citet{Kennicutt2009}:
\begin{equation}
SFR(\msun\, yr^{-1}) =7.9 \times 10^{-42} (\lha+0.0024\ltir) (erg s^{-1}).
\end{equation}
\ltir\, could be also used  to estimate the total SFR by adopting the  conversion factors  by \cite{Kennicutt1998}  that assumes  a Salpeter IMF in the mass range 0.1--100\,\msun\,, solar metallicity  and for continuous bursts of age  10-100\,Myr:
\begin{equation}
SFR (\msun\, yr^{-1}) =4.5 \times 10^{-44} \ltir  (erg s^{-1}) .
\end{equation}
where \ltir\,  refers to the  IR luminosity  integrated over the full-IR spectrum (8--1000 $\mu$m).
We have to stress that this relation applies only to starbursts with age less than $10^{8}$ years, where the approximations applied by  \cite{Kennicutt1998} are valid. In more normal star-forming galaxies  the relation between \ltir\, and SFR is more complicated since the  IR emission  is  still dominated  by dust heated by  the currently star-forming populations but the contribution from evolved stellar populations  could be non-negligible  \citep{Kennicutt1998}.  Moreover if the galaxies are not completely obscured in the UV, part of UV emission emerges from the galaxy leading to an underestimate of SFR based on \ltir\, \citep{Kennicutt1998}.
Estimating  the contamination of evolved populations and the fraction of unabsorbed UV photons  is a challenge so we did not  adopt the SFR based on \ltir\ in our analysis.

\subsection{$B$ and $K$ band luminosities}
While we now have direct SFR indicators within 11Mpc and a SN rate for
comparison, one would like to put this SN rate into context with
previous results within 60-100Mpc (e.g. Leaman et al. 2011, Cappellaro et
al. 1999). Complete and direct SFR measurements  (from $H\alpha$ or $FUV$) 
are not available in these more distant galaxy samples hence use of 
 the total $B$-band  and $K$-band galaxy luminosities is necessary. 
Trasitionally,  SN rates have been normalised to the total luminosity
($B-$band) or to the total mass.

 For each galaxy in the samples A, B and C we  determined the $B$ band luminosity (\lb)  from the observed magnitude and distance adopting M$_{B,\odot}=+5.48$\,mag.
 We correct \lb\,  for foreground reddening assuming $A_{B} =1.64 \times  A_{\rm{H}\alpha}$\,mag  and for internal reddening assuming $A_{B}=0.72  \times A_{\rm{H}\alpha}$ \,mag to take into account differential reddening between gas and stars  ($E(B-V)_{\rm stars}=0.44 \times E(B-V)_{\rm gas}$) as discussed in \citet{Calzetti2001}.
 The number of galaxies, the total  luminosity in different bands and  total SFR for the three samples are summarised in Table~\ref{galsample} while the distributions of  \lha, \lb\,  for the three samples  are  illustrated in  Fig.~\ref{distlblha}.  Luminosity distributions for samples A and B are quite similar, whereas group C subsamples higher luminosity galaxies.

  For each galaxy in the sample C  we also  determined the $K$ band luminosity (\lk)  from the observed 2MASS flux \citep{Dale2009} and distance adopting M$_{K,\odot}=+3.28$\,mag.
  The galaxy mass  can be also estimated in this sample with the method developed by \cite{Bell2001} and  based on the use of the $K$ luminosity and $B-K$ colour  which is an indicator  of the mean age of the stellar population in a galaxy:
\begin{equation}
 log( \frac{M / \lk}{\msun / \lksun}) = 0.22 (B-K) - 0.59.
\end{equation}
This relation has been obtained by adopting the values from Table 4 in \cite{Bell2001} and a Salpeter IMF.
Obviously, this method gives a rough estimate of the mass but it can be applied to large samples of galaxies
with data available for a limited number of filters.  A similar
equation has been adopted by \cite{Mannucci2005} and \cite{Li2010}
assuming a ``diet'' 
Salpeter to normalise the CC SN rates per unit mass in  a  larger volume.

\begin{table}
\caption{ The number of galaxies and CC SNe discovered in the last 13 years, the total  \lha,  \lb\,  and SFR for the three samples  we have analysed.  The SFRs have been estimated in the three samples by using equation 10, 11, 12,  respectively.}\label{galsample}
\begin{tabular}{lrrr}
\hline\hline
Parameter              &                                 A                               &   B                         &              C\\
\hline
 N$_\mathrm{gal}$  &                                 383     &                         312     &                               167 \\
 N$_\mathrm{CC}$             &                   14                            &       13                       &                       12\\
 \hline
$\mathrm{L}_{\rm{H}\alpha}$   (10$^{43}$\,erg s$^{-1}$ )      &   $1.1\pm 0.05$  &       $ 0.9\pm 0.05$                       &  $0.7\pm0.04$\\
 \hline
$\lb$    (10$^{10}$ \lbsun)     &   $140\pm7$  &          $123\pm7$                   &  $85\pm 5$ \\
$\lk$  (10$^{10}$ \lksun) &         ...               &           ...              &   $154 \pm 9 $\\
M  (10$^{10}$ \msun) &         ...               &           ...              &   $187\pm 11$\\
\hline
$\mathrm{SFR}_{\rm{H\alpha}}$ (\msun yr$^{-1}$ ) &    $ 87\pm 4$        &        $78\pm 4$                     & $58 \pm 4$ \\
$\mathrm{SFR}_{\rm{FUV}}$  (\msun yr$^{-1}$) &     ...      &            $123 \pm 8$                  & $ 94 \pm  6$ \\
$\mathrm{SFR}_{\rm{H\alpha+TIR}}$  (\msun yr$^{-1}$)   &      ...     &         ...                     & $ 62\pm  4$ \\
\hline
R$_\mathrm{CC}$ (yr$^{-1}$) &     $1.1_{-0.3}^{+0.4}$    &      $1_{-0.3}^{+0.4}$         &         $0.9_{-0.3}^{+0.4}$                         \\
R$_\mathrm{CC}$   (SNu)     &      $0.7_{-0.3}^{+0.4}  $              &      $ 0.8_{-0.2}^{+0.3} $                       &   $1.1_{-0.3}^{+0.4}$\\
R$_\mathrm{CC}$ (SNuK) &         ...                &      ...                  & $0.6_{-0.2}^{+0.2}  $     \\
R$_\mathrm{CC}$ (SNuM) &        ...                 &      ...                   &  $0.5_{-0.1}^{+0.2}  $   \\
\hline
\end{tabular}
\end{table}

\section{SN sample}\label{sn}
 To estimate the CC SN rate we initially identified SNe known to have occurred in the galaxies of the Sample A from the Asiago SN catalogue\footnote{http://graspa.oapd.inaf.it/} \citep{Barbon2008} from 1885 to 2010:  38 CC SNe  and 10 type Ia SNe  (Table~\ref{clSNe}). 
 SN~2008iz  was discovered in NGC3034 (M82) in the radio \citep{Marchili2010}  and its SN nature was confirmed with
identification of the expanding ring \citep{Brunthaler2010}. The extinction is extremely high towards this event, and it has not been 
detected at optical or IR wavelengths, hence we  leave it out of our analysis since we are considering only the CC SNe discovered in the optical bands. 
SN~2008jb, a type II SN,  was  discovered in archival optical images obtained by
the Catalina Real-time Transient Survey and the All-Sky Automated Survey by \cite{Prieto2011}. This SN was missed by galaxy-targeted SN surveys  and by amateur astronomers mainly because the host galaxy, ESO 302-14 at 9.6\,Mpc, is a low-luminosity dwarf  galaxy that was not included in the catalogs of galaxies that are surveyed for SNe.  We did not consider this SN in our sample but discuss the bias to large star-forming galaxies present in the sample of nearby SNe in the Sect. \ref{ct}.
Additionally there have been discoveries of  7 Luminous Blue Variables (LBVs) in outburst and three optical transients  whose nature is still debated (SN~2008S, NGC 300-2008OT, SN2010da).  A SN origin from a massive star ($>7-8$\,\msun) has been proposed for SN~2008S and NGC 300-2008OT  by a  number of authors \citep{Prieto2008,Thompson2009,Botticella2009,Pumo2009} but is disputed by others who favour an outbursting massive star event \citep{Smith2009,Berger2009,Bond2009,Humphreys2011}. To be conservative, we will not consider these two transients as genuine CC SNe in our
main analysis  but we will include them in our discussion of the detectability of CC SNe, since their faint detection magnitudes (with peak magnitudes $M_{R}\sim-14$\,mag) illustrate
the depth and completeness of nearby SN searches no matter what their nature. 
SN 2010da seems to be a LBV-like outburst of a dust enshrouded massive
star with bluer colours than those of the progenitors of SN 2008S and
NGC 300 OT2008-1. The light curve and spectrum also seem to be different from SN 2008S and NGC 300 OT2008-1 \citep{2010ATel.2632....1K,2010ATel.2636....1E,2010ATel.2637....1C,2010ATel.2639....1I,2010ATel.2640....1B,2010ATel.2660....1P}.
\footnote{There are 10 further events that have not been spectroscopically
classified but are likely SNe of some sort, and an additional faint
transient in NGC 4656 discovered in 2005 of unknown origin
\citep{2005IAUC.8497, 2005IAUC.8498} that we leave it out of our
analysis. However none of these are in our 11Mpc and 13yr SN sample.}

 We restrict our comparison between the CC SN rate and the SFR estimates  to the last 13 yr (1998--2010), assuming a constant and continuous intensity level of surveillance (i.e., a control time of 13 years).  This period is well justified as since 1998 we have witnessed a large increase in the discovery rate of SNe in the Local Universe. This is due to the 
 start of Lick Observatory Supernova Search (LOSS) in 1998 that monitored about 15,000 galaxies with $z<0.05$ for 13 years and discovered about 1000 SNe \citep{Leaman2010,Li2010}  and the high number of amateurs searching SNe in the nearby galaxies who have  been using telescopes of 20-50\,cm and modern CCDs for the last $\sim$15 years.

The  majority of  the CC SNe within 11 Mpc in the last 13 years  (Table~\ref{SNecarat}) have been discovered by amateur astronomers  (60\% of events) with 40\%  coming from the LOSS professional searches.
The distribution of the discovery epoch with respect to the maximum light,  of the  discovery magnitude and the absolute magnitude at maximum light of our SN sample  are illustrated in Fig.~\ref{SNdist}. 

\begin{figure}
\resizebox{\hsize}{!}{\includegraphics{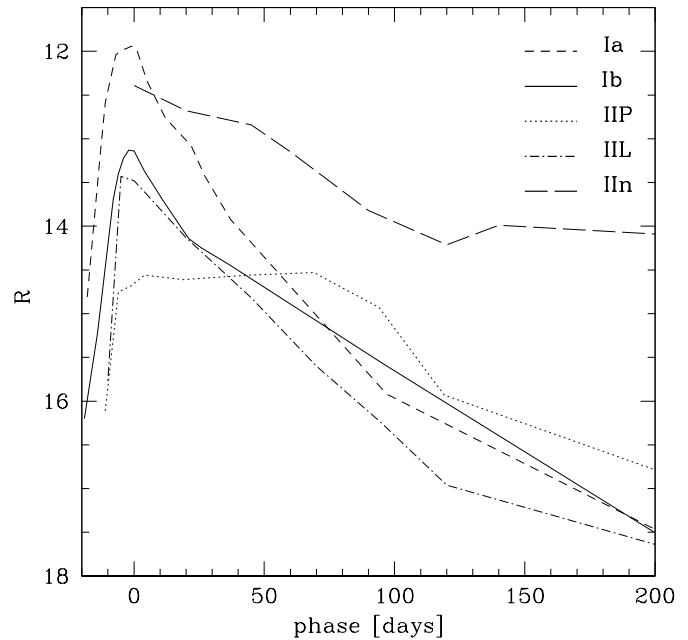}}
 \caption{The observed light curves of  different  unreddened CC SN sub-types at 11 Mpc. The absolute magnitudes in $R$ band at maximum are -19.16\,mag for type Ia SNe,  -17.92\,mag for type Ib SNe,  -16.38\,mag for type IIP SNe,  -17.70\,mag for type IIL SNe and -17.14\,mag for type IIn SNe. Phases are relative to the maximum epochs.} \label{lc}
\end{figure}

\begin{figure}
\begin{center}
$
\begin{array}{c@{\hspace{.1in}}c@{\hspace{.1in}}c@{\hspace{.1in}}c}
\includegraphics[width=6cm]{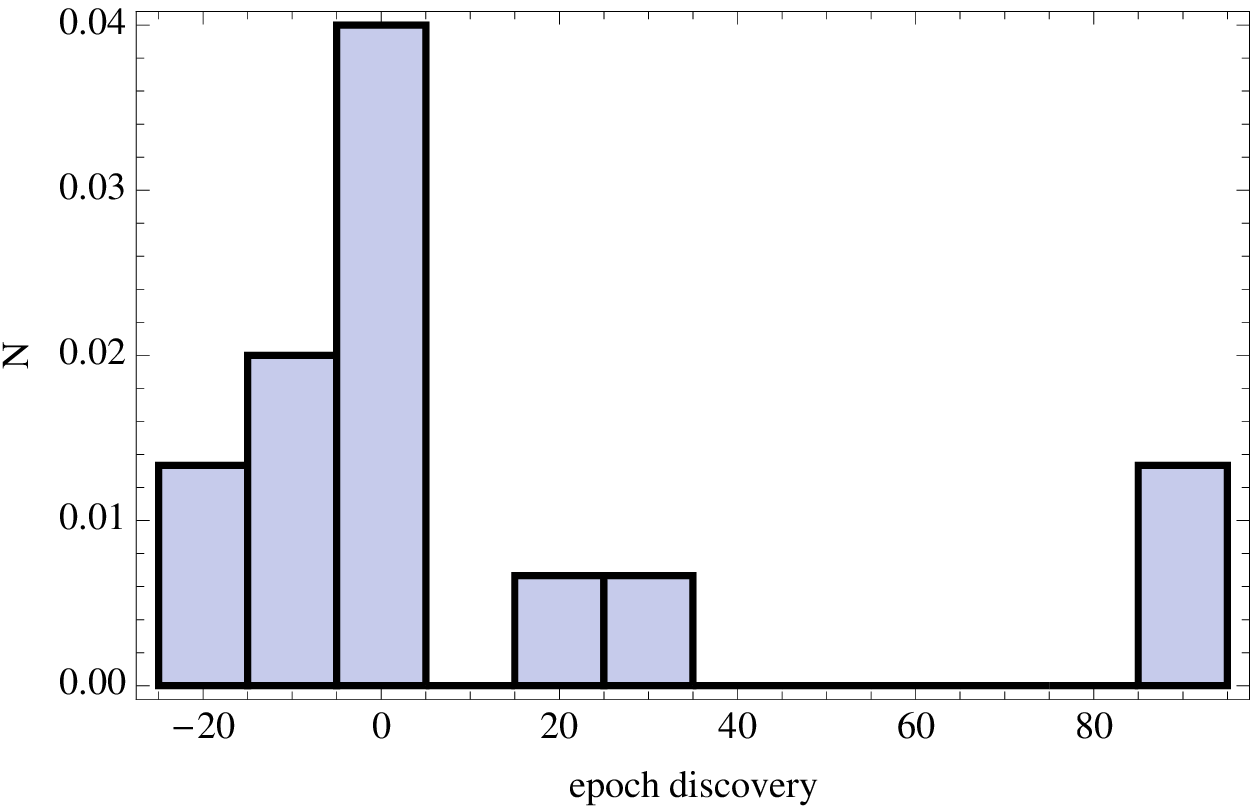}\\
\includegraphics[width=6cm]{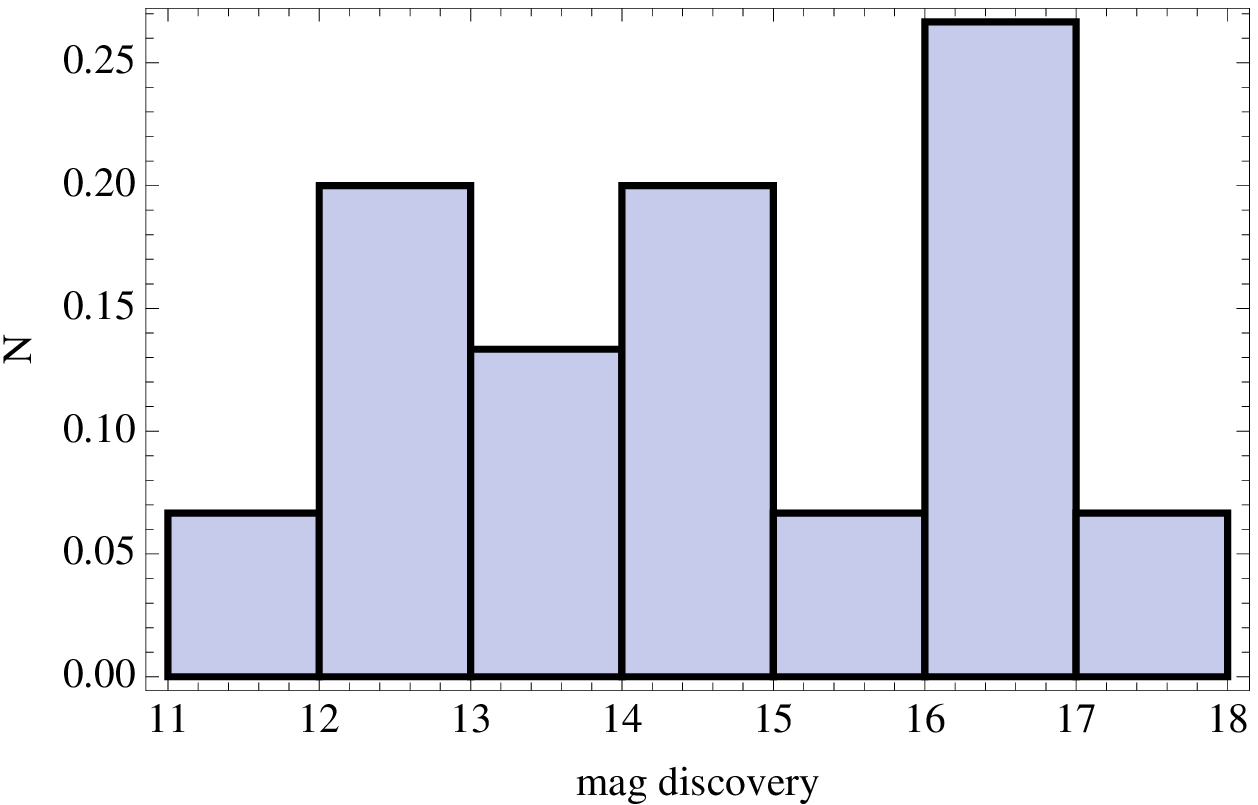} \\
\includegraphics[width=6cm]{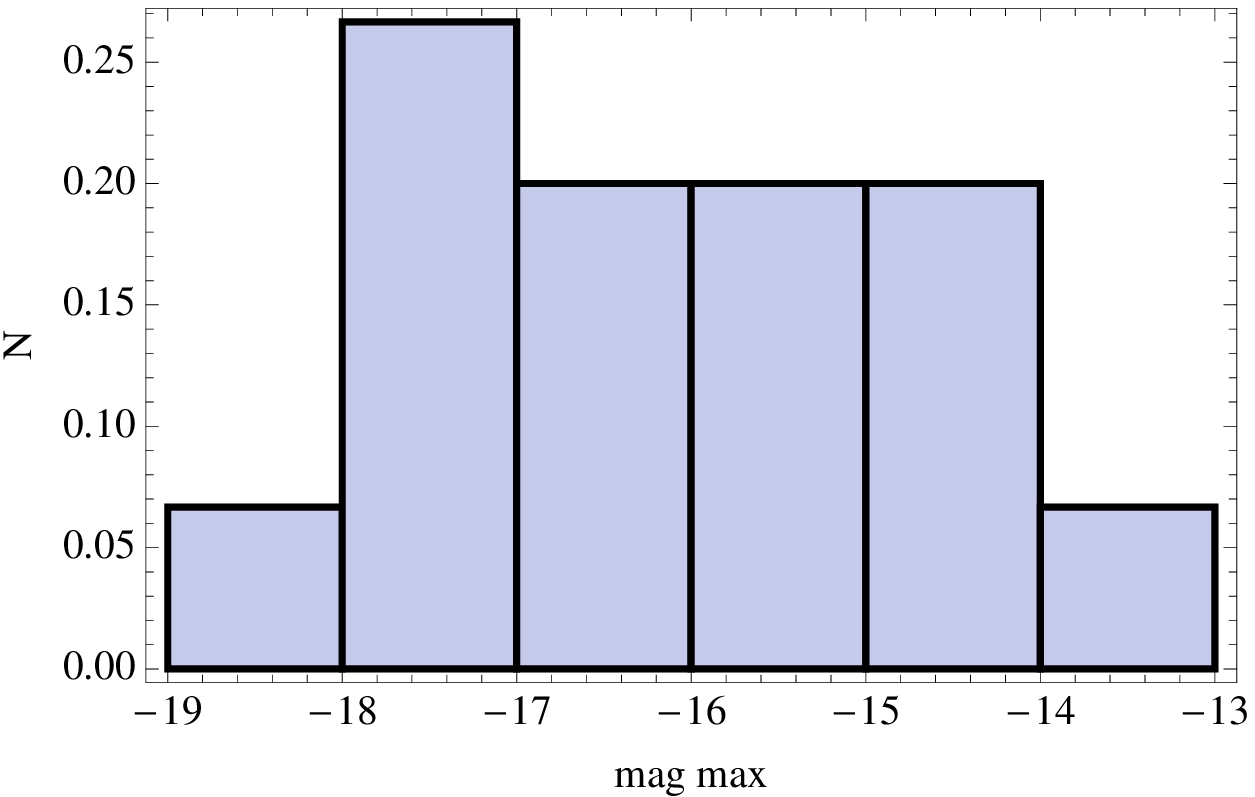}\\
\end{array}
$
\end{center}
\caption{The distribution of the discovery epoch with respect to the maximum light,  of the discovery magnitude and  absolute magnitude at maximum for our SN sample. The values for each CC SN and sources are reported in Table \ref{SNecarat}.}
\label{SNdist}
\end{figure}

At a typical distance modulus of 31\,mag for the most distant galaxies in our sample, the limiting magnitude of $\sim$18-19\,mag in the SN searches results in detections down to $M_R\sim-12$\,mag for unreddended events. 
CC SNe which are not heavily extinguished or intrinsically faint stay above 18\,mag  for  about 200 days  (Fig.~\ref{lc}).  Hence we can be fairly sure that {\em significant} numbers 
have not been missed due to solar conjunction or a lack of searching by the global community. 
One observation every few months is sufficient to ensure that the normal SN population is well surveyed.
Of course, there may be a population of significantly extinguished SNe, or intrinsically faint SNe.  
For a typical IIP, with $M_R\sim-16$\,mag, one might expect the surveys
to be sensitive to SN obscured by about 4\,mag. Indeed two SNe (2002hh and 2004am) have been discovered by LOSS 
\citep{Leaman2010,Li2010}  which were faint but located in quite nearby galaxies. The expected dust obscuration was
confirmed  with extinction estimates of $A_{V} = $5.2 and 3.7\,mag respectively \citep{Pozzo2006,Smartt2009}. It is not implausible to argue that there are more obscured local events
evading detection in the optical bands. 

In our CC SN sample four events (SNe 2002ap, 2005cs, 2008ax, 2008bk) were discovered soon after the explosion (within a few days of shock breakout). Five of the events were discovered either before light-curve maximum (SNe 2007gr, 2008S and NGC300-2008OT), or early in the plateau phase (SNe 2002hh, 2004et). The other four events (SNe 2003gd, 2004am, 2004dj, 2005af) are type IIP SNe discovered during mid-plateau.  Early discoveries of SNe 2003gd and 2004dj were missed simply due to the galaxies being in solar conjunction at SN explosion epoch. All of this supports our view that the A, B and C galaxy samples have been systematically surveyed during the last 13 years, and the CC SN rate is at least reliable enough for a meaningful comparison with the SFR estimates now available. It is of course  a robust lower limit. 
 
Restricting our CC SN sample to that discovered in the last 13 years has another advantage. All of the SNe are spectroscopically classified and the majority have extensive photometric and spectroscopic coverage of their evolution.    We do not subdivide our small sample into different CC SN sub-types since individual bins contain only few objects. \citet{Smartt2009} compiled all SN discoveries in a fixed period (10 years) within a fixed distance (28 Mpc) and estimated the relative frequency of all subtypes (58.7\% type IIP, 2.7\% type IIL, 3.8 \% type IIn, 5.4 \% IIb, 9.8 \% Ib, 19.6\% Ic). In a larger 60\,Mpc volume, the LOSS survey \citep{Li2010} has estimated :
48.2\% type IIP, 6.4\% type IIL, 8.8 \% type IIn, 10.6 \% IIb, 26 \% Ibc.
Our smaller sample has 60\%  type IIP SNe, and the other 40\%  as such it is too small
to attempt any further  meaningful subdivision, but the overall ratios are similar
to the LOSS and 28\,Mpc volumes. 

\begin{table*}
\caption{The SN  type, SN magnitude at the discovery epoch and at maximum light, phase (days) between the discovery epoch and the maximum light for the CC SNe discovered within 11 Mpc in the last 13 years, the host galaxy's name, morphological type, $B$ band absolute magnitude, SFRs (\msun yr$^{-1}$), $B-K$ colour, mass  ($10^{10} \msun$), specific SFR (yr$^{-1}$)and $\eha$.References:(1)~\citet{Nakano2002}; (2)~\citet{Puckett2002}; (3)~\citet{Foley2003};(4)~\citet{Li2002}; (5)~\citet{Pozzo2006}; (6)~\citet{Evans2003}; (7)~\citet{Hendry2005}; (8)~\citet{Singer2004}; (9)~Mattila priv comm; (10)~\citet{Nakano2004}; (11)~\citet{Vinko2006}; (12)~\citet{2004IAUC.8413....2Y}; (13)~\citet{Maguire2010}; (14)~\citet{Jacques2005}; (15)~\citet{M2005}; (16)~\citet{Kloehr2005}; (17)~\citet{Pastorello2006}; (18)~\citet{Madison2007}; (19)~\citet{Hunter2009}; (20)~\citet{Arbour2008}; (21)~\citet{Botticella2009}; (22)~\citet{Mostardi2008}; (23)~\citet{Pastorello2008}; (24)~\citet{Monard2008CBET.1315....1M}; (25)~ Pignata priv comm; (26)~\citet{Monard2008IAUC.8946....1M}; (27)~\citet{Monard2009CBET.1867....1M}.}\label{SNecarat}
\begin{tabular}{lllllllllllllll}
\hline\hline
  SN  & type &mag disc. &mag max. & phase & ref. & gal.& T &  Host M$_{B}$ & SFR$_{H\alpha}$&SFR$_{UV}$ &  B-K & M &sSFR  &$\eha$ \\
  \hline     
2002ap & Ic &  14.5 (V)  &  12.39 (V) & -9&   1& NGC 628   &    5 &   $-19.58$ & 1.3&2&2.3 & 2 & 11&35\\   
2002bu& IIn &    15.5      &   14.77 (R) & -5&  2, 3&   NGC 4242  &  8 &  $-18.18$ & 0.1&0.17& 1.9 &0.2&9 &18 \\
2002hh &IIP  &   16.5     &  15.53 (R) &-4  & 4, 5 &   NGC 6946  &  6  &$ -20.79$ &5.7&9.1& 2&3.6 &26&33 \\
2003gd &IIP  &   13.2    & 13.63 (R) & +90  & 6, 7 & NGC 628  &  5  & $-19.58$& 1.3&2& 2.3& 2 &11&  35\\
2004am& IIP &    17.0   &  $\sim$16 (R) &+90 & 8, 9 &   NGC 3034  &   7 & $ -18.84$ & 1.9 & 	5.6 &3.6 &6.4 &9& 64  \\
2004dj & IIP  &   11.2    &  11.55 (R) & +21&  10, 11 &   NGC 2403  &   6 &  $-18.78$ &0.8 &  1.0& 2.2 &0.6 &17&50\\
2004et &IIP &    12.8    &  12.2 (R)  & -18 &12, 13 &   NGC 6946  &  6 & $ -20.79$&  5.7&9.1& 2&3.6 &26&33 \\
2005af &IIP  &   12.8    & 12.8 (R) & +30 & 14 &    NGC 4945  &6 & $ -19.26$ &	0.9&-- & -- &-- &--&17\\
2005at &Ic   &   14.3    &   14.3  & 0  & 15  &   NGC 6744 & 4 &  $-20.94 $& 3.3 & 12 & --& --&--&15 \\      
2005cs& IIP &    16.3 (V) &  14.50 (V)  & -2   & 16, 17 &  NGC 5194  & 4 &$  -20.63$ & 4.5&  7.6&2.6 & 8.5&9&28\\
2007gr& Ic   &   13.8    &  12.76 & -13 &   18, 19 &  NGC 1058   &  5 & $-18.24$ &0.3 &  0.5& 2&0.3 &14&29\\
2008S &  ... &   16.7 (R) & 16.26 (R)  & -11  &  20, 21  &  NGC 6946  &  6 & $ -20.79$&   5.7&9.1&2&3.6 &26&33  \\         
2008ax& IIb  &   16.1   & 13.38 (r) &  -23  &22, 23 &  NGC 4490 & 7 &  $-19.37$ & 1.9& 	2.5 &2.3 &1.3&18& 66 \\
2008bk &IIP  &  12.6   & 12.5   & &  24, 25 &  NGC 7793   &  7 &$-18.41$ &0.5& 0.7& 2.4&0.5&13&40\\
2008OT & ... &   14.3  &14.3  & 0 &  26 &  NGC 300   & 7  & $-17.84$ & 0.2& 0.3&-- &-- &--&24\\
2009hd & IIP & 17.2&  16 (R)  &   &  27 & NGC 3627&  3 & $ -20.44$ & 2.6& 4.9 & 3&12 &4&19\\
\hline
\end{tabular}
\end{table*}

\begin{figure*}
\begin{center}
$
\begin{array}{c@{\hspace{.1in}}c@{\hspace{.1in}}c}
\includegraphics[width=6cm]{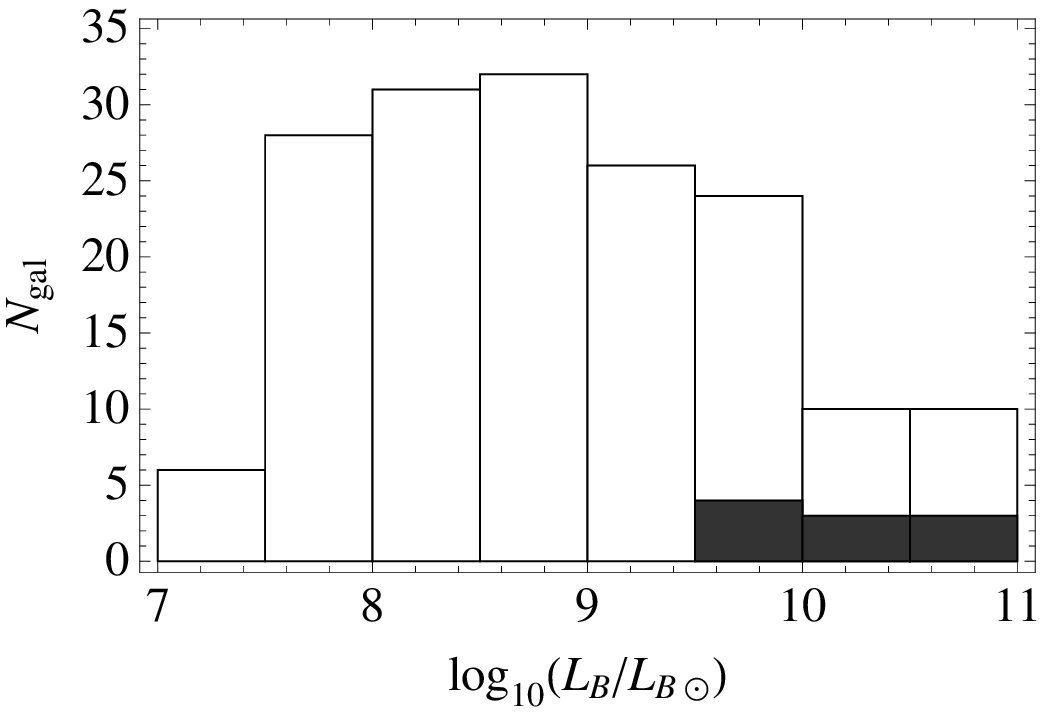} &
\includegraphics[width=6cm]{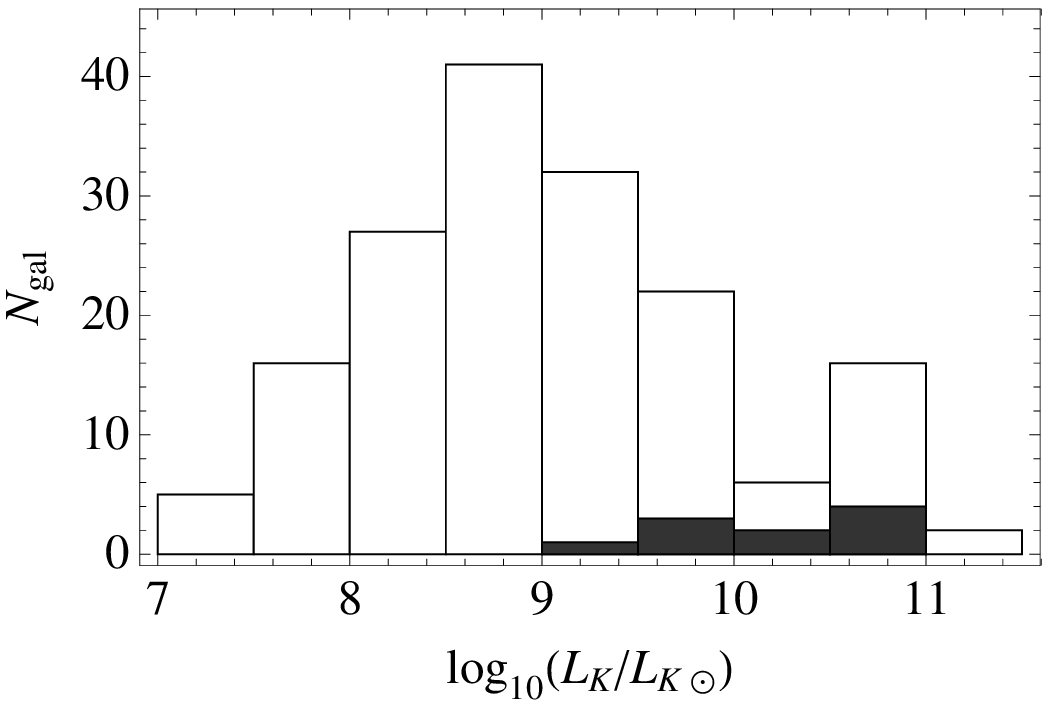}\\
\includegraphics[width=6cm]{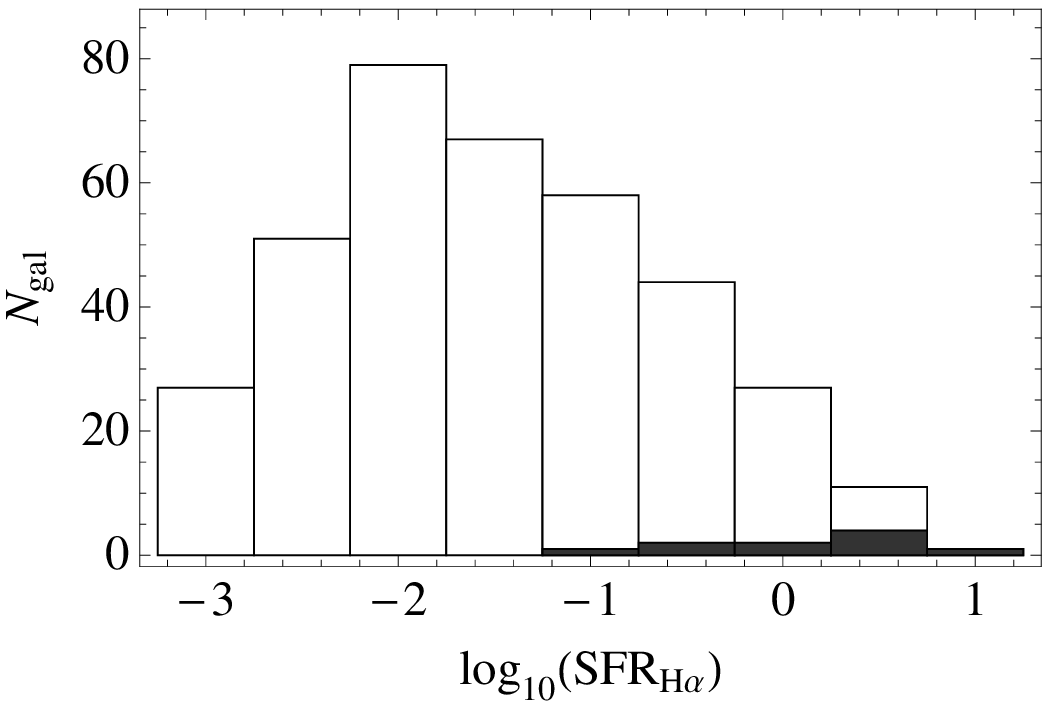} &
\includegraphics[width=6cm]{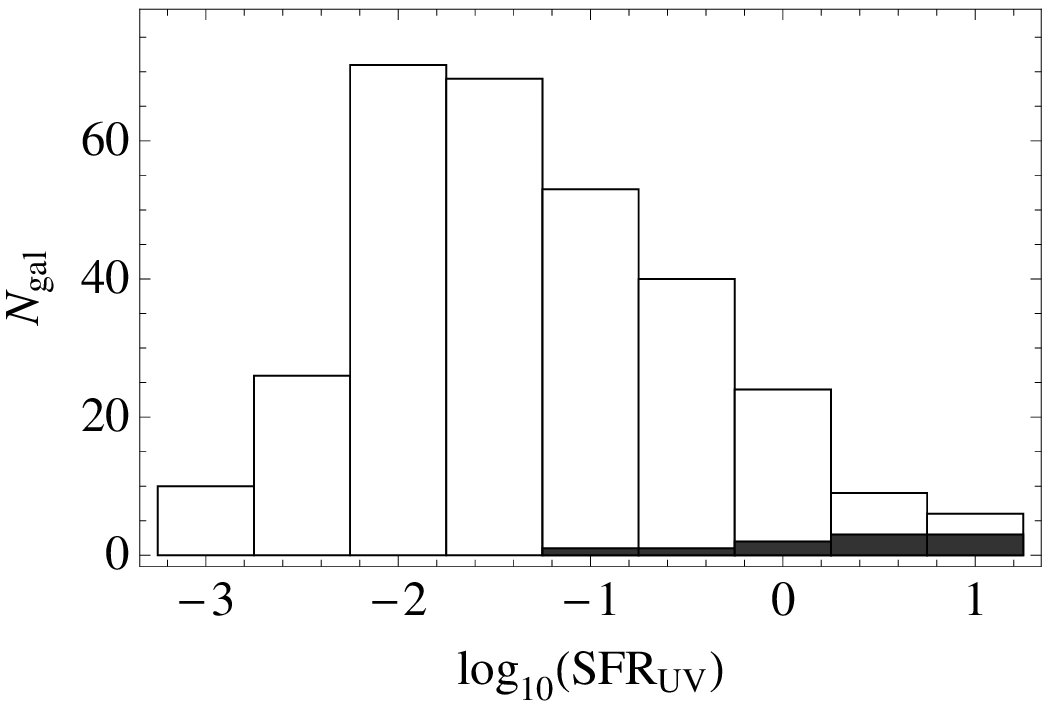}\\
\includegraphics[width=6cm]{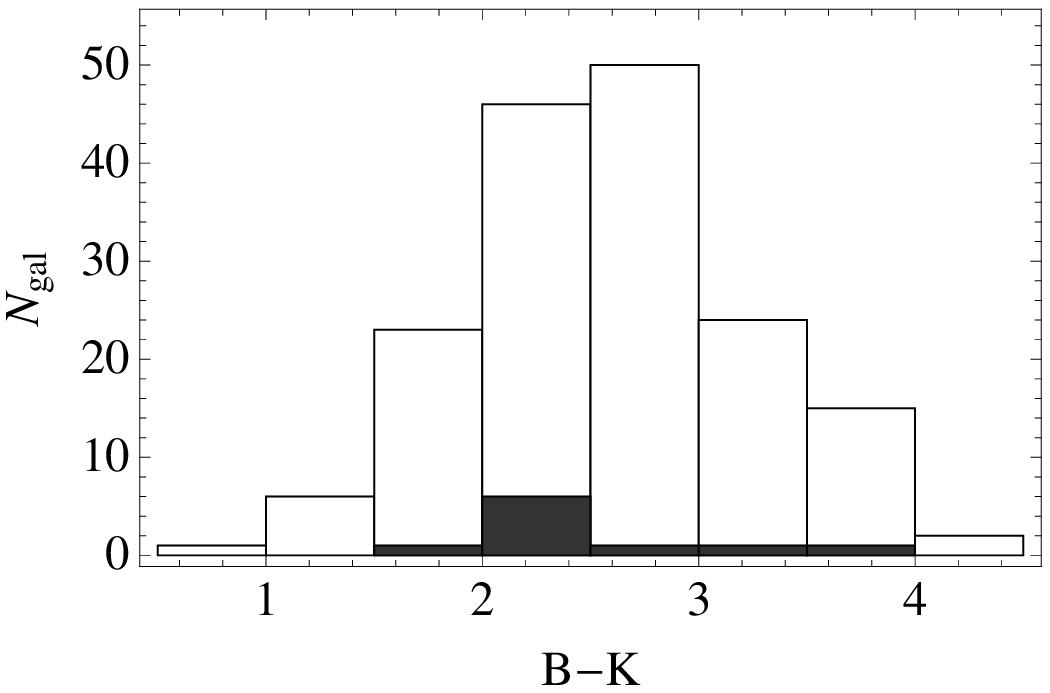}&
\includegraphics[width=6cm]{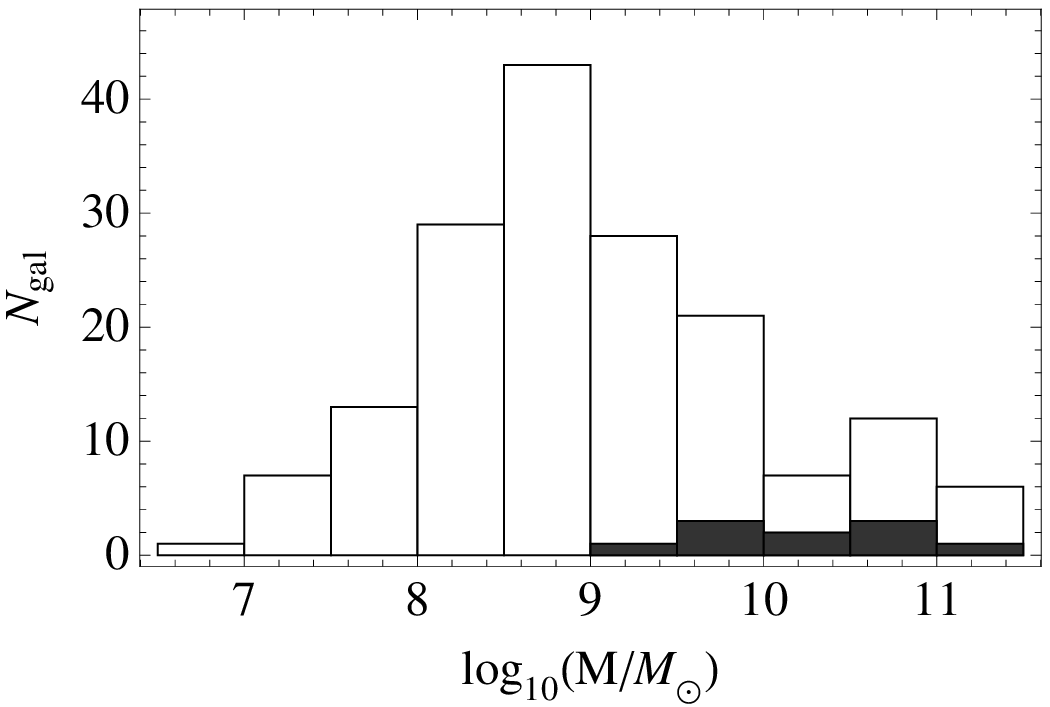}\\
\includegraphics[width=6cm]{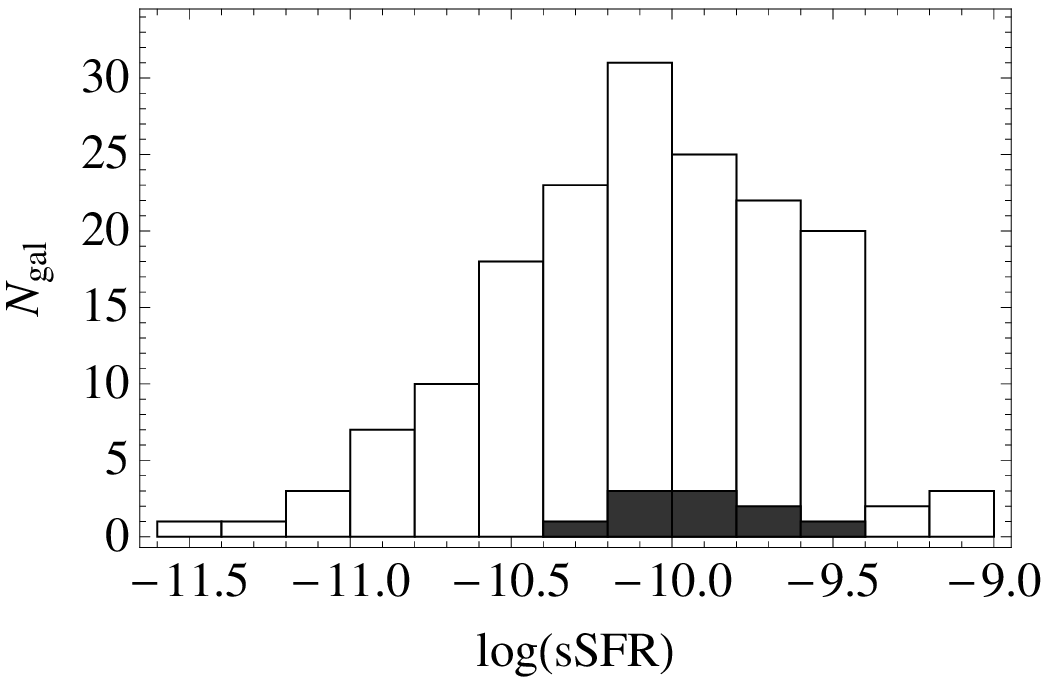}&
\includegraphics[width=6cm]{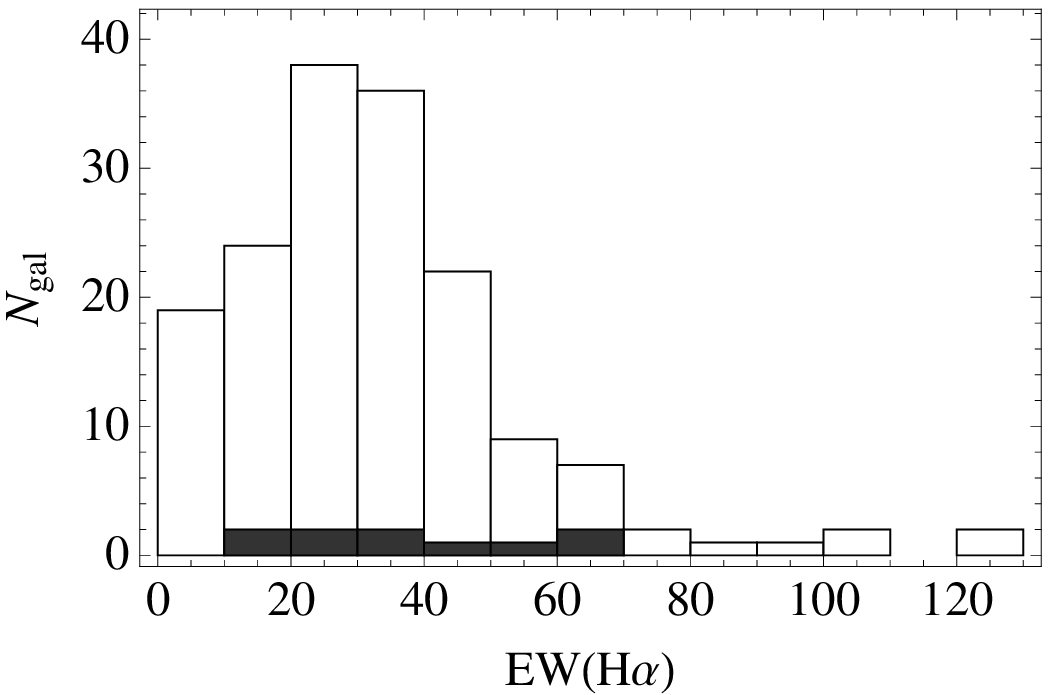}\\
\end{array}
$
\end{center}
\caption{The distribution of the \lb, \lk\, (in \lbsun and \lksun  unit), SFRs, $B-K$, mass, specific SFR and \eha\, for the galaxies  that hosted SNe (full histogram) and all the galaxies (empty histogram) in the sample C.}
\label{fighost2}
\end{figure*}

The distribution of   \lb, \lk ,  SFRs, $B-K$, mass  specific SFR and \eha\, for the galaxies in the sample C that hosted CC SNe  are illustrated in Fig.~\ref{fighost2}. 
The host galaxies have highest SFRs, luminosities and masses, while their distribution in $B-K$, sSFR and \eha\, is more shallow.

\section{Comparison of SFR indicators}\label{compSFR}
The number of SN discoveries within the 11\,Mpc volume makes for an
interesting comparison between  the SFRs obtained from the observed CC SN rate 
and those  based on multi-wavelength flux measurements. 
Each provides an independent measurement which suffers from different uncertainties and biases. 
The  CC SN rate is likely biased towards the brighter SNe and maybe systematically misses a population of  SN explosions
(due to either modest intrinsic brightness or large extinction) so it gives a lower limit for the  current SFR.

Dust extinction is probably the largest source of systematic uncertainty in the direct measurements of SFRs.
Different SFR tracers  are affected  by extinction to different extents:  typical dust attenuation is of order 0--2\,mag in \ha\, and 0--4\,mag in UV continuum \citep{Kennicutt2009}.
The resulting systematic error in the overall SFR measurements is generally removed by applying a statistical correction for dust extinction \citep{Kennicutt1983,Calzetti1994,Calzetti2000} or  by combining observations in UV and \ha\, with those in the IR wavelength range \citep{Kennicutt2009}.

\begin{figure}
\begin{center}
$
\begin{array}{c@{\hspace{.1in}}c@{\hspace{.1in}}c@{\hspace{.1in}}c}
\includegraphics[width=6.2cm]{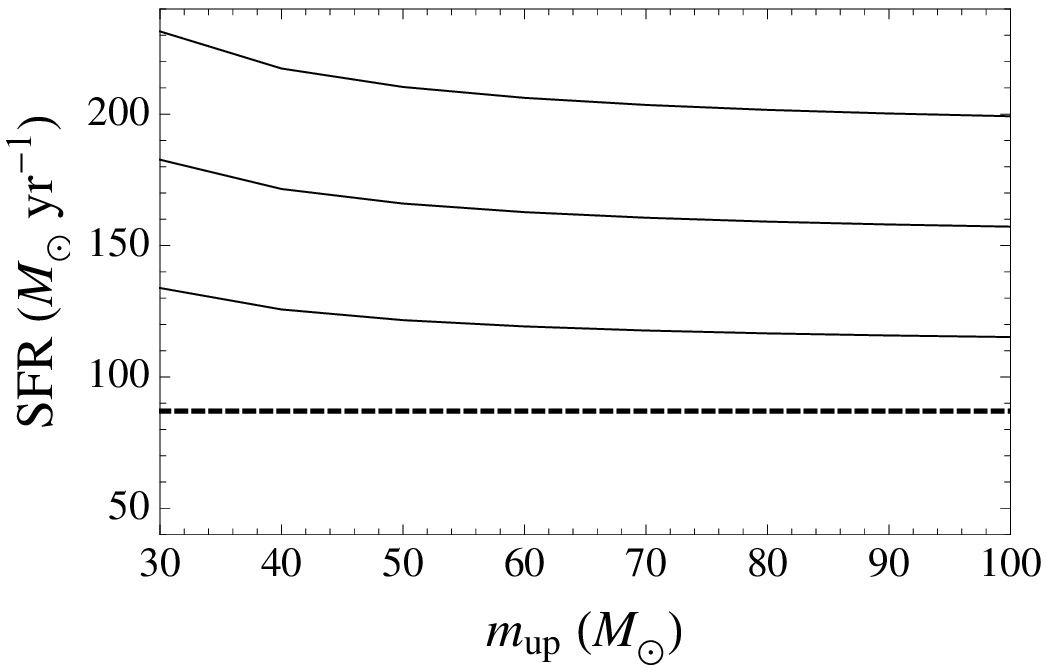} \\
\includegraphics[width=6.2cm]{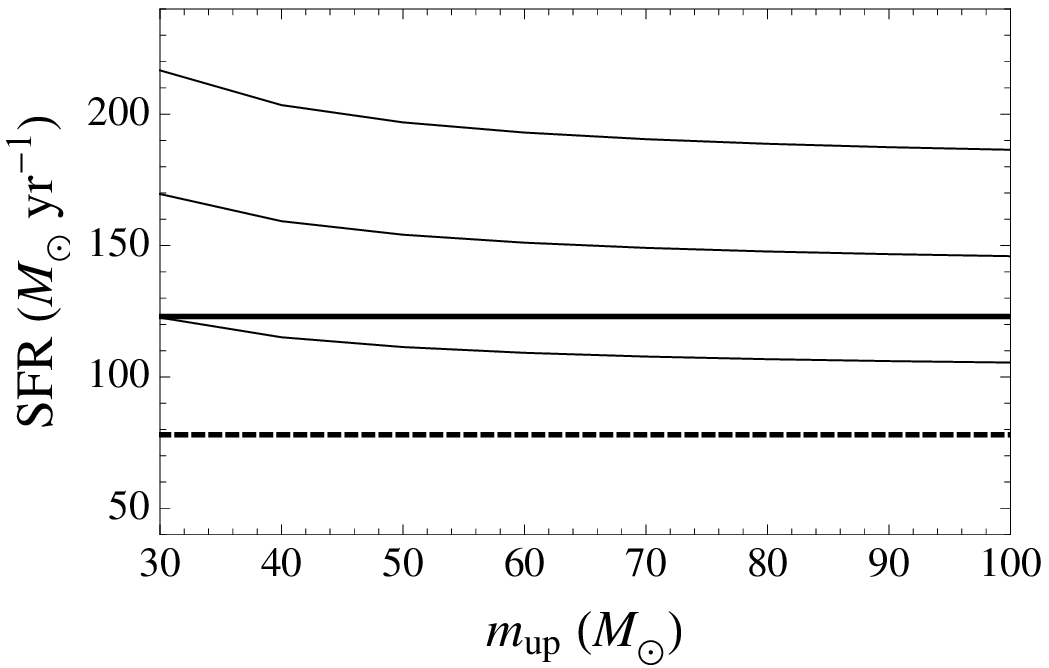}\\
\includegraphics[width=6.2cm]{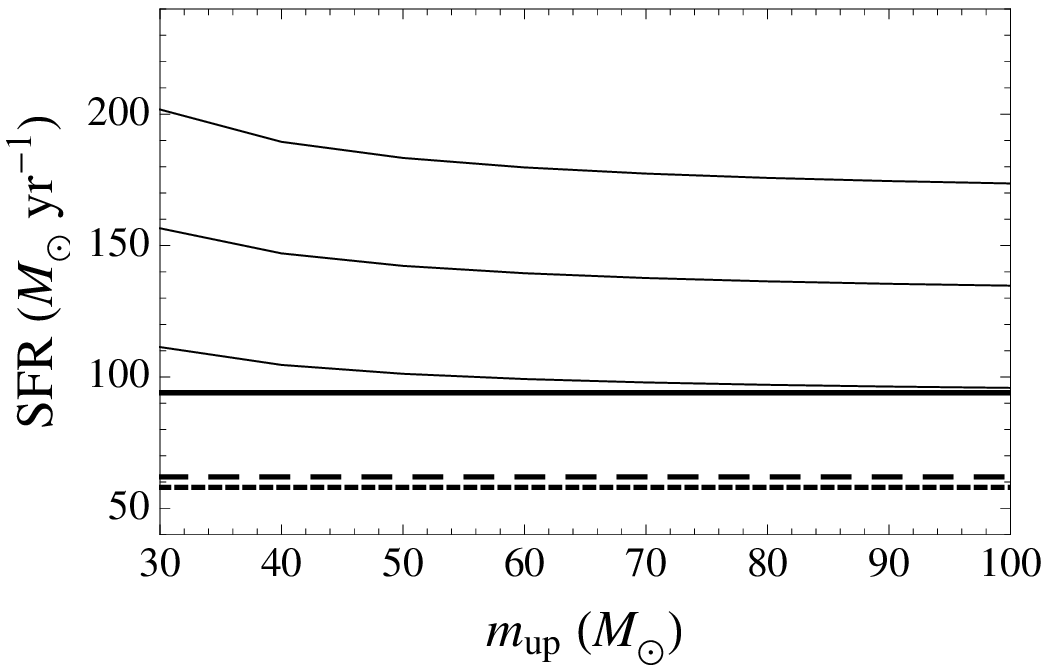}\\
\end{array}
$
\end{center}
\caption{
From the top to the bottom: the SFR expected from the CC SN rate in
the sample A, B and C (thin lines) as a function of the upper mass
limit for CC SN progenitors. We adopted a lower mass for CC SN
progenitors of  8\,\msun\,.  We plot the central value and 1$\sigma$
confidence limits.
The short dashed line indicates the value of SFR
based on \lha\,  (in all panels: sample A, B and C), the short dashed line the SFR based on \lfuv\, (middle and bottom panels: samples B and C) and long dashed line the SFR based on  $\lha\, +  \ltir\,$(bottom panel: sample C).}
\label{sfrcom}
\end{figure}

In order to estimate the SFR from CC SN rate measurements we have to assume the mass range of CC SN progenitors and to correct the rates for the fraction of  the extinguished CC SNe that are missed in optical searches.
 The lower mass limit for CC SN progenitors from direct detections of progenitor stars in high-resolution images
has arrived at a best estimate of $8.5^{+1}_{-1.5}$\,\msun\, \citep{Smartt2009}, which 
is in reasonable agreement with the most massive white dwarf progenitors 
\citep{Williams2007,Williams2009}. This has led   \citet{2009ARA&A..47...63S}
to suggest that the current best estimate from these two methods is  $8\pm1$\,\msun. 
If we assume this value of 8\,\msun\,   the observed CC SN rates in the galaxy samples A, B and C
imply SFRs  are plotted in Fig~\ref{sfrcom}.  The observed CC SN rate is of course only a robust lower limit  since we have not applied any correction for 
undetected SNe. 
The SFR from CC SNe  is 
higher by a factor two compared with those in Sample A and C based on \lha\,  while there is  good agreement with SFR based  on \lfuv\,  that suggests we are 
not missing a large number of CC SNe within 11\,Mpc due to dust
extinction, intrinsically faint magnitudes, or over-estimating the control time. 

The main source of  the difference
between the corrected SFRs  based on \lfuv\, and  \lha\,  is due likely to the attenuation corrections.
A few of the galaxies with the highest SFRs tend to
show especially large discrepancies between \lfuv\, and  \lha\,
derived SFRs, and we suspect that some of
these may arise from spurious causes such as extremely
heavy extinction in edge-on systems (e.g., M82),
very large foreground Galactic extinction (e.g., NGC 6946),
or poorly measured \ha\, fluxes (e.g., NGC 6744).
These galaxies carry disproportionate weight in the
total SFRs for the samples, but even taking them into
account the \lfuv\, based SFRs remain systematically larger.

A small part of the offset comes from the adoption of the \cite{Buat2005} formula for
estimating  \lfuv\, extinction corrections. \citet{Kennicutt2009} compared attenuations derived from
that method with those from  \ha\,+TIR and \ha\,+24 $\mu$m schemes, and found that the former are systematically
larger, by about 0.1--0.2\,mag. 
There is also a more important systematic offset  (30-40\%) in TIR luminosity between MIPS, which
was used for nearly all of our sample, and IRAS, which was
used to calibrate the \cite{Buat2005} relation \cite[for more details see Figures 1-2 of][]{Kennicutt2009}.  
This difference is only important for galaxies with cold
IRAS colours (where basically the IRAS wavelength coverage
is not sufficient to integrate the IR emission reliably).
Unfortunately that colour regime applies to most of our galaxy
sample.

The comparison  between CC SN rate and SFR based on other diagnostics has also been done at larger volumes \citep{Dahlen2004,Botticella2008,Bazin2009,Horiuchi2011} and  points out a discrepancy in the 
opposite direction with respect to the local Universe since the observed CC SN rate is lower  than the predicted one from SFR measurements.
It is interesting to note that this discrepancy (about a factor two) is constant  in a large range of redshift \citep{Botticella2008,Horiuchi2011}.
\cite{Dale2010}   estimated the SFR density  in four different redshift bins ($z \sim 0.16, 0.24, 0.32, 0.40$) exploiting the data from the Wyoming Survey for \ha\, (WySH). \lha\,  has been corrected for dust extinction by using the luminosity dependent prescription of \citet{Hopkins2001} and 
the volume averaged SFR has been estimated by integrating under the fitted Schechter function and adopting the \cite{Kennicutt1998}  conversion factor.
The evolution  of the cosmic SFR density suggested by these measurements  are well fitted by a power law  $\rho_{SFR}=\rho_{SFR}(0) (1+z)^{4.5\pm0.7}$, while if we consider also the other results of recent emission line surveys for the SFR density over $0 \le z \le 1.5$  the evolution is  given  by  $\rho_{SFR}=\rho_{SFR}(0) (1+z)^{3.4\pm0.4}$ \citep{Hopkins2006,Horiuchi2009,Dale2010}. 
The evolution with redshift of the volumetric CC SN rate can be fitted with a power law  $ (1+z)^{3.6}$ \citep{Botticella2008,Bazin2009} so the CC SN rate evolution seems to be consistent with that of SFR in a wide range of redshift but there is a problem in the normalisation \citep{Hopkins2006,Botticella2008,Beacom2010,Horiuchi2011}. 
We also emphasise that  the prediction  of the stellar mass  density based on the integrated SFH  also exceeds the observed value at the present epoch by a factor of two  and  remains systematically  higher  with cosmic time evolution\footnote{An IMF with a high-mass slope shallower ($\gamma=2.15$) than the Salpeter slope  can reconcile the observed stellar mass density with the cosmic SFH, but only at low redshifts \citep{Wilkins2008b}.} \citep{Wilkins2008b}.
\cite{Horiuchi2011} have analysed  the normalization  discrepancy   between predicted and measured  CC SN rates  in the local Volume (d $\le 100$\,Mpc)
  exploring whether the cosmic  CC SN rate predicted from the
cosmic SFR is too large, or whether the measurements underestimate
the true cosmic  CC SN rate, or a combination of both.  They suggested three main possible outcomes:  half of stars\footnote{ This is possible if some SN impostors, as SN 2008S, 
are true CC SNe.} with masses 8--40 M\sun\, are producing dim CC SNe, either due to dust
obscuration or being intrinsically weak, and  the fraction of dim CC SNe could explain the normalization  discrepancy; there is  a high fraction of
optically dark CC SNe while the dim CC SN fraction is only slightly
higher than the most recent SN luminosity function of
LOSS \citep{Li2011};  the normalization discrepancy could be explained by systematic
changes in our understanding of  SF or CC SN formation.
The main concern in the CC SN rate measurements at higher redshift  is the dust extinction correction  since the level of  dust obscuration is expected to be higher.
\citet{Mannucci2007} derived that the fraction of missing CC SNe is only $\sim 15\%$ at intermediate redshift, far too small to fill the gap between observed and predicted rates.
To obtain an acceptable agreement between the measurements of CC SN rate  and the predictions with the SFH as compute in
\citet{Hopkins2006}  requires an average $E(B-V)=0.3$\,mag at $z=0.3$ and $E(B-V)=0.5$\,mag at $z=0.7$  \citep{Dahlen2004,Hopkins2006} or the high extinction scenario  adopted in \citet{Botticella2008}.
These values are quite extreme and require an  extremely high dust content in galaxies which is not favoured by present measurements or inferred from the
luminosity-dependent obscuration corrections  for UV  and \ha\, data at similar redshifts. Moreover, we have to stress that the extinction in a region nearby a CC SN can be higher than the average attenuation in the host galaxy.

\section{Estimate of the CC SN progenitor mass range}\label{res}
The mass range of  CC SN progenitors can be observationally constrained by comparing  the  birth rate of  stars and the rate of  CC SNe   in the same  galaxy sample  assuming a distribution of the masses with which stars are born. 
The simplest  Poisson model formulation is to compare the
total observed number of CC SNe(\ncc) to the expected number ($\NCCExp$):
\beq
\NCCExp = \frac{\int_{\mlcc}^{\mucc} \phi(m) dm}{\int_{\ml}^{\mup} m \phi(m) dm}  \psi CT .
\eeq
The posteriori density function (PDF) can then be expressed as the Poisson probability with a prior accounting for the total observed SFR in the galaxy sample:
\beq
{\cal{L}}  \propto \NCCExp^{\ncc} 
\exp \left[ -\NCCExp \right]  \times
\exp \left[ -\frac{(\psi-\psi_\mathrm{obs})^2}{2 \delta \psi^2 } \right].
\lab{like2}
\eeq

As estimators for the central value and scale, we used the mean and
the standard deviation of PDF. We
considered flat, rather than  informative priors for the initial masses, i.e.,  $4 \le \mlcc/\msun \le 20$ and $20 \le \mucc/\msun \le 100$. 
A different approach would have been to consider each galaxy as the source of a random process  and to obtain the joint probability function as a combination of Poisson distributions. We found no statistical improvement using this variant, so we prefer to illustrate the simpler approach. 

The SNe discovered by the old local  SN surveys  (Asiago, Crimea, Evans, OCA and Calan Tololo searches) and exploited by \citet{Cappellaro1999} to obtain SN rate measurements at z$<0.01$ have been collected from 1960 to 1997 so  it is possible to merge this SN sample with that  of SNe discovered in the last 13 years.
 We cross matched our galaxy samples with the galaxy sample  from \citet{Cappellaro1999} (7773 galaxies) and found 201 common galaxies  with the sample A  and  8 CC SNe,  3 type Ia and 2 unclassified SNe\footnote{(1969L, 1969P, 1973R, 1980K, 1982F, 1983N, 1984R, 1985F, 1986G, 1987A, 1989B, 1993af, 1996cb)} discovered in these galaxies, 167 common galaxies with the sample B  and 8 CC SNe,  112  galaxies in common with the sample C and  7 CC SNe.
The unclassified  SNe have been redistributed among the three SN types according to the observed distribution: 100\%  type Ia in  E--S0, 35\% type Ia, 15\%  type Ib and 50\% type II in spirals. By  taking into account SNe discovered from 1960 we  increase the number of CC SNe  from 14 to 23.3 in the sample A, from 13 to 21.3 in the sample B and from 12  to  19.7 in the sample C.
Obviously to use both SN samples  we have to  properly combine, for the galaxies in common,  the control times of the past SN surveys and the assumed control time in the last 13 years.

In our analysis we considered for each galaxy sample two SN counts: the SNe discovered from 1998 to 2010 ( this gives $m_\mathrm{l,13}^\mathrm{CC}$) and the SNe discovered from 1960 to 2010 (this gives $m_\mathrm{l,13+old}^\mathrm{CC}$).  When considering the extended period (1960--2010) we should add a third variable to the likelihood function to account for the number of actual CC SNe which are within the unclassified group and we should marginalise over.  However, we verified that the simpler approach gave the same results i.e. we considered an effective number of CC SNe in the Poissonian likelihood, as given by the mean number of CC SNe expected in an unclassified sample (taking the local estimates for spiral galaxies).  Results for the different galaxy and SN counts are reported in Table \ref{tab_m_l} and are statistically consistent.

\begin{figure}
\resizebox{\hsize}{!}{\includegraphics{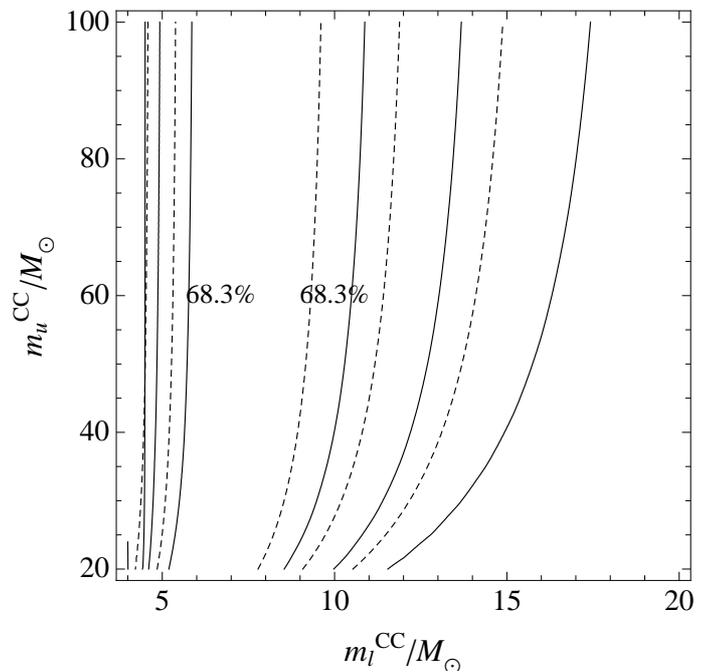}}
 \caption{Probability density function of $\mlcc$ and $\mucc$, after marginalization over the star formation rate, for the B sample. The contours show the 68.3\%, 95.4\% and 99.7\% confidence limits for two parameters, according to a frequentist approach. Thin and dashed lines are for $\ncc=14$ and $16$, respectively. Values of masses are in solar units.} \label{fig_like_m_l_m_u}
\end{figure}

The likelihood in the $\mlcc-\mucc$ plane after marginalization over $\psi$ is plotted in Fig.~\ref{fig_like_m_l_m_u} for the sample B. The contours in the figure correspond to the 68.3\%, 95.4\% and 99.7\% confidence limits for two parameters, obtained as ${\cal{L}}/{\cal{L}}_\mathrm{max}=\exp{-2.30/2}$, $\exp{-6.17/2}$ and  $\exp{-11.8/2}$, respectively. These values, which make sense only in a frequentist statistical analysis, are plotted only to illustrate parameter degeneracies.  As expected contours are very elongated and no significant constraint can be put on the upper limit mass of CC SN progenitors. 

\begin{figure}
\resizebox{\hsize}{!}{\includegraphics{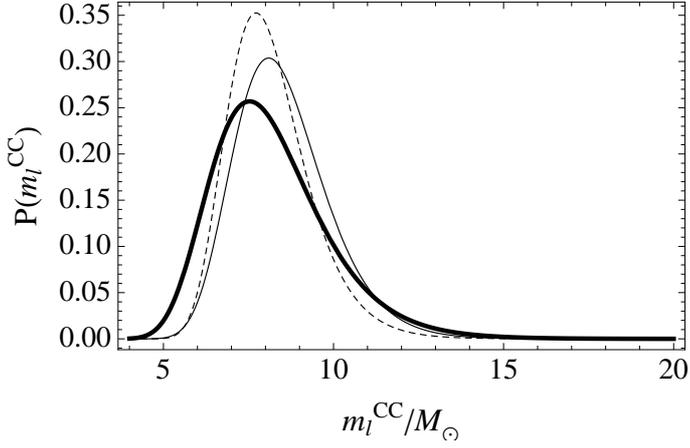}}
\caption{Posterior probability density function for $\mlcc$ as derived from sample B. The thick and thin lines refer to the PDF as derived considering a control time of 13 years and $\rm{\ncc}=13$ and an extended control time, summing the control times of the past searches for the galaxy in common, and  $\rm{\ncc}=21.3$, respectively. The dashed line  accounts for a statistical prior inferred from the local rate from \citet{Cappellaro1999}.}
 \label{fig_like_m_l}
\end{figure}

The PDF for $\mlcc$ is plotted in Fig.~\ref{fig_like_m_l}. We note a sharp decline at low masses, whereas the tail at larger values is quite shallow. 

An independent  measurement of the CC SN rate in a larger galaxy sample, i.e. , the estimate by \citet{Cappellaro1999} at z$<0.01$, can be used as a statistical prior to further constrain the lower mass of CC SN progenitors when  we take into account only the CC SNe discovered in the last 13 years (Table \ref{tab_m_l}, column $m_\mathrm{l,13+prior}^\mathrm{CC}$). 
In general adding such a prior brings about two effects (Fig.~\ref{fig_like_m_l}):  the peak of the PDF shifts towards higher values of $\mlcc$ and the tail for large masses is reduced. The overall effect on the final estimate  is that the mean is nearly unchanged whereas the standard deviation is lowered.

\begin{table}
\caption{Estimated minimum mass in different galaxy samples and for different SN samples. $m_\mathrm{l,13}^\mathrm{CC}$  has been derived counting the CC SNe discovered from 1998 to 2010 and assuming a  control time of 13 years.  $m_\mathrm{l,13+old}^\mathrm{CC}$ has been derived counting the CC SNe discovered from 1960 to 2010  and summing the control time of  13 years  (1998--2010) and that of the past searches  (1960--1997) for the galaxy in common. $m_\mathrm{l,13+prior}^\mathrm{CC}$ has been obtained by using the estimate of the local rates as a statistical prior and the CC SNe discovered from 1998 to 2010.}\label{tab_m_l}
\begin{tabular}{lllll}
\hline\hline
Sample& SFR  &$m_\mathrm{l,13}^\mathrm{CC} $  & $m_\mathrm{l,13+old}^\mathrm{CC}$ & $m_\mathrm{l,13+prior}^\mathrm{CC}$  \\
 & ($\msun yr ^{-1}$)& (\msun)  &  (\msun) & (\msun)  \\
\hline
A& SFR$_{H\alpha} = 87\pm 4 $ &   $6\pm 1$  & $5.8\pm 0.9$  & $6.2\pm0.8 $ \\
\hline
B &SFR$_{UV} = 123\pm8$ & $8\pm 2$  & $8.2 \pm 0.9$  &$8.7\pm 1.2$  \\
\hline 
C& SFR$_{H\alpha} = 58\pm3$ &  $5\pm 1$  & $4.0\pm 0.5$  & $6.0\pm 0.6 $ \\
& SFR$_{H\alpha +TIR} = 62\pm3$&  $5\pm 1$  & $4.5\pm 0.4$  & $6.0\pm 0.8$ \\
& SFR$_{UV} = 94\pm6$&  $7\pm 2$   & $6.2\pm 0.7$  & $ 7.5\pm 1.1$ \\
\hline
 \end{tabular}
\end{table}

The results in the three galaxy samples are in agreement within the uncertainties however the  mean value obtained in the sample B  is higher than those in the other two samples.  In fact in the sample B  we have a  total SFR  a factor of 1.4 higher than that in the sample A and a very similar number of SNe. If we consider \lha\, based SFR in the sample B  with a dust extinction correction via Balmer decrement  and a control time of 13 years and $\rm{\ncc}=13$ we  still obtained   $m_\mathrm{l,13}^\mathrm{CC}=6\pm1$\,\msun.

\section{Systematic errors}\label{uncert}
The method to estimate  the mass cutoff for CC SN progenitors described in the previous section needs  a  well defined  galaxy sample with  accurately measured SFRs  and a systematic SN search for which all information required to calculate the CC SN rate  is available. 
There are several possible sources of  error in our analysis:
a systematic underestimate of  the CC SN rate,  systematic errors in the SFR estimate,  systematic errors in the adopted IMF and  distance scale. The effect of  such errors  on the derived minimum mass for the CC SN progenitors will be discussed in turn.

\subsection{SN rate}\label{ncc}
There are  two effects that would depress  the absolute CC SN  rates:  the underestimate of the SN number and the overestimate of the total CT of  the galaxy sample. 
It is difficult to accuratley determine the degree of uncertainty due to both these effects since the surveys that discover local SNe  are a combination  of professional and amateurs  with complicated and unquantified selection functions. However we can estimate  both uncertainties. 

\subsubsection{SN sample}
The incompleteness  of our SN sample depends both on the extinction  suffered  by CC SNe and on the fraction of
intrinsically faint CC SNe that are  missed by local SN surveys.

Dust extinction is the largest source of systematic uncertainty in measurements of CC SN rates.
The fraction of CC SNe  (about 5\%)  that can be missed in the optical searches in the local Universe  derived by \citet{Mannucci2007}  is much smaller than the uncertainties in the measured rates since  SN statistics  is still confined to small numbers  in the local Universe.

How many nearby SNe are missed owing to their intrinsically faint luminosities  is still  uncertain.
Two intrinsically faint transients which have  dust-embedded progenitors (SN 2008S, NGC300-OT2008)   have been recently discovered and two plausible scenarios have been suggested to explain the characteristics of their progenitors and explosions:   outbursts of massive stars  \citep{Smith2009,Berger2009,Bond2009} or EC SNe in super-AGB stars \citep{Prieto2008,Thompson2009,Botticella2009,Pumo2009} .
\citet{Thompson2009}  estimated that the transients like SN~2008S are the 9\% of  all optical transients discovered within 10 Mpc when averaged over the last 10 yr and estimated a correction for incompleteness  is close to  a factor 2. Including the two dubious SNe (SN 2008S and NGC300-OT2008), we have 16, 15 and 14 CC  SNe in the sample A, B and C respectively. This has the expected effect of pushing 
the $m_\mathrm{l}^\mathrm{CC}$ value to lower masses. 
In this case  less massive progenitor are favoured, but the change is not very significant  (about 10\%, Table \ref{tab_systerr} and Fig.\ref{fig_like_m_l_m_u}).

Finally  some massive stars are expected to produce  weaker explosion  with a black hole formed by fallback  (25 -- 40\,\msun ) or collapse into a black hole directly ($>$ 40\,\msun) without any optical signature (failed SNe),  contributing to the UV or \ha\, luminosity of the galaxies but not the observed CC SN rate.
 The predicted mass range of failed SNe depends on  rotation and metallicity \citep{Heger2003,Limongi2003}.
 The lower mass cutoff  for CC SN progenitors is not strongly dependent on  the choice of the maximum mass since the steep power law nature of the IMF guarantees that a majority of the progenitors have masses within a factor 2 of the lower limit mass regardless of whether the mass spectrum extends to very high mass.
  If we conservatively restrict  the upper mass limit of  detectable CC SN progenitors  to $\rm{m_{up}} < 30$\,\msun\,  and considering  the 14 CC SNe discovered in the last 13 years  this reduces the 
 minimum mass estimate   by  about 5\% and 11\%  in  the sample A and B respectively (Table \ref{tab_systerr}). In other words, the value of $m_\mathrm{l}^\mathrm{CC}$ is not particularly 
sensitive to the highest mass that can produce a CCSNe. 
 
\begin{table}
\caption{Estimated minimum mass of CC SN progenitors  in different galaxy samples derived considering a control time of 13 years and different assumptions about the number of  CC SNe discovered in the last 13 years, the maximum mass for  progenitors of  CC SNe with optical signature, the IMF and the distance scale. }\label{tab_systerr}
\begin{tabular}{ccccc}
\hline\hline
Sample & \multicolumn{4}{c}{$m_\mathrm{l,12}^\mathrm{CC} (\msun)$}   \\
  \hline
 & dubious SNe     &      $m_\mathrm{up}^\mathrm{CC} <30$    &                         Kroupa IMF        &   distance   \\
   \hline
 A & $5.3 \pm 0.9$   &$5.4\pm 0.9$  & $5.8\pm 1.2 $ &$6.3\pm 1.3$\\
 B &  $7.4\pm 1.5$  & $7.3\pm 1.3$  &$8.4\pm 1.8$ &$9.0\pm 1.9$   \\
 \hline
 \end{tabular}
\end{table}

\subsubsection{Control Time}\label{ct}
Some galaxies in our sample might have been monitored only for a short period or might have not been monitored at all. 
For example SN 1996cr\footnote{SN 1996cr was originally identified as a variable x-ray source but later
discovered to be a young SN candidate via archival optical and
radio data.} was missed at a distance of 3.8\,Mpc \citep{Bauer2008,Dwarkadas2010}. 

\begin{figure}
\includegraphics[width=6cm, angle=-90]{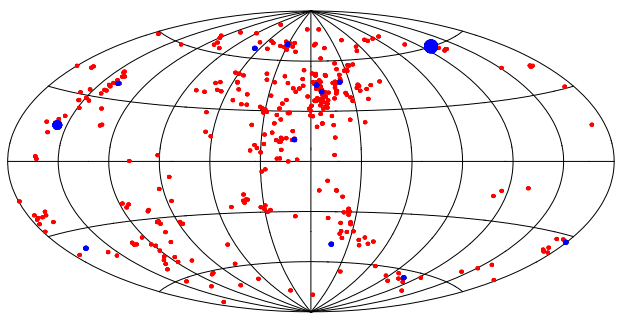}
\caption{The sky distribution of the galaxies in our sample A (red points). Blue points  represent the galaxies that hosted  CC SNe. The dimension of points is proportional to the CC SN number.}\label{SNsky}
\end{figure}

To investigate the influence of different control times of different regions of the sky the galaxy sample  has been divided  in three different groups located in  regions of celestial sphere with similar area.
We considered  the fraction of the galaxies,  the fraction of $B$ band luminosity   and  the fraction of discovered SNe for each group (Table~\ref{SNefrac}).  
The fraction of  discovered SNe is very similar to  the fraction of  galaxies and \lb\, so we can exclude that the SN search coverage in the  northern hemisphere is significantly better than in the southern one (Fig \ref{SNsky}).
 
 \begin{table}
\caption{The fraction of galaxies, $B$ band luminosity and discovered SNe in three different regions of the sky.}\label{SNefrac}
\begin{tabular}{crrrrrr}
\hline\hline
Dec  (degree)& N$_\mathrm{gal}$  &       N$_\mathrm{SNe}$    & f$_{gal}$  &f$_{L_{B}}$ &   f$_{SNe}$ \\
\hline
$d > 20$      &       224   &     9     &       58\%  &     51\%    &    56\% \\
$-20< d< 20 $  &   54   &         3   &         14\%  &     17\%  &     19\% \\
$d<-20$    &        108    &         4    &        28\% &      32\%  &      25\% \\
\hline
\end{tabular}
\end{table}

Different types of galaxies and SNe  seem to be scattered randomly over the sample area so the incompleteness factor depends very little on the position of the galaxy within the sample.


A selection effect  may be due to  galaxy luminosity since the galaxies targeted by  either local SN surveys  or amateurs are predominantly  large, luminous and metal-rich.
It is very difficult to quantify the bias against faint dwarf galaxies. However, the number of  missed  CC SNe in dwarf galaxies in the last 13 years should be not  high due to the their low SFRs. Only two out of 14 SNe in our sample were discovered in dwarf galaxies.   SN 2008jb, which exploded in the southern dwarf irregular galaxy ESO 302?14  and it is not included in our sample, was  recently discovered  in archival optical images obtained by the Catalina Real-time Transient Survey and the All-Sky Automated Survey by \citet{Prieto2011}.  The statistical  error due to the SNe discovered in dwarf galaxies ($\sim 1-2$) is negligible with respect to the Poissonian error of the overall SN sample ($\sim 3-4$)  so  this does not affect our results.

\begin{figure}
\resizebox{\hsize}{!}{\includegraphics{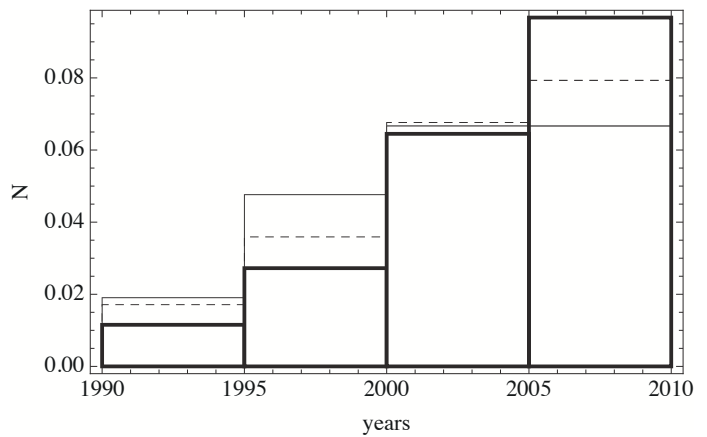}}
\caption{The discovery rate of CC SNe from 1990 to 2009. The thick, dashed and thin lines are for CC SNe discovered in the whole Universe, within 60\,Mpc and  within 11\,Mpc, respectively. }\label{histdisc1}
\end{figure}

Finally we compared the SN discovery rate  within 11\,Mpc, 60\,Mpc and in the  whole Universe  from  1997 to  2009  (Fig.\ref{histdisc1})   to search for any possible fluctuation in the discovery rate within 11\,Mpc. The  SN discovery rate  within 11\,Mpc is  almost constant during the last 13 years.

Hence the SN sample and CT uncertainties imply that any corrections in the future would 
decrease  our estimate of the lower mass limit to produce a CC SN. It is unlikely 
that any of these uncertainties could work in the opposite direction. 

\subsection{SFR}
The systematic uncertainty in the determination of the SFR  is comparable with the statistical fluctuation in the Poissonian distribution. In Sample C, 
we found a 30\% difference between the lower mass limit calculated with SFRs from \lha\, and \lfuv\ . 
A similar result is obtained by comparing the low mass estimates in the sample A and B.
The results  obtained by adopting SFRs  based on \lha\, but with different dust extinction corrections, derived from the Balmer decrement and TIR luminosity, are in excellent agreement.
This is not be surprising since the TIR+\ha\, recipe was partially calibrated using the \lha\, for galaxies  that had Balmer decrements available.

\subsection{IMF}
Subsequent studies to the early measurement of IMF by \citet{Salpeter1955} found  that there is a clear flattening in IMF slope around 0.5\,\msun\, and a further flattening near the sub-stellar mass limit so the IMF  at  the low mass end  is well represented by  either a series of broken power laws \citep{Kroupa2001} or with a log-normal function \citep{Chabrier2003}. 
While there is still some debate regarding the peak and the turnover in  the IMF,  the high mass end is typically well described by the Salpeter slope for a wide variety of environmental conditions \citep{Massey2003, Elmegreen2009}.
Uncertainties of the IMF slope for massive stars are due to uncertainties in theoretical stellar models but variation around the Salpeter value is limited to approximately  $\pm 0.5$.
If we adopt  in our analysis a \citet{Kroupa2001} IMF with $\gamma_0 = 0.3$  for  $0.01  \le m/\msun <0.08$, $\gamma_1 =1.3$ for  $0.08  \le m/\msun <0.5$ and $\gamma_2 =2.3$ for  $0.5 \le m/\msun \le 100$ and the corresponding scale factors between luminosities and SFR
we obtain  very similar results for the lower mass of CC SN progenitors  (Table \ref{tab_systerr}). 

However, recent researches  indicate that  the supposedly universal IMF within young star clusters  does not necessarily yield the same average IMF over a whole galaxy, refered to as the integrated galaxial IMF (IGIMF). 
The IGIMF  could be deficient in high mass stars compared to the  IMF  since the maximum stellar mass in a cluster seems to be limited by the embedded total cluster mass \citep{Kroupa2003, Weidner2005}.
A further complication arises from the possibility that the maximum embedded star cluster mass steepens with decreasing SFR so it
 is  constrained by the current SFR \citep{Altenburg2007,Altenburg2009}.
The combination of these two effects gives a  IGIMF which is steeper in the massive range than the Salpeter IMF and is dependent on the SFH of the galaxy. 
If the IMF varies between galaxies  there could be substantial variations in the conversion of \ha\, or UV flux to SFR \citep{Hoversten2008}, 
the number of  CC SNe per stellar generation  could be suppressed relative to that expected for a Salpeter IMF and dwarf galaxies could have a suppressed number of  CC SNe per formed stellar generation relative to massive galaxies.  Given the uncertainties, more extensive
analysis of the issue is beyond the scope of this paper.

\subsection{Binarity}
The study of young stellar populations revealed that most stars are in binary or higher order multiple systems and
that the binarity fraction among stars is a function of mass \citep{Zinnecker2007}.
In particular, the
observed fraction of O stars in massive multiple systems lies between
at least 20 and 80\% \citep[][and reference therein]{Weidner2009}.
For typical models of binary statistics, 50--70\% of CC SN progenitors  are members of a binary system at the time of the explosion \citep{Kochanek2008}.

The Salpeter IMF is not corrected for binarity. A limited number of studies have addressed
how the large proportion of binaries affect IMF  measurements since  multiple system that are not resolved into individual stellar companions  hide the less  luminous members \citep{Kroupa2002,Apellaniz2008,Weidner2009}.
\citet{Kroupa2002} suggested that  the slope value of IMF may be artificially large at high masses  ($\gamma = 2.7$) due to the effect of unresolved binaries.
\citet{Weidner2009} studied the influence on the IMF of large
quantities of unresolved multiple massive stars  and found that  even
under extreme circumstances (100\% binaries or higher order multiples), the difference
between the power-law index of the mass function of all stars and the observed IMF is
small  ($\sim 0.1$). They concluded that  if the observed IMF has the Salpeter index $\gamma =2.35$, then the true stellar
IMF has an index not flatter than $\gamma= 2.25$.

However, the binarity affects also  other parameters involved in our calculations.
The presence in a binary system can increase the mass loss and mass transfer and dramatically affect the stellar evolution.
This has two important effects: a different scaling factor of SFR tracers  predicting the correct ionising flux and  a different structure of the core of a massive star at the time of core collapse.
We did not consider these effects in our analysis.

\subsection{Distance scale} 
The typical distance uncertainties in our galaxy sample are of 5--6\% for a galaxy with direct  stellar measurements 
and 15\% for a galaxies with distances obtained from secondary indicator  or  with flow based distances \citep{Kennicutt2008}.
If  the distances were systematically underestimated by 10\%  we would find a difference of  $\sim 10\% $ in the mass cutoff for CC SN progenitors (Table~\ref{tab_systerr}).\\

In summary the uncertainties in CC SN rate act to decrease the
lower mass limit of  CC SN progenitors,  the uncertainties in dust obscuration correction for SFR estimates can
raise it,  while the choice of  the  IMF from \citet{Kroupa2001} systematically  increases the lower mass limit. Individually these are each less than a 10\% effect.

\section{Comparison with other observational estimates}\label{comp}
The first time that the linear correlation between \lha\, and CC SN rate was exploited to determine
a lower mass limit for CC SN progenitors was by \citet{Kennicutt1984}. This study used a  sample of  $80$ nearby Sc-SBc  galaxies and  combined them with an estimate of the CC SN rate ($1.4 \pm 0.2$  SNu)  in face-on Sc-SBc  from \citet{Tammann1982}   assuming an "extended" Miller-Scalo IMF ($\gamma=2.5$ between 1--100\,\msun) and $\rm{H}_{0}=50$\,$\rm{kms}^{-1} \rm{Mpc}^{-1}$.
Although \citet{Kennicutt1984}  used one of the earliest estimates of the CC SN rate based on few events and  a different IMF  his result, scaled  to H$_{0}=75$\,$\rm{kms}^{-1} \rm{Mpc}^{-1}$,  of $\mlcc =5 \pm 0.8$\,\msun\,  is consistent with our estimate in the sample A. 

\citet{Blanc2008} compared the redshift evolution of the CC SN rate
with a parametric form of the SFH found  a 
minimum progenitor mass greater than 10\,\msun. 
Although they admit that incompleteness in
the observed CC SN rate would imply a lower 
mass.   \citet{Maoz2010} used 119 SNe from the the LOSS survey and 
compared this rate to the SFHs of individual galaxies. 
They derived the SFHs for 3505 galaxies with SDSS spectra by
using the VESPA code and assuming a dust model. The CC SN rate 
of $0.010 \pm 0.002$ SNe
per \msun\,  is in agreement with expectations if all stars more
massive than 8\,\msun\, give CC SNe.  \citet{Maoz2010} also suggest that their
SN rate estimates argue against a significant fraction of massive
stars collapsing without producing a visible and detectable SN. In
other words, that the upper mass limit must be quite high.  However as
shown in Fig.~\ref{sfrcom} the upper mass limit for CC SNe is quite
unconstrianed from SN rate measurements above 20\,\msun.  The CC SN
rates themselves can't quanitatively constrain the upper mass limit
within the $\sim$20-150\,\msun\ range (simply due to the steep power
law nature of the IMF).

The direct detection of progenitor stars in pre-discovery images has provided 
identifications, mass estimates and mass limits for over 20 CC SNe (see Smartt 2009 for a
review).  \citet{Smartt2009}  carried out a detailed and homogeneous analysis of  all 
II-P SNe progenitor searches within 28 Mpc. 
 A maximum-likelihood analysis gives  the best fitting minimum and maximum masses for the SNe IIP progenitors,  
$8.5^{+1}_{-2}$\,\msun\, and $16.5 \pm 1.5$\,\msun\, respectively,  assuming a Salpeter IMF.
The minimum mass is consistent with our estimate within the errors. This lower
mass limit is consistent with other studies of type II progenitor stars
\citep{2005MNRAS.364L..33M,2006ApJ...641.1060L,2008ApJ...688L..91M,2010arXiv1011.5873V,2010arXiv1011.6558F}.  However some estimates of the  hydrodynamic mass of the ejected enevlopes of IIP SNe give systematically higher results,  e.g. 9--12\,\msun\,  in \citet{Zampieri2007} and 15--30\,\msun\,  in \citet{Utrobin2009}.  While there is 
no systematic study of a large enough sample to produce an estimate of the
lower mass limit, the discrepancy should be taken seriously in attempts to 
determine masses from both methods. 

The mass dividing CC SN progenitors from WD progenitors is theoretically expected to lie in the mass
range 7--11\,\msun\, depending on metallicity and the degree of overshooting \citep{Siess2007}. 
Observations of most massive WD progenitor  in young star clusters provide a lower limit on the value of this mass
\citep{Koester1996}.  \citet{Williams2007} and \citet{Williams2009} studied the WD populations in the open clusters of NGC 6633, NGC 7063 and NGC 2168 and found a lower limit on the maximum mass of WD progenitors  between 6.3--7.1\,\msun.  
This result is also consistent  with our estimates of the minimum mass for CC SN progenitors.

The minimum mass for CC SN progenitors is an important factor in the study of  \cite{Keane2008}, to estimate the birth rate of the Galactic neutron star population. They took an estimate of the Milky Way CC SN rate of  $1.9 \pm 0.9$\,per century from 
the measurements  of the Galactic 1.809\,MeV emission line from the radioactive decay of   $^{26}$Al 
\citep{Diehl2006}. Alternatively, they assumed a lower mass limit of
11\,\msun, and a Milky Way star-formation rate of 4\,\msun\,yr$^{-1}$ to arrive at the same CC SN
rate ($1.9 \pm 0.9$\,per century). 
  This appears to be significantly
lower than the observed population of pulsars, rotating radio transients, X-Ray dim isolated neutron stars and magnetars, which imply a
neutron star birth rate of 10.8$^{+7}_{-5}$ per century.  Our results in this paper and those on the direct progenitor
detections and WD progenitor limits, argue for lower values of $\mlcc$. It appears that a value of 10-12 \msun\, for $\mlcc$ is disfavoured by the 
combination of all these studies.  A value of  7\,\msun\,  would not be inconsistent within the three independent estimates and  that would increase the Milky Way SN rate to  $4.4 \pm 2$. This is still below the high neutron star birth rate estimate, but just within the 1$\sigma$ error.

\section{Conclusions}\label{concl}

The massive star birth and death rates are tightly correlated due to
their short lifetime. We can  
exploit the CC SN rate as a diagnostic of the current SFR by assuming
an IMF and a  mass range of the CC SN progenitor. 
Conversely we can obtain a significant constrain on the CC SN progenitor mass range by assuming a SFR inferred through the galaxy luminosity.

Only the estimate of CC SN rate in a well defined galaxy sample can  provide a direct link between SN rates and different stellar populations. Complete and volume-limited SN and galaxy samples are crucial to perform a  statistically meaningful analysis and 
the advent of large sets of multi-wavelength observations of nearby galaxies from the 11HUGS and LVL programmes provide us,  for the first time, the opportunity  to compare SFRs based on  CC SN rate and more established  tracers in the same galaxy sample.
The data are complete enough that we can take into account the different uncertainties and biases that affect these  SFR diagnostics. 
Assuming a lower mass limit cut-off of 8\,\msun\, for CC SN progenitors and a Salpeter IMF
for massive stars, we find that the  SFR based on \lha\, can not reproduce the observed CC SN rate while there is a good agreement with SFR  based on \lfuv\,  in our galaxy sample. 
The multi-wavelength data allow \lha\ to be corrected by adopting different dust extinction corrections,  from either the Balmer decrement or  by combining TIR and \ha\, luminosity.
Even with this correction, our analysis suggests that \lha\,  may under-estimate the total SFR in our galaxy samples, by nearly a factor of two.  

A future prospective of this analysis is to study the connection between SFR tracers and CC SN rate on a galaxy-by-galaxy basis and  to compare the spatial distribution of CC SNe with that of the SFR in spiral arms.
Multi wavelength data  also will also allow us to better constrain the dust effect and in principle to discriminate  the fraction of missed SNe  due to the dust extinction and failed SNe.

Conversely, we assumed  that the SFRs based on \ha\, and FUV luminosity are reliable and  obtained an observational constraint on the mass range  of CC SN progenitors by comparing the expected number of CC SNe from SFR measurements with the observed number in our  galaxy samples. Our analysis suggests that the minimum mass to produce a CC SN  is $8 \pm 1$\,\msun\,  or $ 6\pm 1$\,\msun\, if we consider  FUV and \ha\, based SFR, respectively. 
The first result is in excellent agreement with that obtained  by analysing a sample of nearby SNe with detected progenitor stars \citep{Smartt2009}.
Obviously, assuming SFRs inferred through \lha\,  a larger mass range is required to fit  the  expected and observed numbers of  CC SNe.

Recent comparisons between the SFR and CC SN rate at redshifts between
$z=0-1$ have suggested a discrepancy between the two, with the numbers
of CC SNe detected being too low by a factor of two
\citep{Botticella2008,Horiuchi2011}. The very local volume of 11Mpc
which we study here does not show that discrepancy. This is likely due
to the faint events (both intrinsically faint, and obscured) being
missed outside the 11Mpc volume and would suggest that even in
the nearby Universe, SN surveys  are incomplete. 
The uncertainties in our calculations (Poissonian and systematic) are not
low enough to rule out that some massive stars collapse
to black holes and produce optically dark SNe. 

\bibliographystyle{aa}
\bibliography{Botticella_11Mpc}

\begin{acknowledgements}
MTB thanks Andrea Pastorello for kindly and quickly providing information on CC SNe in Table~\ref{SNecarat} and  is indebted  with Laura Greggio for interesting and fruitful discussions about SN rates.
This work, conducted as part of the award "Understanding the lives of
massive stars from birth to supernovae" (S.J. Smartt) made under the
European Heads of Research Councils and European Science Foundation
EURYI (European Young Investigator) Awards scheme, was supported by
funds from the Participating Organisations of EURYI and the EC Sixth
Framework Programme. MTB is also supported by the PRIN-INAF 2009 with the project "Supernovae Variety and Nucleosynthesis Yields".
\end{acknowledgements}

\begin{appendix}
\section{SNe discovered from 1885 to 2010}
We have been identified SNe known to have occurred in our galaxy samples  from 1885 to 2010 from the Asiago SN catalogue \citep{Barbon2008}. 
\begin{table*}
\caption{SNe discovered in the galaxy sample A.}\label{clSNe}
\centering
\begin{tabular}{llccccc}
\hline\hline
   SN & type  &host gal.& T &  M$_{B}$ & $\lha$& $\eha$ \\
\hline
1885A & ... &  NGC 0224  &   3 &  $-20.31$&  40.43&  4  \\
1895B & Ia &   NGC 5253 &  11&  $-16.83$& 40.34&  120 \\ 
1909A & IIP: & NGC 5457 &   6 &  $-20.84$ & 41.33 & 31\\
1917A & II &   NGC 6946 &  6 &  $-20.79$&  41.46 & 33 \\ 
1923A & IIP:&  NGC 5236 &  5 & $ -20.26$ & 41.25 & 33 \\         
1937C & Ia &   IC 4182  &  9&  $ -15.89$&  39.48&  28 \\
1937D & Ia &   NGC 1003 &   6 &  $-18.08$&  40.40 &  42 \\   
1939C &  ...  &NGC 6946 &  6&   $-20.79 $& 41.46&  33 \\
1940E & ... &  NGC 253     & 5& $-20.00$ & 40.99& 16 \\
1945A  & ... &NGC 5195     & 2& $-19.14$ & 39.79& 4\\
1945B & ... &     NGC 5236 &  5 & $ -20.26$ & 41.25 & 33  \\  
1948B & IIP &   NGC 6946  &  6 &  $-20.79 $& 41.46&  33   \\ 
1950B & ... &      NGC 5236 &   5&   $-20.26$&  41.25&  33   \\  
1951H&  II&    NGC 5457 &  6 &  $-20.84$&  41.33&  31  \\ 
1954A&  Ib &   NGC 4214  &  10 &  $-17.13$&  40.19&  62  \\  
1954J&  LBV& NGC 2403 &  6 & $ -18.78$ & 40.78&  50   \\
1957D &   ... &   NGC 5236  &  5 &  $-20.26$ & 41.25&  33   \\ 
1960H & Iapec& NGC 4096 & 5&  $ -18.40$ & 40.31&  20    \\
1961V & LBV ?&NGC 1058 &   5 & $ -18.24$&  40.26&  29   \\  
1962M & IIP &  NGC 1313 &  7 &  $-19.14$ & 40.60 &  35  \\  
1963L &  ... &  M+06-07-09&   8&  $ -14.92$ & 39.76&  28  \\   
1968D & II  &  NGC 6946  &  5&  $ -20.79 $& 41.46&  33   \\   
1968L & IIP &  NGC 5236 &  5 & $ -20.26$&  41.25&  33  \\   
1969L & IIP &  NGC 1058 &  5 & $ -18.24$ & 40.26 & 29   \\  
1969P  & ..   &  NGC 6946 &  6 & $ -20.79$ & 41.46 & 33   \\  
1970G & IIL&   NGC 5457 &  6 & $ -20.84$& 41.33 & 31   \\ 
1971I & Ia &   NGC 5055 &  4 & $ -20.12$ & 40.87 & 20   \\  
1972E & Ia &   NGC 5253 & 11 &  $-16.83$ & 40.34 & 120  \\  
1973R & IIP&   NGC 3627 &  3 &  $-20.44$ & 41.11&  19   \\  
1978K & LBV ? &   NGC 1313 & 7 &  $-19.14$ & 40.60&  35   \\  
1980K & IIL &  NGC 6946 &  6 &  $-20.79 $& 41.46&  33   \\ 
1981K & II  &   NGC 4258 & 4&  $ -20.44$ & 41.21&  15   \\   
1982F & IIP &  NGC 4490& 7 &  $-19.37$ & 41.09&  66   \\  
1982L & II: &  NGC 7713 & 7 & $ -18.18$&  40.39 & 47   \\    
1983N & Ib &   NGC 5236 &  5 &  $-20.26$ & 41.25 & 33   \\ 
1984R  & ... &   NGC 3675  & 3& $-19.16$ &40.58&15  \\ 
1985F & Ib/c&  NGC 4618 &   8&  $ -18.25$ & 40.39 & 33   \\  
1986G & Iapec& NGC 5128 &  $-2$ & $ -20.47$&  40.81 & 10   \\  
1986J & II  &  NGC 891  &   3 & $ -19.29$ & 40.59&  15   \\  
1987A & IIpec& LMC    &  9 & $ -17.87$&  40.49&  38   \\  
1989B & Ia  &  NGC 3627 &  3 & $ -20.44$ & 41.11 & 19   \\  
1993J & IIb &  NGC 3031 &  2&   $-20.15$ &  40.77 & 10   \\  
1993af& Ia &   NGC 1808 &  1 &  $-19.66$ & 41.14&  29   \\  
1994I & Ic &   NGC 5194 &  4 & $ -20.63$ &  41.28 & 28    \\
1996cb& IIb &  NGC 3510  &  8 &  $-15.47$ &  39.81&   42    \\   
1996cr& IIn:&  E097-G13 &   3 &    $-17.15$ &  40.21&   22  \\  
1997bs& LBV &  NGC 3627&    3&  $ -20.44$&   41.11 &  19    \\
1998bu &Ia &   NGC 3368&   2 &  $ -20.08$ &   40.50 &   5    \\  
2000ch &LBV &  NGC 3432 &    9 &  $ -18.06$ &   40.55&   64   \\  
2002ap &Ic& NGC 628   &    5 &   $-19.58$ &  40.87 &  35  \\   
2002bu &IIn&   NGC 4242  &  8 &  $-18.18$ & 39.95 &   18  \\  
2002hh &IIP&   NGC 6946  &  6  &$ -20.79$ & 41.46&  33   \\
2002kg & LBV&   NGC 2403  &  6  & $-18.78$ &  40.78&  50   \\
2003gd &IIP&    NGC 628  &  5  & $-19.58$&  40.87 & 35  \\   
2004am &IIP &   NGC 3034  &   7 & $ -18.84$ & 41.07 & 64   \\  
2004dj &IIP&   NGC 2403  &   6 &  $-18.78$ & 40.78 & 50    \\
2004et &IIP &   NGC 6946  &  6 & $ -20.79$&  41.46 & 33   \\  
2005af &IIP&    NGC 4945  &6 & $ -19.26$ & 40.75 & 17   \\  
2005at &Ic &   NGC 6744 & 4 &  $-20.94 $& 41.21 & 15    \\
2005cs &IIP &  NGC 5194  & 4 &$  -20.63$ & 41.28 & 28    \\
2007gr &Ic&  NGC 1058   &  5 & $-18.24$ & 40.26 & 29    \\ 
2008S & II:   &  NGC 6946  &  6 & $ -20.79$&  41.46 & 34   \\      
2008ax &IIb   &  NGC 4490 & 7 &  $-19.37$ & 41.09&  66   \\   
2008bk &IIP   &  NGC 7793   &  7 &$-18.41$ &40.58&40\\
2008OT&II:  &  NGC 300   & 7  & $-17.84$ & 40.18& 24\\
2009hd&IIP & NGC 3627&  3 & $ -20.44$ & 41.11 & 19   \\  
2010da&LBV& NGC 300&7  & $-17.84$ & 40.18& 24\\
 \hline
\end{tabular}
\end{table*}
\end{appendix}

\begin{appendix} 
\section{CC SN rate}\label{rate} 
\begin{table*}
\caption{The number of the galaxies and  CC SNe discovered in the last 13 years, the total SFR based on \lha\, and \lfuv, the CC SN rate (R$_\mathrm{CC}$), the ratio between CC SN rate and SFR  (K$_{CC}$) and the total \lb, \lk\, and  mass  in different sub-samples based on SFRs.}\label{KccTab}
\begin{tabular}{llccccccccc}
\hline\hline
sample&     sub-sample                                   &N$_\mathrm{gal}$ &N$_\mathrm{CC}$   & SFR$_{H\alpha}$ & SFR$_{UV}$ & $\rcc $  & \kcc &$L_{B}$ & $L_{K}$ &M\\
             &                                        &                                      &                                 &  (\msun yr$^{-1}$) &   (\msun yr$^{-1}$)   &   ( yr$^{-1}$)  & (\msun$^{-1}$)                       &         (10$^{10}$ \lbsun)          &       (10$^{10}$ \lksun)    &       (\msun)        \\
  \hline
A & SFR$_{H\alpha} < 1$         &      360                        &       4                       &    $35\pm1$             &   --              &    $0.3_{-0.1}^{+0.2}$               &      $0.009_{-0.004}^{+0.007}$   & $69\pm5$           &    --  &    --  \\
   &SFR$_{H\alpha} \ge 1$            &      23                             &   9                            &      $52\pm4$                    &    --             &         $0.7_{-0.2}^{+0.3}$           &  $0.013_{-0.004}^{+0.006}$    &  $71\pm5$                           &      --     &    --        \\
   \hline
   &SFR$_{H\alpha,1}$                &       367                        &        7                       &             $43\pm2$                      &       --          &             $0.5_{-0.2}^{+0.3}$       & $0.012_{-0.005}^{+0.007}$&       $85\pm5$                            &          --    &    --     \\
   &  SFR$_{H\alpha,2}$              &       16                            &       7                        &             $44\pm3$                      &      --           &            $0.5_{-0.2}^{+0.3}$    &  $0.012_{-0.004}^{+0.006}$  &             $55\pm5$                           &         --    &    --      \\
 \hline\hline
  B & SFR$_{UV} <1$                 &         287                            &      3              &         $27\pm1$                          &      $32\pm1$            &        $0.2_{-0.1}^{+0.2}$        &  $0.007_{-0.004}^{+0.007}$ &           $52\pm4$                        &      --     &    --        \\
      & SFR$_{UV} \ge 1$                    &           25                          &         10               &            $51\pm4$                       &      $90\pm7$            &        $0.8_{-0.2}^{+0.3}$        &   $0.009_{-0.003}^{+0.004}$  &           $71\pm5$                        &      --     &    --        \\
    \hline
    &SFR$_{UV,1}$                        &              304                       &        8                       &          $47\pm2$                         &        $65\pm3$          &         $0.6_{-0.2}^{+0.3}$     &  $0.010_{-0.003}^{+0.005}$   &              $86\pm5$                     &        --   &    --      \\   
    &SFR$_{UV,2}$                        &                  8                   &             5                  &           $31\pm1$                        &         $58\pm7$         &         $0.4_{-0.2}^{+0.3}$         &  $0.007_{-0.003}^{+0.004}$  &          $36\pm5$                     &          --      &    --    \\
\hline\hline 
 C & SFR$_{UV} <1$                  &        147                            &       3                       &        $17\pm1$                           &       $23\pm1$           &         $0.1_{-0.1}^{+0.2}$           &  $0.010_{-0.005}^{+0.009}$ &       $26\pm2$     &    $47\pm6$      &       $58\pm5$         \\
    & SFR$_{UV} \ge 1$                     &          20                    &          9                     &          $41\pm4$                         &       $72\pm6$           &          $0.7_{-0.2}^{+0.3}$           &    $0.010_{-0.003}^{+0.004}$&      $60\pm5$        &     $107\pm7$     &       $129\pm9$        \\
    \hline
    &SFR$_{UV,1}$                         &        160                      &            8                   &            $33\pm2$                       &       $48\pm3$           &          $0.6_{-0.2}^{+0.3}$       &  $0.013_{-0.004}^{+0.006}$ &            $56\pm3$      &        $109\pm8$   &    $136\pm8$      \\    
    &SFR$_{UV,2}$                         &         7                       &               4                &            $25\pm3$                       &       $47\pm5$           &          $0.3_{-0.1}^{+0.2}$          &  $0.007_{-0.003}^{+0.005}$ &         $29\pm4$         &        $44\pm5$        &    $51\pm4$      \\
 \hline    
\end{tabular}
\end{table*}

The CC SN rate is  $1.1_{-0.3}^{+0.4}$,  $1.0_{-0.3}^{+0.4}$   and $0.9_{-0.3}^{+0.4}$  SNe yr$^{-1}$ in the sample A, B and C, respectively. 
It is interesting to note that the ratio between type Ia and CC SNe in our  galaxy sample is 1/14\footnote{SN 1998bu in NGC 3368 is the only type Ia SN discovered in the last 13 years within 11 Mpc (Table \ref{clSNe}).},  much lower than that observed by \citet{Cappellaro1999} ($\sim 0.4$) or by \citet{Botticella2008}  at redshift z=0.2 ($\sim 0.2$). 
This  can be explained by a different galaxy content within 11 Mpc  in comparison with larger volumes and by an increasing number of  faint CC SNe  missed from SN searches at higher redshift.

We have  estimated K$_{CC}$) in different  galaxy sub-samples  to analyse a possible dependence on the SFR.  
We have split  the galaxy samples A, B and C both by using a cut value of 1 \msun yr$^{-1}$ and by  measuring the same total SFR. In both cases we found that  K$_{CC}$ is constant  within uncertainties (Table~\ref{KccTab}) consistently with the assumption that the mass range of CC SN progenitors and IMF are universal.

It is interesting to estimate K$_{CC}$ by comparing the observed  CC SN rate and SFR density in the same volume.
The volumetric CC SN rate  within 11 Mpc is about $(2\pm 0.5) \times 10^{-4} \rm{Mpc}^{-3} \rm{yr}^{-1}$. Recently \citet{Bothwell2011}  estimated a SFR  density   of  $0.025\pm 0.0016$\,\msun\, Mpc$^{-3}$  yr$^{-1}$  at $z\sim0$ confirming the result by \citet{Karachentsev2008}. The distribution function  of the SFR density shows that this is dominated by bright, late type, modest star forming spiral galaxies,  with about 20\% occurring in starburst galaxies. Early type spirals provide only a small contribution in spite of including many of the highest mass galaxies while the fraction of the SFR  in dwarf galaxies is very low. 
The value of  \kcc\, obtained by using the SFR density estimate by \citet{Bothwell2011} and our estimate of the CC SN rate  is  0.008  \msun$^{-1}$  that is consistent with the values estimated above.
\citet{Blanc2008} obtained an estimate of \kcc\, by fitting the volumetric CC SN rate measurements at different redshifts  with two different laws for the SFH  \citep{Chary2001,Cole2001} and assuming a Salpeter IMF.  Their results ($ \kcc\,= 0.00333\pm0.00089  h^{-2}_{70}$\,\msun$^{-1}$ and  $ \kcc\,= 0.00326\pm0.00172  h^{-2}_{70}$\,\msun$^{-1}$)  suggest a lower number of  CC SNe per unit mass of the parent stellar generation and, as a consequence, a narrow mass range  for CC SN progenitors.
The discrepancy between measurements of K$_{CC}$ in different volumes is connected to the  difficulty  in matching CC SN rate and SFR at higher redshifts  that we discussed  in the section \ref{compSFR}.

\section{ CC SN rate per unit luminosity and mass}
The CC SN rate is proportional to the number of progenitor stars in a galaxy and hence to its SFR.  When no information about SFR is available,  a CC SN rate measurement  in a galaxy sample can be normalised to some parameter related to the SFR, such as the total  luminosity in an opportune band. 

The $B$ band luminosity has been for a long time the only available parameter  to normalise the SN rates in nearby galaxies (SNu unit\footnote{$1 \rm{SNu}=  1 SN/10^{10}$\,\lbsun / century}). The CC SN rate normalised per unit $B$ band luminosity  shows a strong dependence on the  galaxy morphological  type \citep{Tammann1974,Cappellaro1999}  since  the $B$  band luminosity is the result of the combined effects of the emission by both old and young stellar populations  and their relative contributions change along the Hubble sequence.
A different tracer of the population of massive stars is provided by the TIR luminosity since  a good correlation between SN rate and \ltir has been observed  \citep{Cappellaro1999,Mannucci2003} but  measurements were available only for a small sample of galaxies until few years ago.  

Whereas $B$ band luminosity  is  un-specific  to  the stellar age the  luminosity in a reddish band, such as  $K$ band,  is good tracer of the old population in a galaxy  and therefore of the mean age of the stellar population.  The combination of the $K$ band luminosity and the $B-K$ colour  can be used as an  indicator  of the  galaxy mass in star  \citep{Bell2001}. 
 The data collected from 2MASS allowed  \citet{Mannucci2005} to normalise the SN rates in a sub-sample of \citet{Cappellaro1999}  per unit $K$ band luminosity and per unit galaxy mass by using the relation of \citet{Bell2001} and adopting a diet Salpeter.  
 
\begin{figure*}
\begin{center}
$
\begin{array}{c@{\hspace{.1in}}c@{\hspace{.1in}}c}
\includegraphics[width=6cm]{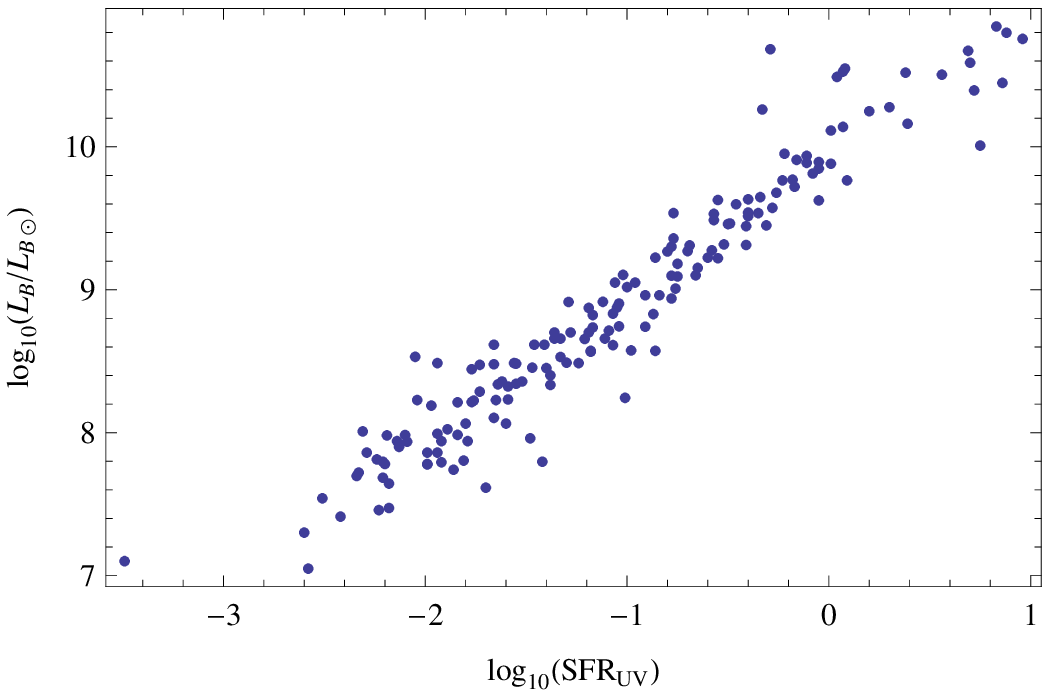} &
\includegraphics[width=6cm]{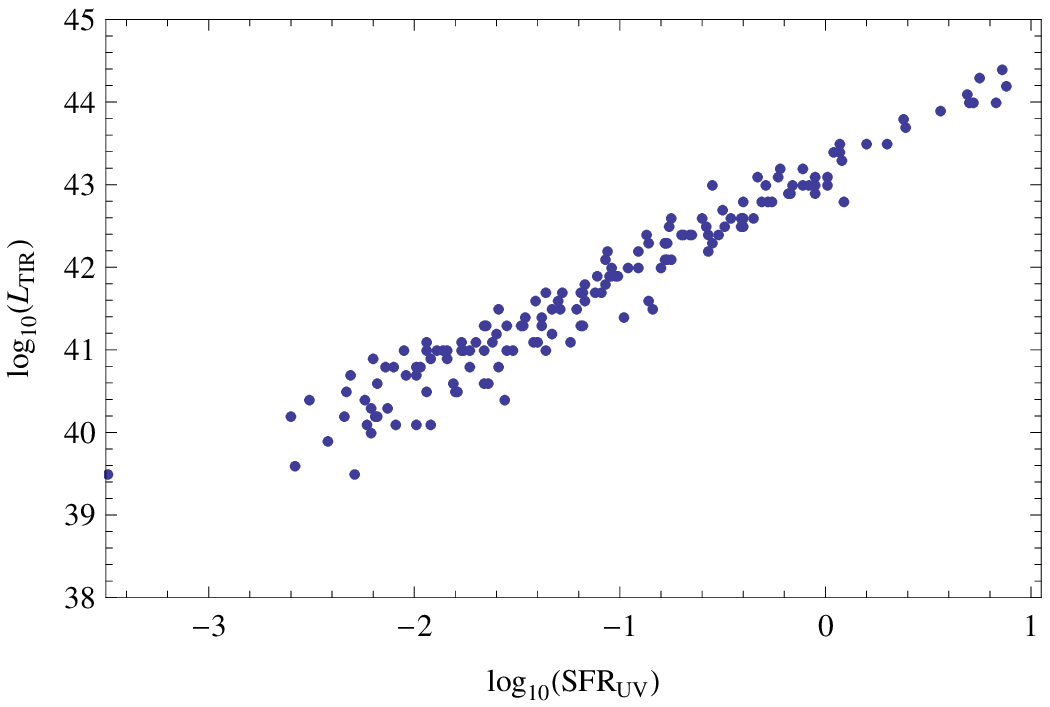} \\
\includegraphics[width=6cm]{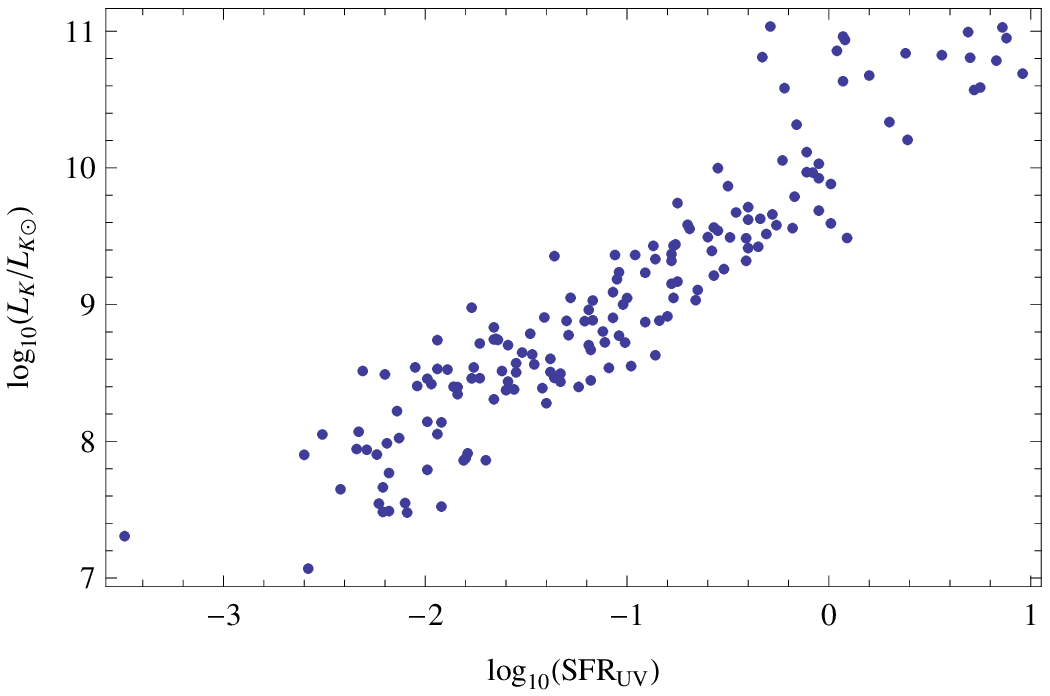} &
\includegraphics[width=6cm]{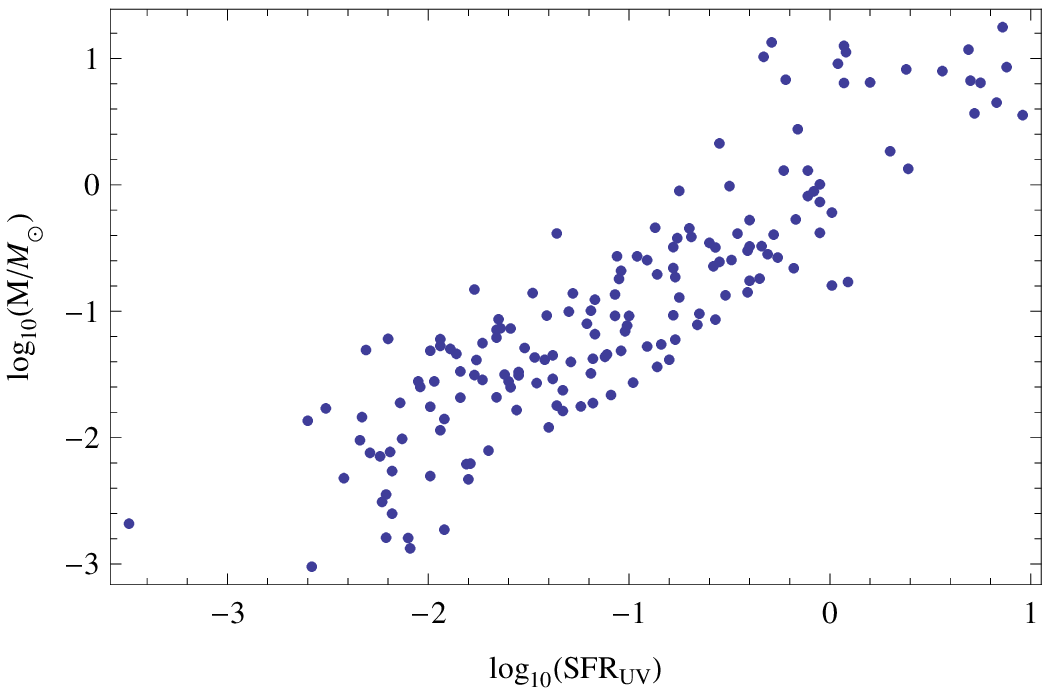} &
\end{array}
$
\end{center}
\caption{The  \lb, \lk, \ltir\, and galaxy mass as a function of  SFR$_{UV}$ in the sample C.}\label{RCC_LB_SFR_sampleC}
\end{figure*}

 We stress that the CC SN rate is related to the current SFR and not to the total stellar mass in a galaxy, while the type Ia SN rate is expected to be proportional to the  $K$ band  luminosity and  the galaxy mass so the normalisation  per unit mass  has a different meaning for different SN types.  
 The relation between \lb, \lk, \ltir, mass and  SFR for the galaxies in the sample C  is illustrated in Fig.~\ref{RCC_LB_SFR_sampleC} and a detailed discussion about  the expected type Ia SN rate as a function of the galaxy luminosity and colours can be find in \cite{Greggio2010}.

According to the estimated rates in \citet{Cappellaro1999}  and the total \lb\, in our galaxy samples we expect to observe $14 \pm 4$, $12 \pm 3$, $9 \pm 3$ CC SNe  for the sample A,  B and C compared to the 14, 13 and 12 discovered in the last 13 years, while  according to  the total  \lk\,  and galaxy mass in our sample C  and  the CC SN  rates from \citet{Mannucci2005}  we expect to observe $7\pm 2$ and $13 \pm 2$ CC SNe  respectively compared to the 12 discovered in the last 13 years.  We found a good agreement between the expected and observed number of CC SNe both for rate normalised per unit luminosity and per unit mass,  i.e. there is no discrepancy in the number of CC SNe detected within the 11HUGS galaxy sample and the rate of CC SNe within $z\lesssim0.01$.    The expected number of type Ia SNe in 13 years by adopting the rate estimate by \citet{Cappellaro1999}  and the $B$ band luminosity of our galaxy sample  A is $4\pm1$.   The probability of finding only 1 or 0  type Ia SN  is $\sim9\%$.  




It is interesting to compare the observed volumetric CC SN rate, $(2\pm 0.5) \times 10^{-4} \rm{Mpc}^{-3} \rm{yr}^{-1}$,  within 11 Mpc with that  expected by  assuming a  luminosity density or a SFR density.  If we adopted a $B$ band luminosity density $j_{B}=1.03 \times 10^{8}$ \lbsun\, Mpc$^{-3}$ obtained by fitting several measurements in the redshift interval $0<z<1$ as in \citet{Botticella2008}   the expected CC SN rate is $\rm{r}_{CC}=0.8 \pm 0.2 \times 10^{-4}$\,Mpc$^{-3}$ yr$^{-1}$  a factor two lower than the observed value.  However, if we normalise the CC SN rate per unit $B$ band luminosity there is a perfect  consistency between the values within 11 Mpc and z$<0.01$ \citep{Cappellaro1999}.

\cite{Beacom2010} estimated a CC SN rate of  $(1.25 \pm 0.25)  \times 10^{-4}$\,Mpc$^{-3}$ yr$^{-1}$ at redshift 0 from the cosmic SFR density measurements assuming a Salpeter IMF, a mass range between 8--50\,\msun\, and $H_{0}=70$ \,km s$^{-1}$ Mpc$^{-1}$.  This estimate becomes $(1.02\pm 0.25) \times 10^{-4}$\,Mpc$^{-3}$ yr$^{-1}$ if we adopt $H_{0}=75$ \,km s$^{-1}$ Mpc$^{-1} $ and it is still a factor two lower than the observed value.  \citet{Horiuchi2011} adopted a SFR density of $0.017$\,\msun\, Mpc$^{-3}$ yr$^{-1}$ for $H_{0}=73$ \,km s$^{-1}$ Mpc$^{-1}$ that becomes $0.018$\,\msun\, Mpc$^{-3}$ yr$^{-1}$ $H_{0}=75$ \,km s$^{-1}$ Mpc$^{-1} $ obtaining a CC SN rate of $1.5  \times 10^{-4}$\,Mpc$^{-3}$ yr$^{-1}$.  The luminosity density estimate $j_{B}$ is independent of the SFR measurements used by \citet{Beacom2010} and \citet{Horiuchi2011}.
 
In both cases the difference between expected and observed rates is due to the local galaxy over-density: the SFR and luminosity density estimated in  the Universe averaged over larger volumes  do not trace the galaxy density within 11 Mpc. 
\citet{Karachentsev2004} found that the local luminosity density exceeds about 2 times the global density in spite of the presence of the Local Void\footnote{The near empty Local Void occupies about a third of the volume at $1<D< 8$\,Mpc and contains just two of the 480 known galaxies.}. 
The galaxy content also differs significantly  from that in a larger volume since 11 of the 19 largest galaxies within 8 Mpc are near pure disks \citep{Karachentsev2004}.   As a consequence, the cumulative distribution of the discovered SNe since 1998 as a function of distance does not follow the expected distribution in a growing volume.

\end{appendix} 

\begin{appendix} 

\section{CC SN rate as a function of galaxy  properties}

\begin{figure*}
\begin{center}
$
\begin{array}{c@{\hspace{.1in}}c@{\hspace{.1in}}c@{\hspace{.1in}}c}
\includegraphics[width=6cm]{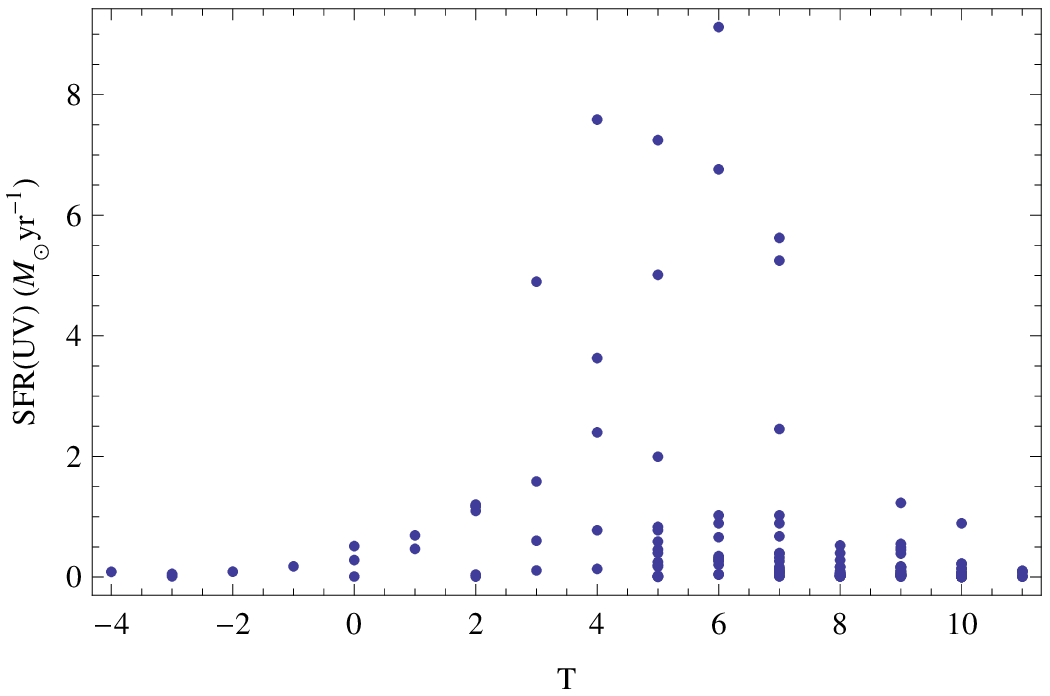} &
\includegraphics[width=5.9cm]{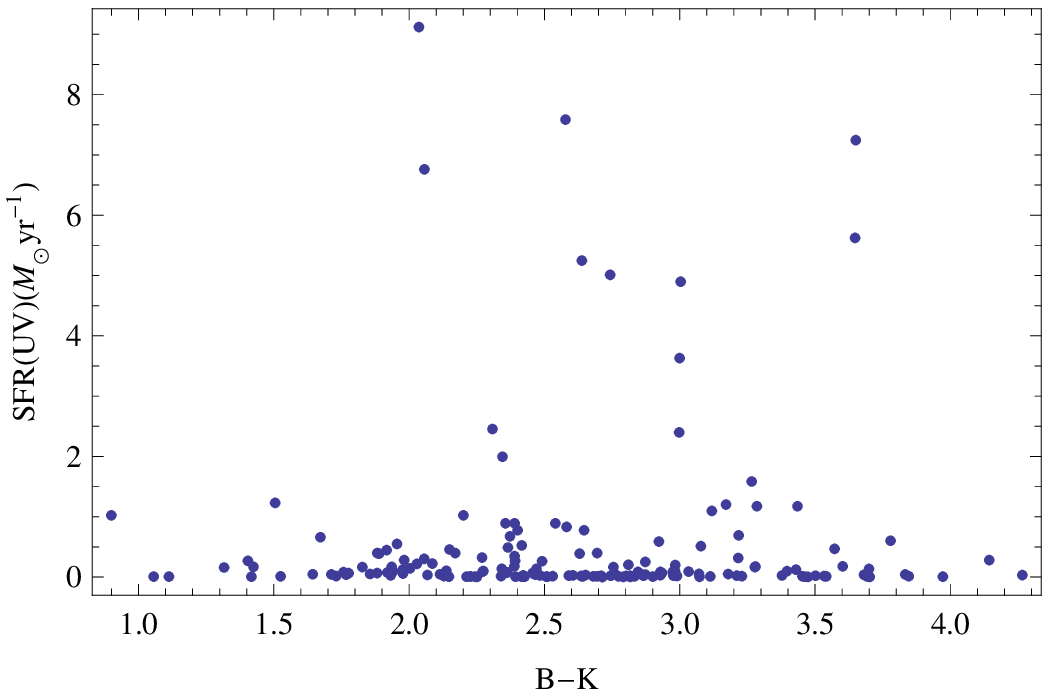} &
\includegraphics[width=6.1cm]{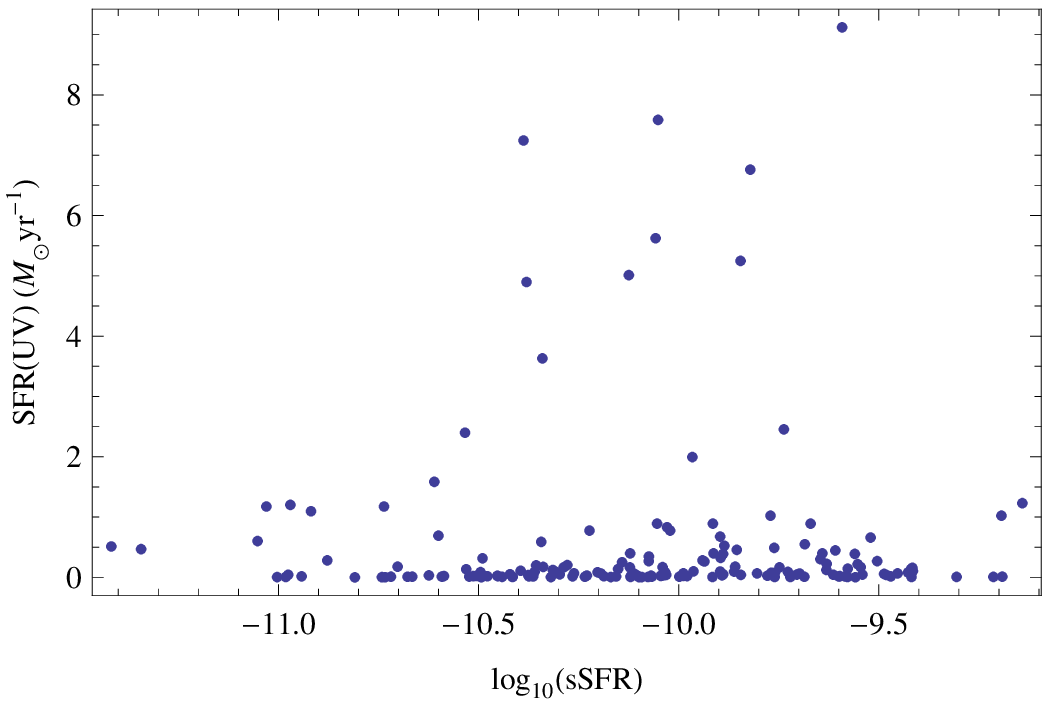} &
\end{array}
$
\end{center}
\caption{The  SFR$_{UV}$ as a function of the morphological type, $B-K$ and sSFR in the sample C.}\label{RCC_MT_SFR_sampleC}
\end{figure*}

We expect that CC SN rate has a dependence on  some  galaxy properties related to the SFR  (Fig. \ref{RCC_MT_SFR_sampleC}) and the trend is different for different normalizations.
In order to investigate some possible dependence we divided our galaxy sample C into sub-samples based on  galaxy morphological type, $B-K$ colour, mass,  sSFR and  \eha\,  and measured the CC SN rate in each sub-sample (Table~\ref{depSampleC}).\\

\begin{figure*}
\begin{center}
$
\begin{array}{c@{\hspace{.1in}}c@{\hspace{.1in}}c@{\hspace{.1in}}c}
\includegraphics[width=6cm]{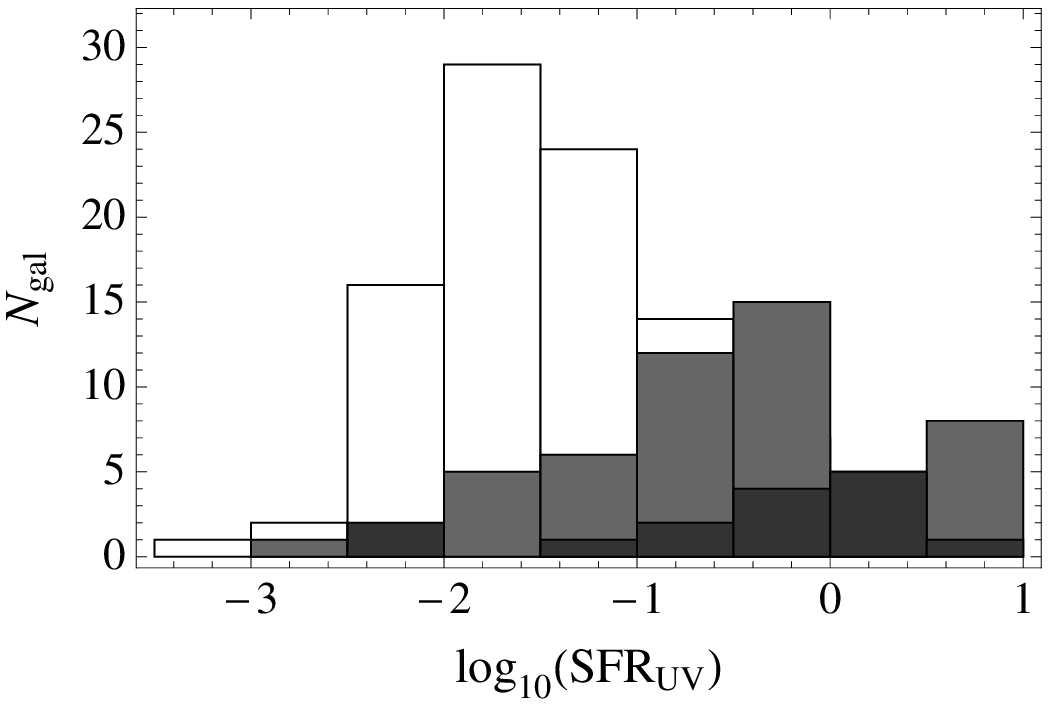}&
\includegraphics[width=6cm]{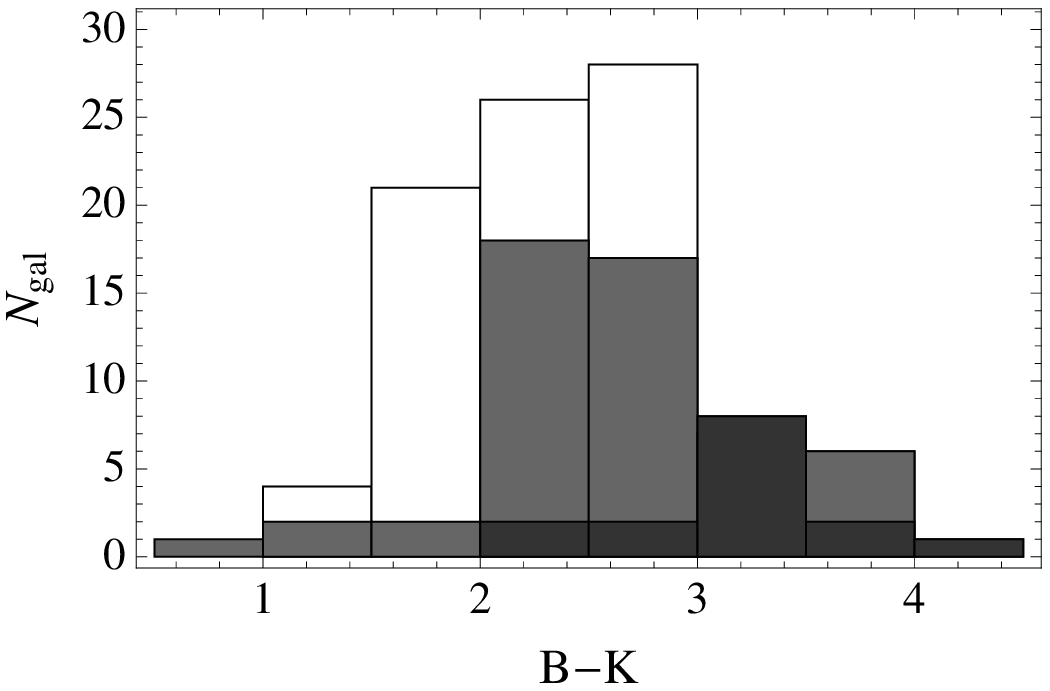}&
\includegraphics[width=6cm]{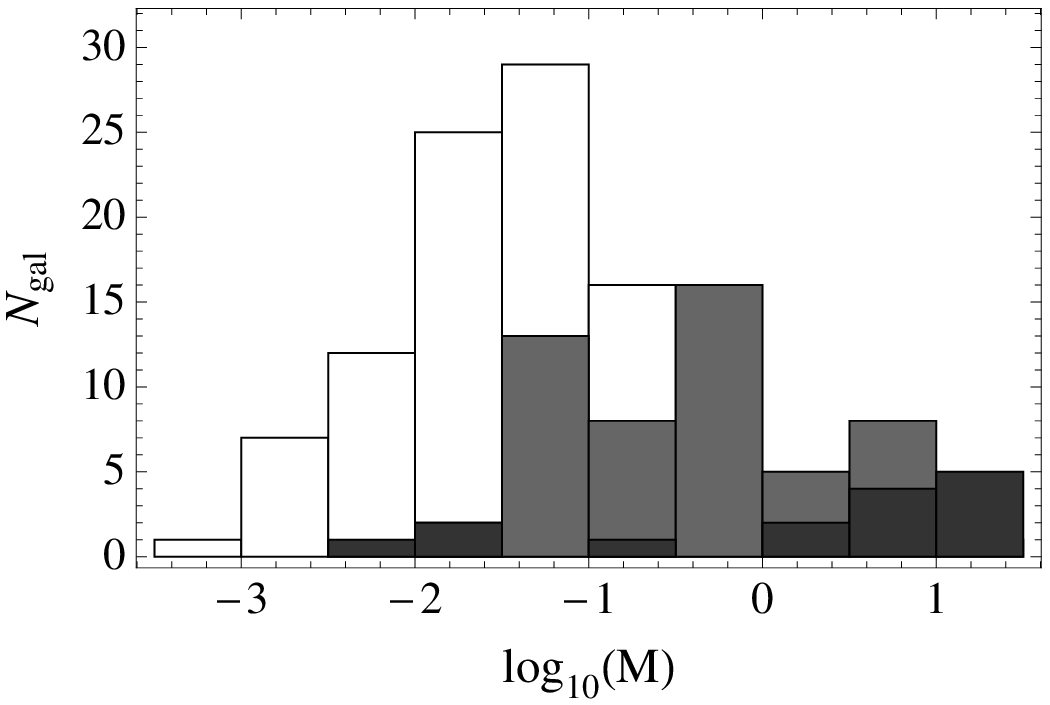} &
\end{array}
$
\end{center}
\caption{The distribution of the SFR, $B-K$ colour and mass in the sub-samples with $0 \le \mathrm{T} < 4$ (black histogram), $4\le \mathrm{T} \le 7$ (grey histogram) and the  $\mathrm{T} > 7$ (empty histogram)}\label{Hist_morph_sampleC}
\end{figure*}

{\em Morphology}\\
It is well known that the morphological type sequence, in a first approximation, corresponds to a sequence in the SFR.
We  split  each galaxy sample in four different sub-samples based on the galaxy RC3 type (Table~\ref{depSampleC}).  The distribution of the SFR, $B-K$ and mass in the $0 \le \mathrm{T} < 4$, $4\le \mathrm{T} \le 7$ and the  $\mathrm{T} > 7$ sub-samples  are illustrated in Fig.~ \ref{Hist_morph_sampleC}.  There is a statistically significant number of  both galaxies  and CC SNe only in  the $4 \le T \le7$ sub-samples with  30\% of galaxies and 80\% of CC SNe, as expected. The total SFR$_{H\alpha}$ and SFR$_{UV}$ in this sub-sample are about 75\%  of the total value in the sample C, while the CC SN rate  is  $0.8_{-0.2}^{+0.3}$ yr$^{-1}$  (Table~\ref{depSampleC}). 
To compare our result with that from  \citet{Cappellaro1999}  we assumed that  the  $\mathrm{T}<0$,  $0 \le \mathrm{T} < 4$, $4\le \mathrm{T} \le 7$ and the  $\mathrm{T} > 7$ sub-samples correspond to E--S0, S0a--Sb, Sbc--Sd  and "Other" in the \citet{Cappellaro1999} sample, respectively.  
The CC SN rate  in the sub-sample $4 \le T \le7$  has the value  of  $1.5_{-0.5}^{+0.6}$\,SNu for the sample C that are in nice agreement with the value of  $1.0\pm 0.4$\,SNu from \citet{Cappellaro1999}.  

\begin{figure*}
\begin{center}
$
\begin{array}{c@{\hspace{.1in}}c@{\hspace{.1in}}c}
\includegraphics[width=5cm]{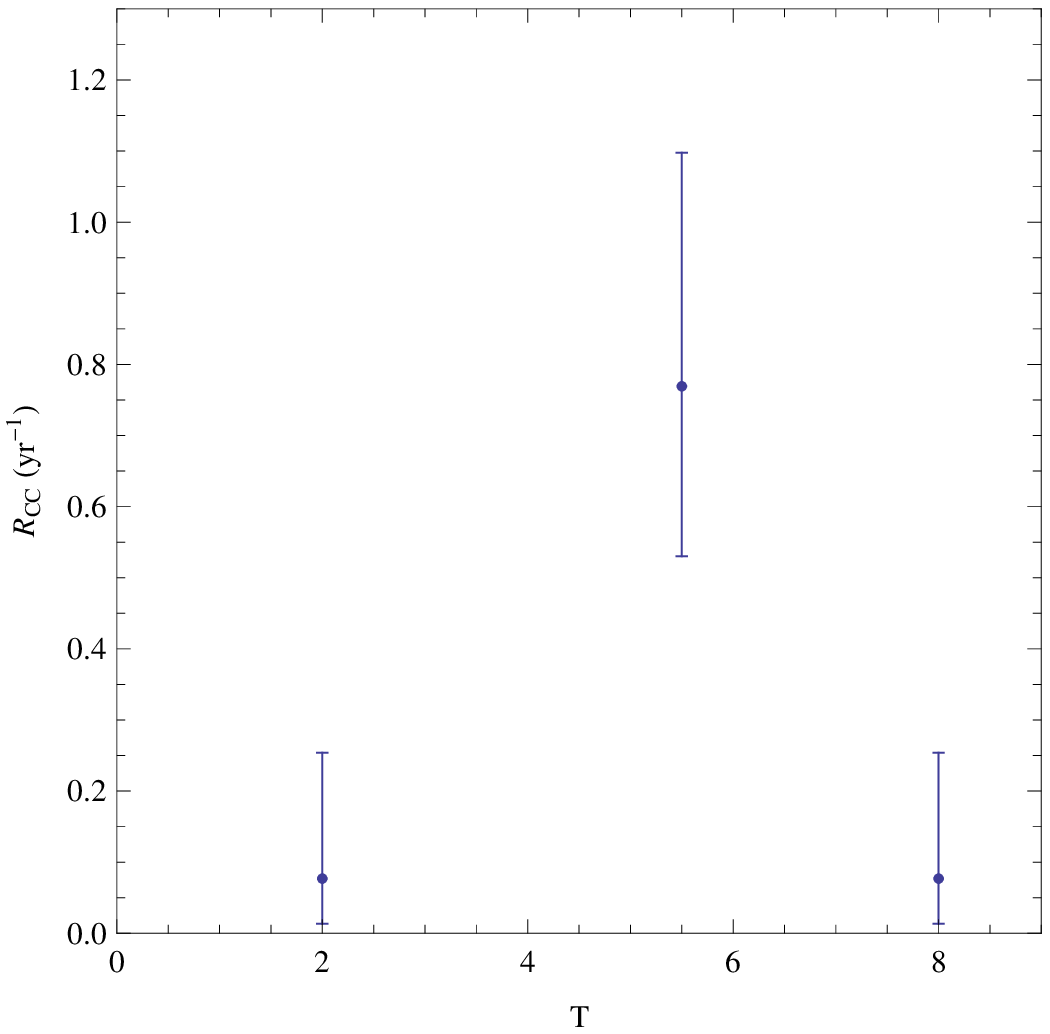} &
\includegraphics[width=5cm]{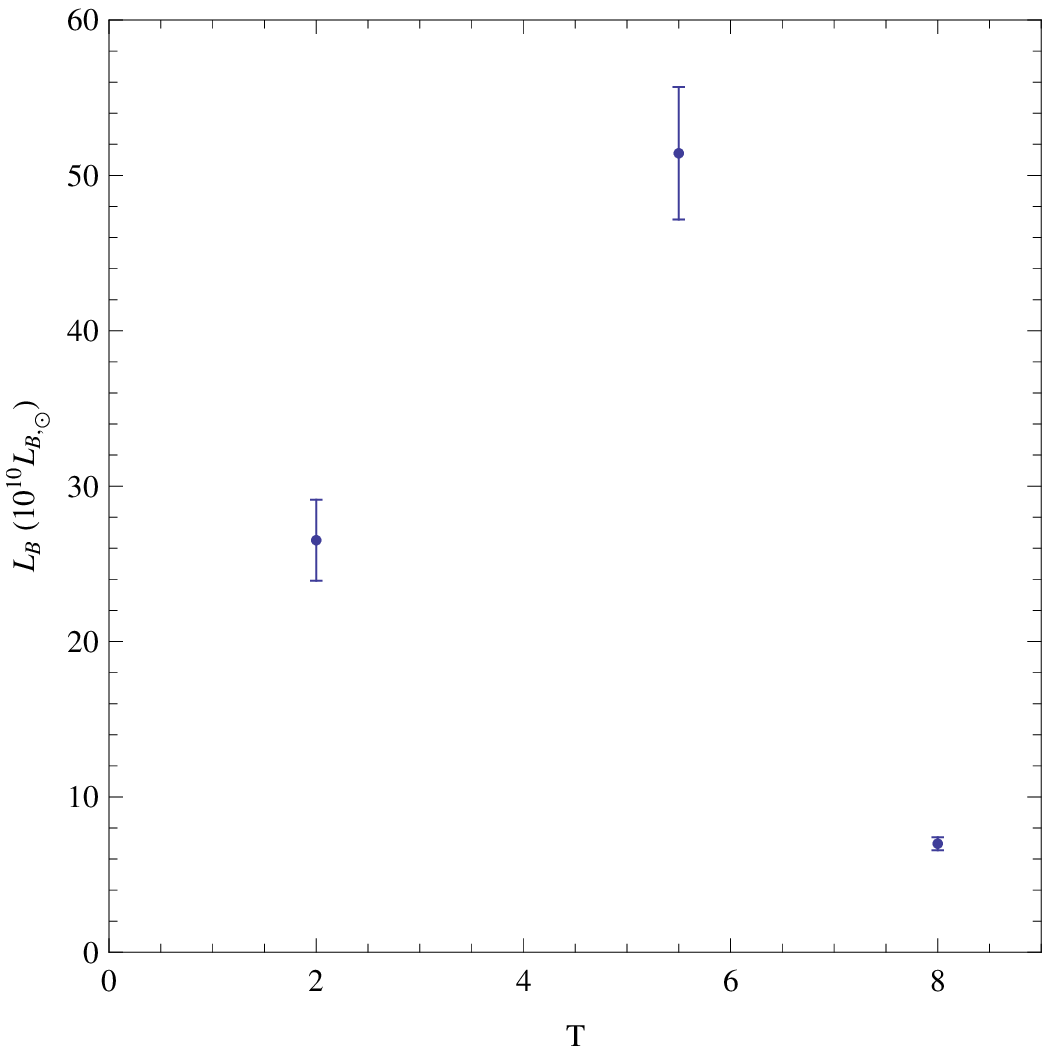} \\
\includegraphics[width=5cm]{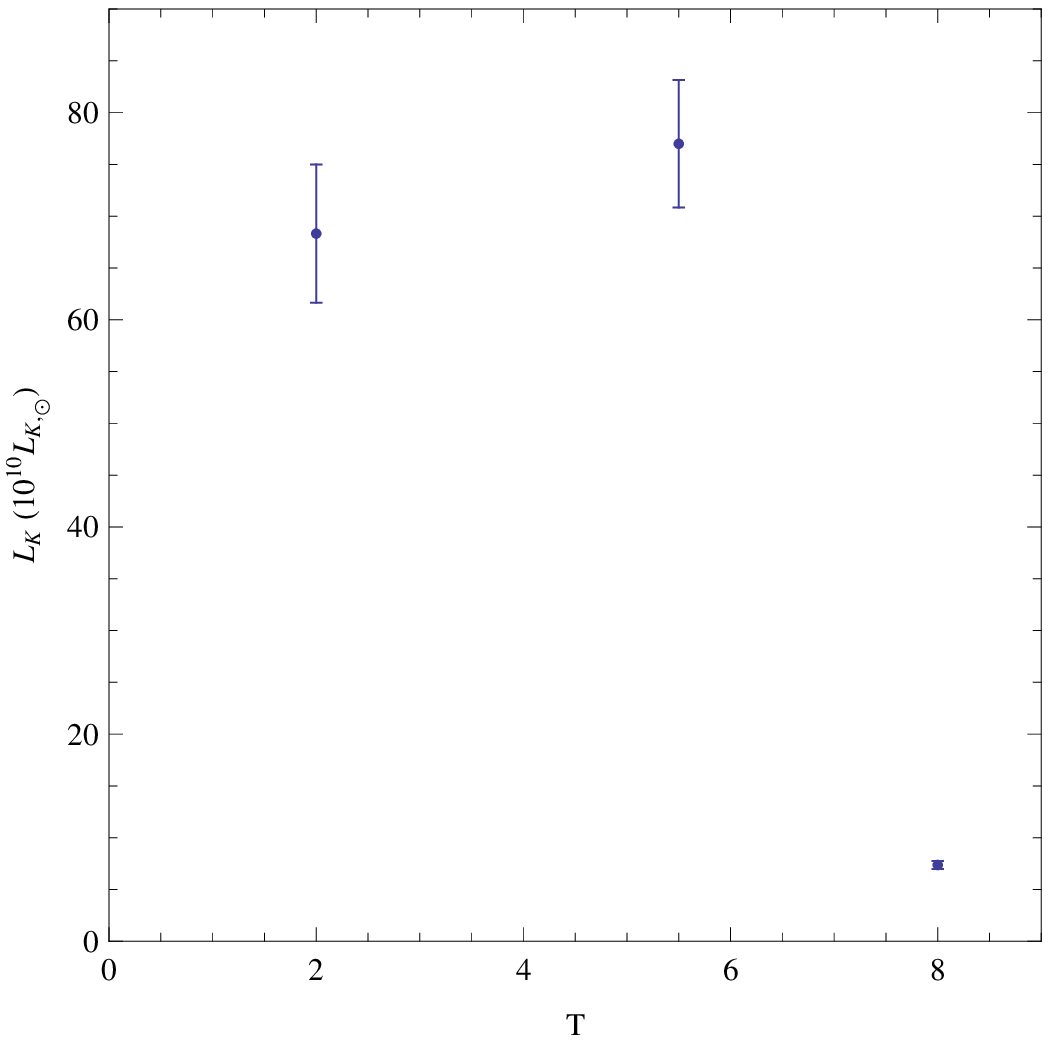} &
\includegraphics[width=5.1cm]{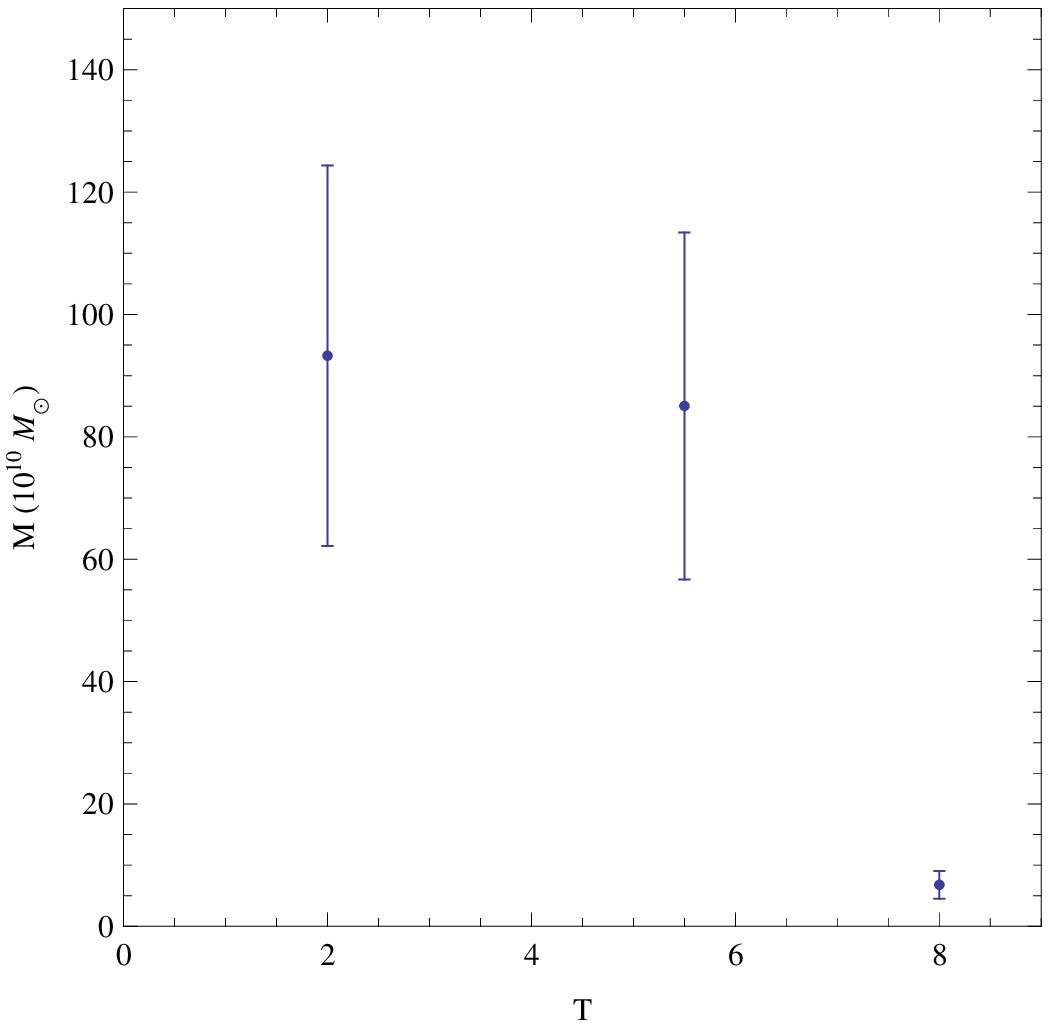}\\
\end{array}
$
\end{center}
\caption{From the top to the bottom: the total SFR$_{UV}$, the  CC SN rate, the total \lb, \lk\, and mass in different sub-samples based on the galaxy morphological type.}\label{T_sampleC}
\end{figure*}

Our result is also consistent with that  from LOSS  of  about $0.7_{-0.07}^{+0.08}$\,SNu\footnote{$0.5_{-0.06}^{+0.07}$\,SNu,  $0.7_{-0.07}^{+0.08}$\,SNu,   $0.9_{-0.09}^{+0.1}$\,SNu,  $0.87_{-0.09}^{+0.1}$\,SNu  in Sb, Sbc, Sc  and Scd galaxies, respectively} \citep{Li2010}.
The CC SN rate normalised per unit $K$ band luminosity and mass  in the  sub-sample $4 \le T \le7$ has the value of  $1.0_{-0.3}^{+0.4}$\,SNuK and $0.9_{-0.3}^{+0.4}$\,SNuM, respectively.   \citet{Mannucci2005}  estimated a CC SN rate of   $0.47_{-0.16}^{+0.17}$\,SNuK  and  $0.86_{-0.30}^{+0.31}$\,SNuM unit  in Sbc--Sd galaxies, while \citet{Li2010} estimated about  $0.3_{-0.04}^{+0.04}$\,SNuK\footnote{$0.16_{-0.015}^{+0.016}$\,SNuK, $0.28_{-0.03}^{+0.03}$\,SNuK, $0.39_{-0.04}^{+0.04}$\,SNuK, $0.46_{-0.05}^{+0.06}$\,SNuK in  Sb, Sbc, Sc, Scd   galaxies, respectively.} 
and  about $0.62_{-0.05}^{+0.06}$\,SNuM\footnote{$0.26_{-0.03}^{+0.04}$\,SNuM,  $0.51_{-0.05}^{+0.06}$\,SNuM, $0.75_{-0.08}^{+0.09}$\,SNuM, $0.96_{-0.1}^{+0.15}$\,SNuM  in  Sb, Sbc, Sc, Scd   galaxies, respectively.}.
Both the  values  from \citet{Mannucci2005}  and \citet{Li2010} are consistent with our estimates, the CC SN rate normalised per $K$ band luminosity is slightly lower but the difference is not statistically significant.

The CC SN rate, the total \lb, \lk\, and mass in  the different sub-samples $0 \le \mathrm{T} < 4$, $4\le \mathrm{T} \le 7$ and the  $\mathrm{T} > 7$ are illustrated in the Fig.~\ref{T_sampleC}.\\

\begin{table*}
\caption{The number of the galaxies and CC SNe discovered in the last 13 years, the total SFRs, the CC SN rate,  the total \lb\, and \lk, the CC SN rates normalised per unit luminosity and mass in galaxy sub-samples  of the sample C based on morphological type,  $B-K$ colour,  mass, sSFR  and \eha} \label{depSampleC}
\begin{tabular}{lcccccccccc}   
\hline\hline
 & N$_\mathrm{gal}$ & N$_\mathrm{CC}$ & SFR$_{H\alpha}$  &SFR$_{UV}$      &R$_\mathrm{CC}$ &$L_{B}$ & $L_{K}$ & R$_\mathrm{CC}$    & R$_\mathrm{CC}$   & R$_\mathrm{CC}$   \\
  &                                  &                                   &  (\msun yr$^{-1}$)  & (\msun yr$^{-1}$)    & (yr$^{-1}$) & (10$^{10}$ \lbsun) & (10$^{10}$ \lksun) & (SNu)   &   (SNuK)  & (SNuM)  \\
  \hline
 $T <0$    & 5& 0 &  $0.4\pm0.1 $  &   $0.4\pm0.1 $&  $\le 0.14 $& $ 0.4\pm0.1 $&  $1\pm 0.3$&  $\le 34 $ &  $\le 13$   &$\le 9$ \\
 $0 \le T  <4$	& 15& 1&  $8\pm 0.6$  & $14\pm 0.9$ &  $0.08_{-0.06}^{+0.2}$ & $ 27\pm3 $& $68 \pm 7 $ & $0.3_{-0.2}^{+0.7}$ & $0.1_{-0.09}^{+0.3}$    & $0.08_{-0.07}^{+0.2}$ \\
$4 \le T  \le7$& 52& 10& $ 42\pm 4 $&  $71 \pm 6$ & $0.8_{-0.2}^{+0.3}$ & $51\pm 4$  & $77\pm 6$ &   $1.5_{-0.5}^{+0.6}$ &   $1_{-0.3}^{+0.4}$  & $0.9_{-0.3}^{+0.4}$ \\
$T > 7$  & 95& 1&  $7 \pm 0.4$  &  $9 \pm 0.7$&  $0.08_{-0.06}^{+0.2}$ & $7 \pm0.4 $&  $7\pm0.4$& $1.1_{-0.9}^{+2}$  &  $1_{-0.9}^{+2}$   & $1_{-0.9}^{+3}$ \\
 \hline
$(B-K) \le 3$\,mag &126 & 10& $45\pm 4$& $66\pm 6 $ & $0.8_{-0.2}^{+0.3}$   &  $54\pm 4$  &  $68\pm 6$ & $1.4_{-0.4}^{+0.6}$  &$1.1_{-0.4}^{+0.5}$   &$1.2_{-0.4}^{+0.5}$ \\
$(B-K) > 3$\,mag   &41 & 2& $13\pm1 $&$28\pm1 $  &  $0.15_{-0.1}^{+0.2}$&   $31\pm 3$ & $86\pm 7$     &$0.5_{-0.3}^{+0.6}$&  $0.2_{-0.1}^{+0.2}$&  $0.1_{-0.08}^{+0.2}$ \\
 \hline
M $ \le 5 \times 10^{10} \msun$&152 & 9& $34\pm 3$& $50\pm 5$ &  $0.7_{-0.2}^{+0.3}$& $40\pm3 $   & $45\pm 3$ & $1.8_{-0.6}^{+0.8}$ & $1.5_{-0.5}^{+0.7}$    & $1.6_{-0.5}^{+0.7}$\\
M $> 5 \times10^{10} \msun$ & 15 & 3& $24\pm 2$&  $44\pm 4$& $0.2_{-0.1}^{+0.2}$& $46\pm 4$  & $108\pm9 $ & $0.5_{-0.3}^{+0.5}$ &  $0.2_{-0.1}^{+0.2}$   &$0.16_{-0.09}^{+0.15}$ \\
 \hline
sSFR $ \le 10^{-10}$ yr$^{-1}$& 94& 4& $30\pm2 $& $54\pm 4 $ &  $0.3_{-0.1}^{+0.2}$& $55\pm 4$    & $125\pm 9 $ & $0.6_{-0.3}^{+0.4}$ & $0.2_{-0.1}^{+0.2}$    & $0.2_{-0.1}^{+0.1}$\\
 sSFR $ > 10^{-10}$ yr$^{-1}$ & 73& 8& $27\pm3 $& $41\pm 5$ &   $0.6_{-0.2}^{+0.3}$& $30\pm3 $  & $29\pm 3$ & $2_{-0.7}^{+1}$ & $2_{-0.7}^{+1}$    &$3_{-0.9}^{+1}$ \\
  \hline
 $\eha \le 30\, \AA$ &83 & 4 & $24\pm 2$&  $43\pm 4$&  $0.3_{-0.1}^{+0.2}$& $50\pm 4$  & $112\pm 9$ & $0.6_{-0.3}^{+0.5}$ & $0.3_{-0.1}^{+0.2}$    &$0.2_{-0.1}^{+0.2}$ \\
 $ \eha > 30\, \AA$&80 & 8&$33\pm3 $ & $51\pm5 $ &  $0.6_{-0.2}^{+0.3}$&  $35\pm 3$   &$41\pm3 $  & $1.8_{-0.6}^{+0.9}$ &  $1.5_{-0.5}^{+0.7}$   &$1.6_{-0.5}^{+0.8}$ \\
  \hline
\end{tabular}
\end{table*}

{\em $B-K$}\\
An alternative indicator of the SFR are galaxy colours. \citet{Cappellaro1999} analysed  for the first time the dependance of CC SN rates in SNu unit on the galaxy $U-V$ colour pointing out  that the CC SN rate is higher in the bluer galaxies. The strong dependence of the CC SN rates  normalised per unit $K$ band luminosity and unit mass on the galaxy  $B-K$ colour has been showed by \citet{Mannucci2005}.
 \citet{Li2010} found a very similar trend of the CC SN rates as a function of galaxy $B-K$ colour: an increase from red to the blue galaxies.
 We considered two different bins in $B-K$ colour ($B-K \le 3$\,mag and $B-K > 3$\,mag)  with about the 80\% and  the 20\% of CC SNe and galaxies, respectively (Table~\ref{depSampleC}). The  total SFR (based on \lha\, and \lfuv) and the total \lb\,  are higher in bluer galaxies while the total  \lk\, and total mass are larger  in redder galaxies. The CC SN rates normalised per unit luminosity and mass are an order of magnitude higher in the bluer galaxies. There is a trend depending on the normalisation, increasing from SNu to SNuM  as expected (Table~\ref{depSampleC} and Fig.~\ref{RCC_sampleC}).\\

{\em Mass} \\
There are several recent works probing the dependence of the SFR, sSFR and SF efficiency as a function of the stellar mass and suggesting that stellar mass plays a fundamental role in determining the fate of a galaxy \cite[e.g.][]{Schiminovich2007,Schiminovich2010}. The environment  seems to be a second order effect compared with the mass. 
  To analyse the CC SN rate as a function of the galaxy mass we  split our galaxy sample in two sub-samples  by using a cut of $5\times 10^{10}$ \msun.  In this case we have in the first sub-sample the 90\% and 75\% of the galaxies and CC SNe, respectively (Table~\ref{depSampleC}).  The  SFR$_{H\alpha}$, SFR$_{UV}$ and \lb\, are similar in both sub-samples  while the CC SN rates normalised per unit luminosity and mass are lower  for the most massive galaxies (Table~\ref{depSampleC}  and Fig.~\ref{RCC_sampleC}).
  If we split the sample C in two sub-samples  by measuring the same total mass (about $93\times 10^{10}$ \msun\,) we have the 95\% of galaxies, the 84\%  of CC SNe, the 70\% of  SFR$_{UV}$, the 65\% of \lb\, in the first sample.  The  CC SN rates are $1.4_{-0.4}^{+0.6}$  and $0.5_{0.3}^{+0.7}$ SNu,  $0.9_{-0.3}^{+0.4}$  and $0.2_{-0.1}^{+0.3}$ SNuK,  $0.8_{-0.3}^{+0.4}$  and $0.2_{-0.1}^{+0.2}$   SNuM in the first and second sub-sample, respectively.
  With a different choice of the two sub-samples the difference between rates is lower.
\\ 



\begin{figure}
\begin{center}
\includegraphics[width=5.7cm]{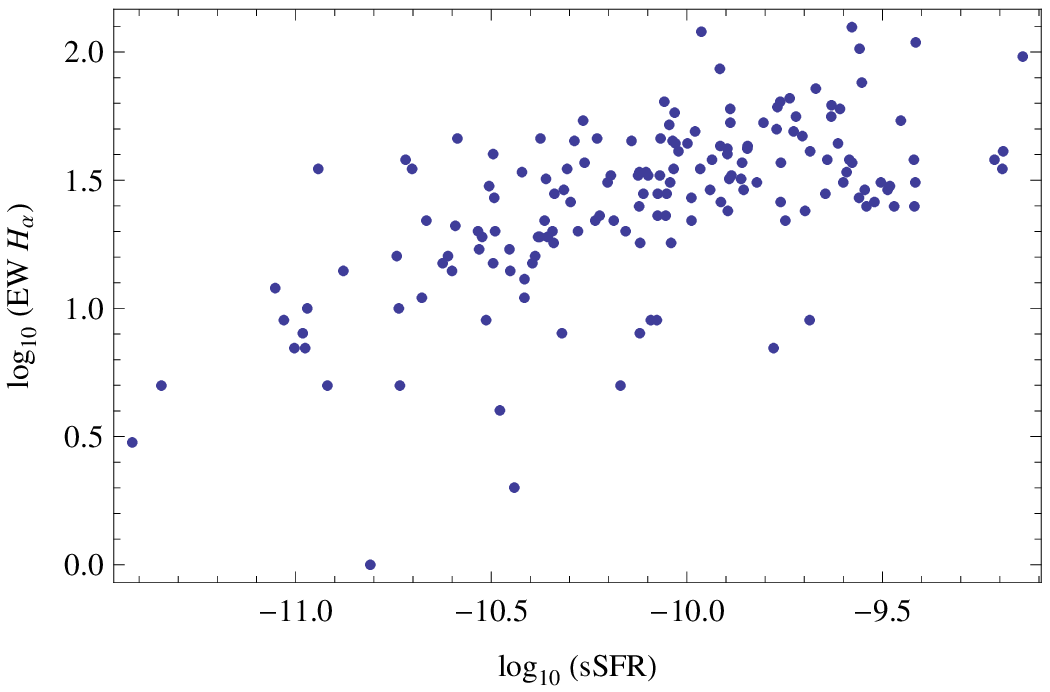} 
\end{center}
\caption{The \eha\, as a function of the sSFR in the sample C.}\label{ewrcc}
\end{figure}

{\em sSFR and $\eha$}\\
We also measured the CC SN rate in two different galaxy sub-samples obtained by assuming a cut value of   $10^{-10}$ yr$^{-1}$ in sSFR (Table~\ref{depSampleC}) and found that about the 70\% of CC SNe is in the galaxies with higher sSFR (44\%). The  fraction of total SFR (based on \lha\, and \lfuv) in each sub-sample has  similar values while  \lk\, and total mass are larger  in galaxies with lower sSFR. The CC SN rates normalised per unit luminosity and mass are an order of magnitude higher in galaxies with lower sSFR. There is a trend depending on the normalisation, increasing from SNu to SNuM (Table~\ref{depSampleC} and Fig.~\ref{RCC_sampleC}).\\
We stress that the sSFR is tightly connected to the CC SN rate normalised per unit mass due  to the relation between CC SN rate and SFR.
\eha\,  is an indicator of the sSFR  so a  tightly correlation between the sSFR, estimated by  using the relation between galaxy mass, $B-K$ colour and \lk\,, and \eha\, is expected  (Fig.~ \ref{ewrcc}).
It is also  interesting to investigate the correlation between CC SN rate and \eha\, as done  for the first time by \citet{Kennicutt1984}  for 171 galaxies and 55 SNe (11 of type II)  that  found a linear relation, supporting the
massive star origin of the CC SN progenitors. 
We divided the galaxies  in the sample C with measured \eha\,  in two sub-samples  by adopting a cut of 30 \AA,  that  roughly corresponds to the same cut adopted in sSFR, and we found as expected very similar results (Table~\ref{depSampleC} and Fig.~\ref{ewrcc}).
If we split the sample C in two sub-samples  by measuring the same total sSFR (about $1 \times 10^{-8} yr^{-1}$) we have the 80\% of galaxies, the 83\%  of CC SNe, the 80\% of  SFR$_{UV}$, the 88\% of \lb\, in the first sample.  The  CC SN rates are $1.0_{-0.3}^{+0.4}$  and $1.5_{0.9}^{+2}$ SNu,  $0.5_{-0.2}^{+0.2}$  and $2_{-1}^{+3}$ SNuK,  $0.4_{-0.1}^{+0.2}$  and $3_{-2}^{+4}$   SNuM in the first and second sub-sample, respectively.

 \begin{figure*}
\begin{center}
$
\begin{array}{c@{\hspace{.1in}}c@{\hspace{.1in}}c@{\hspace{.1in}}c}
\includegraphics[width=6cm]{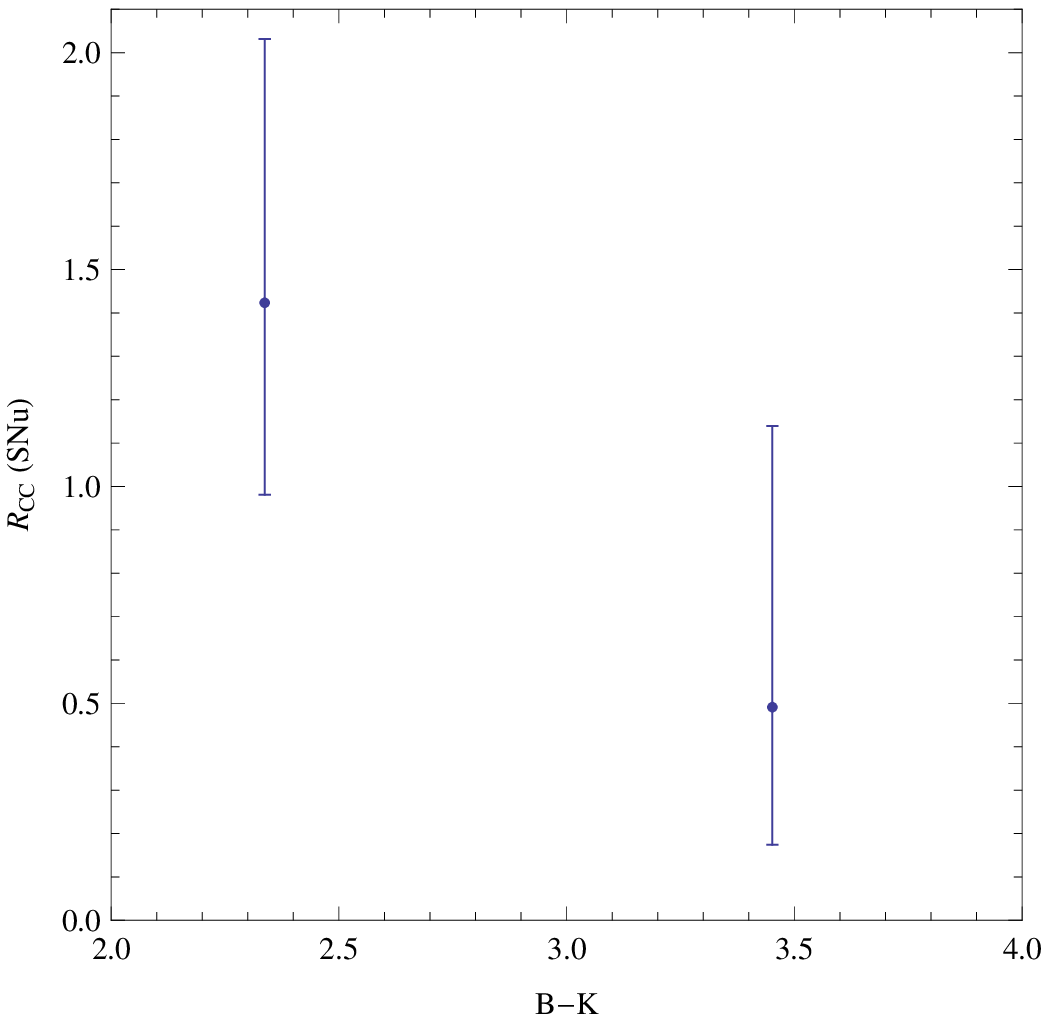}&
\includegraphics[width=6cm]{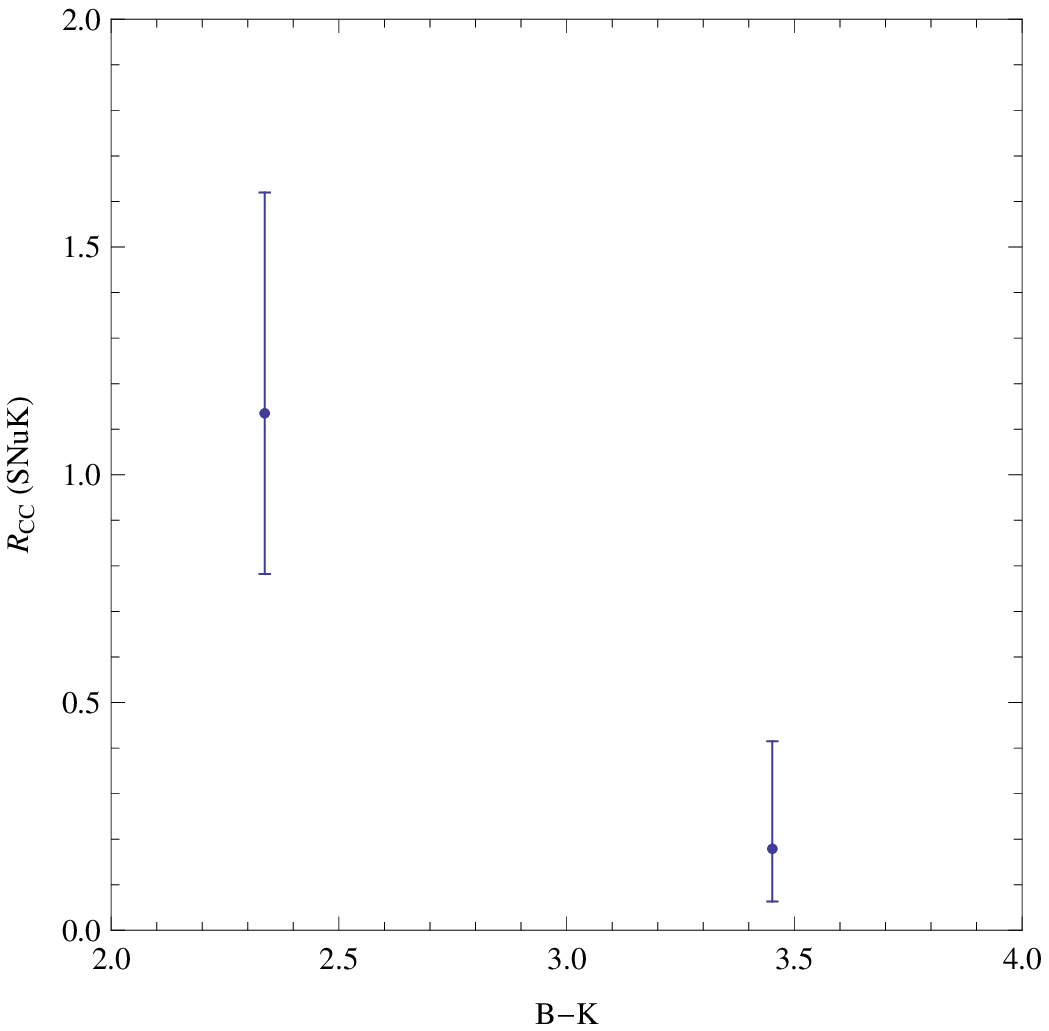}&
\includegraphics[width=6cm]{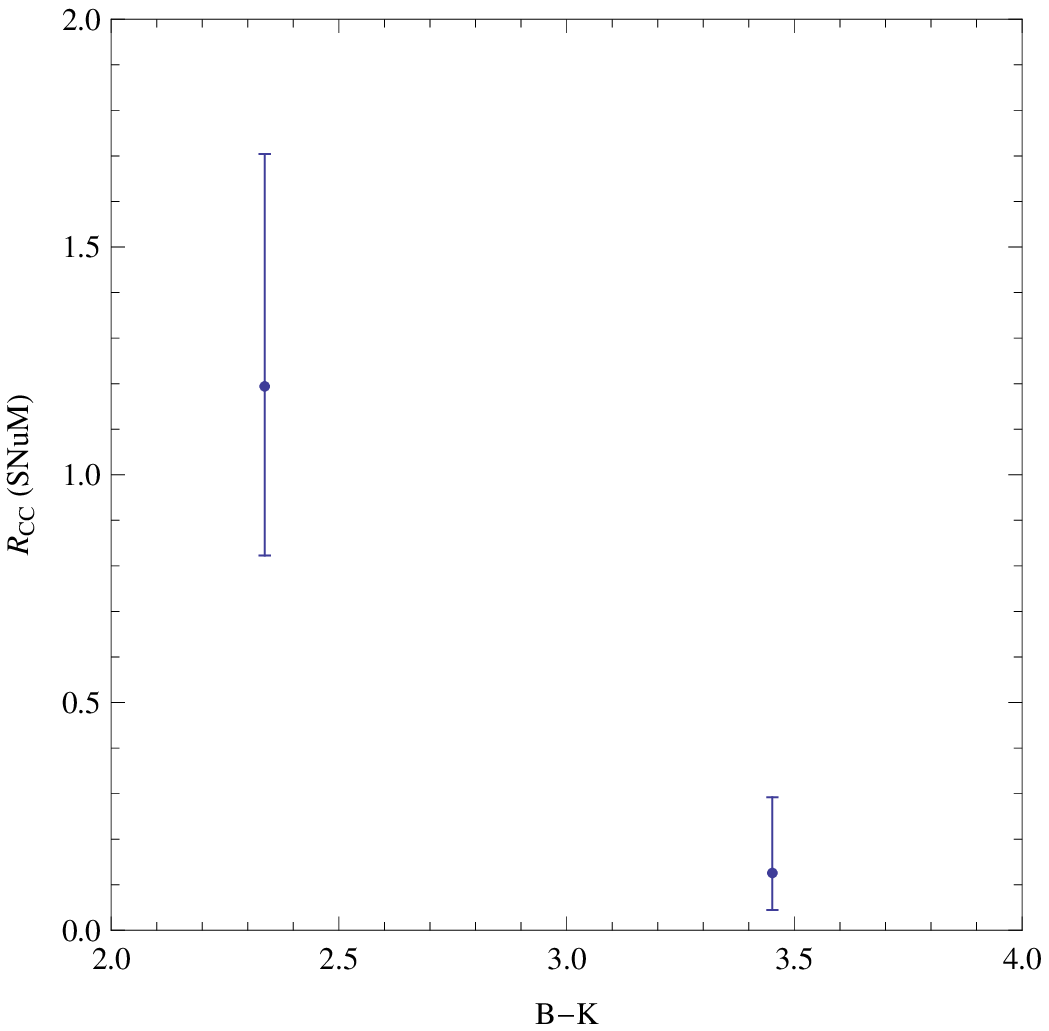} &
\\
\includegraphics[width=6cm]{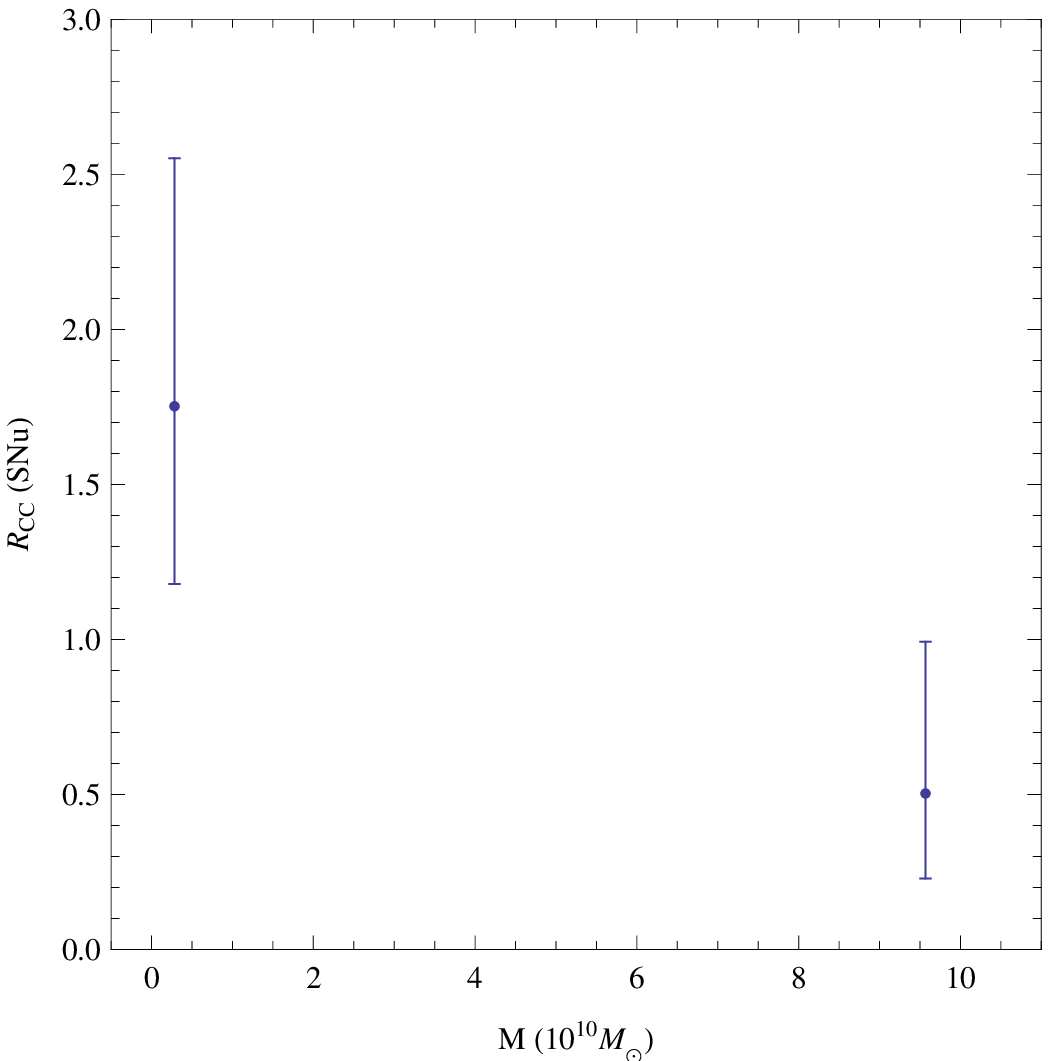} &
\includegraphics[width=6cm]{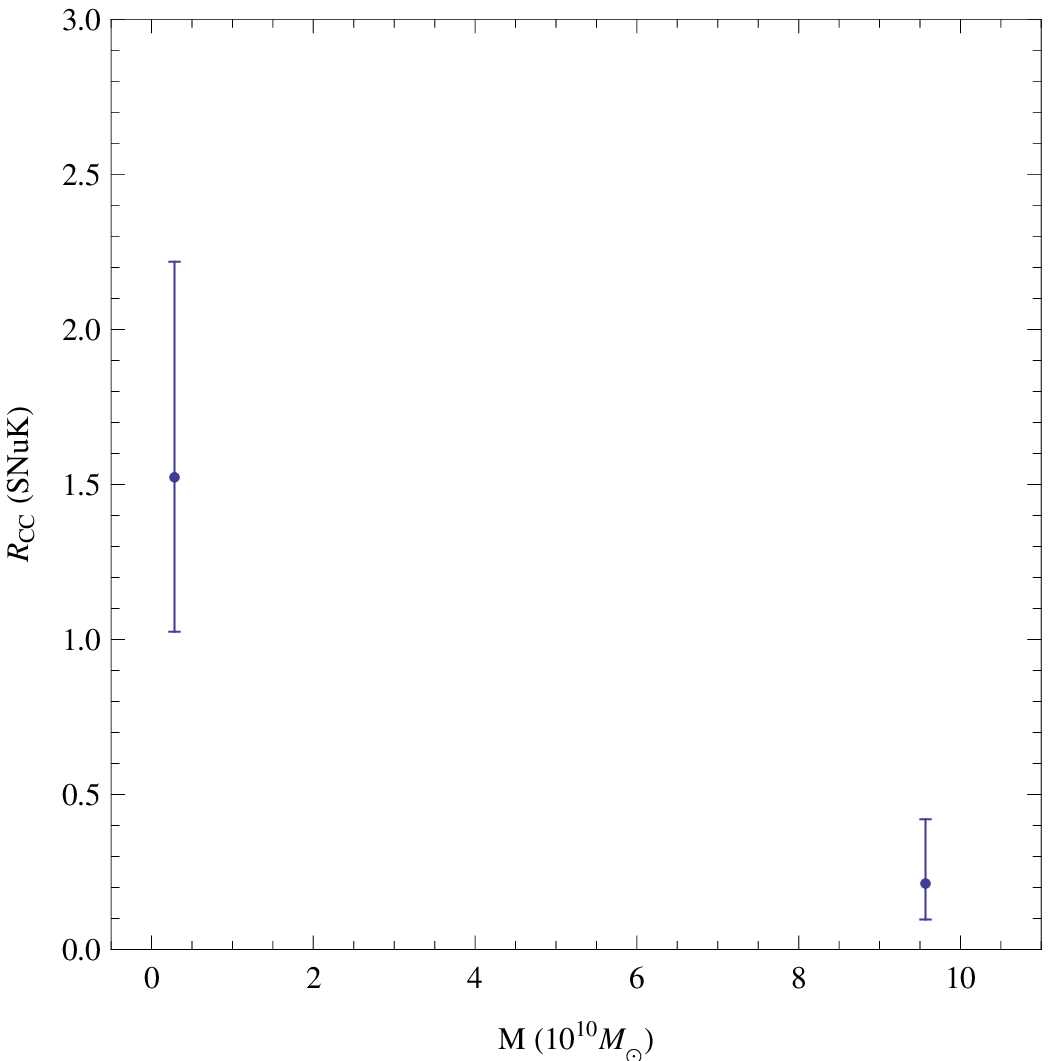} &
\includegraphics[width=6cm]{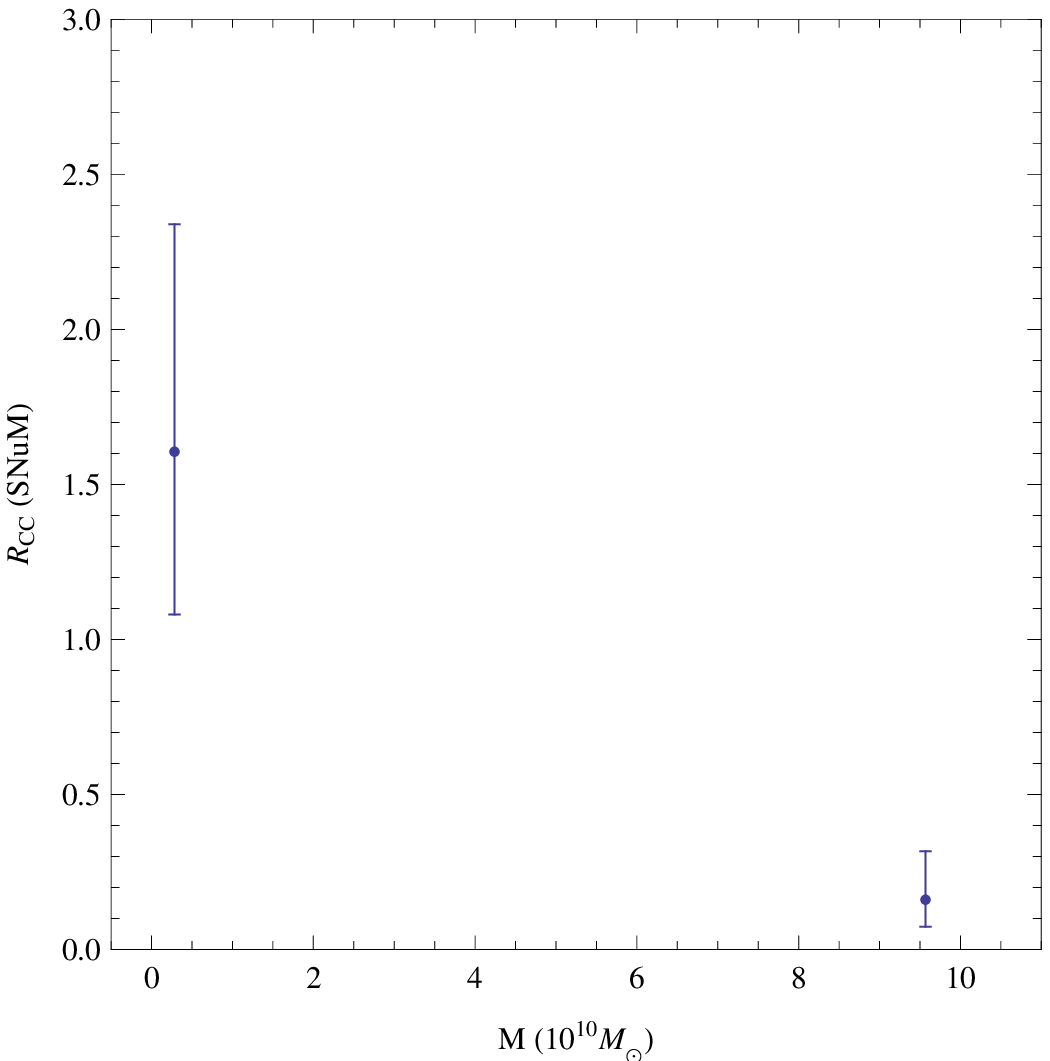}&
\\
\includegraphics[width=6cm]{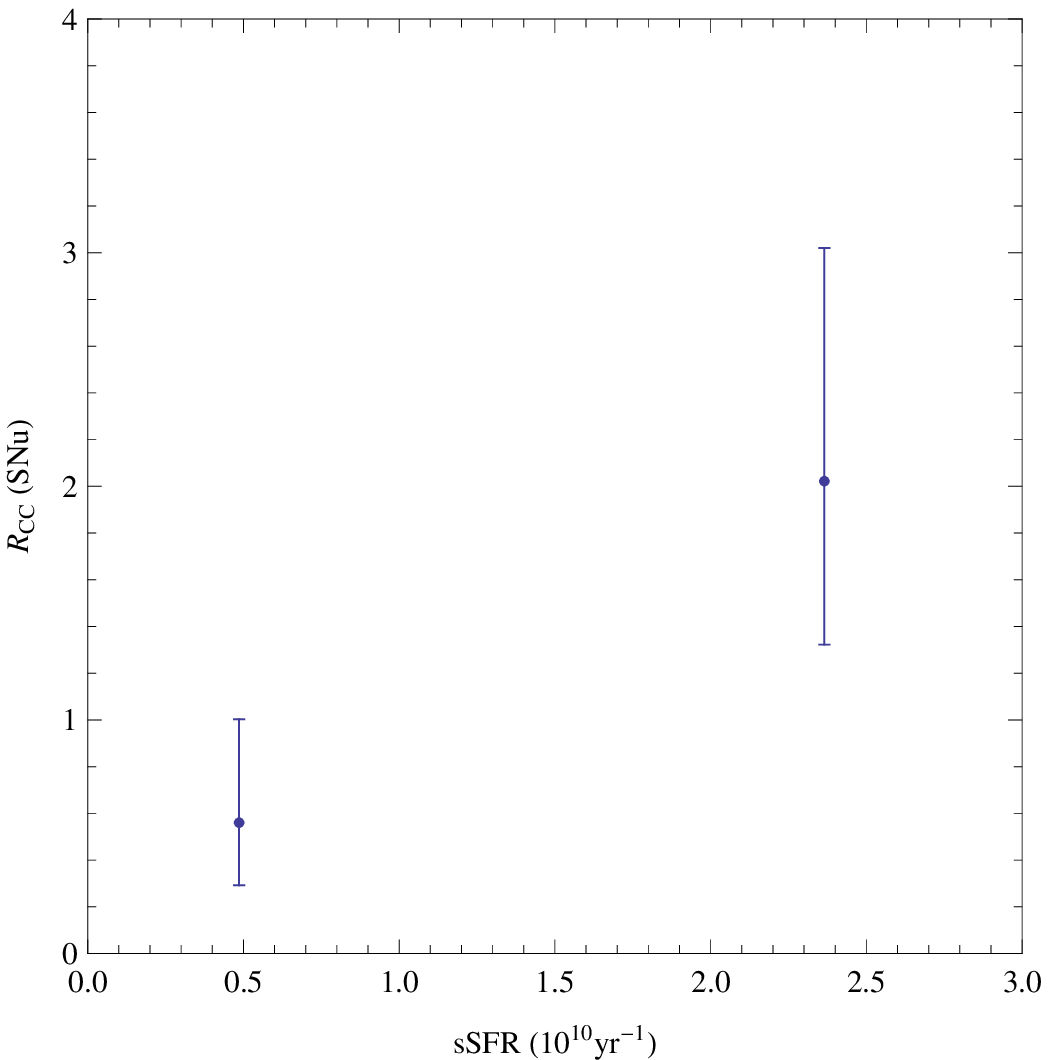} &
\includegraphics[width=6cm]{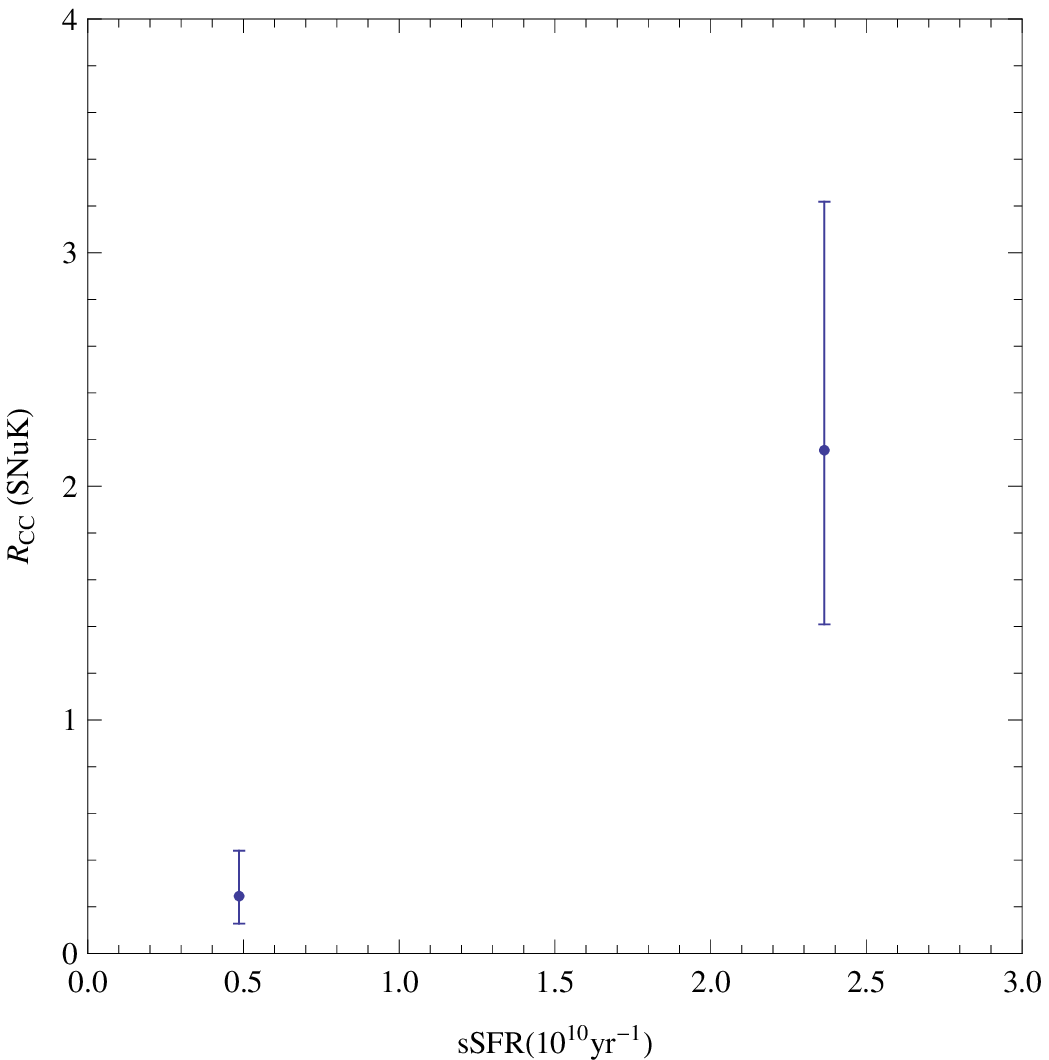} &
\includegraphics[width=6cm]{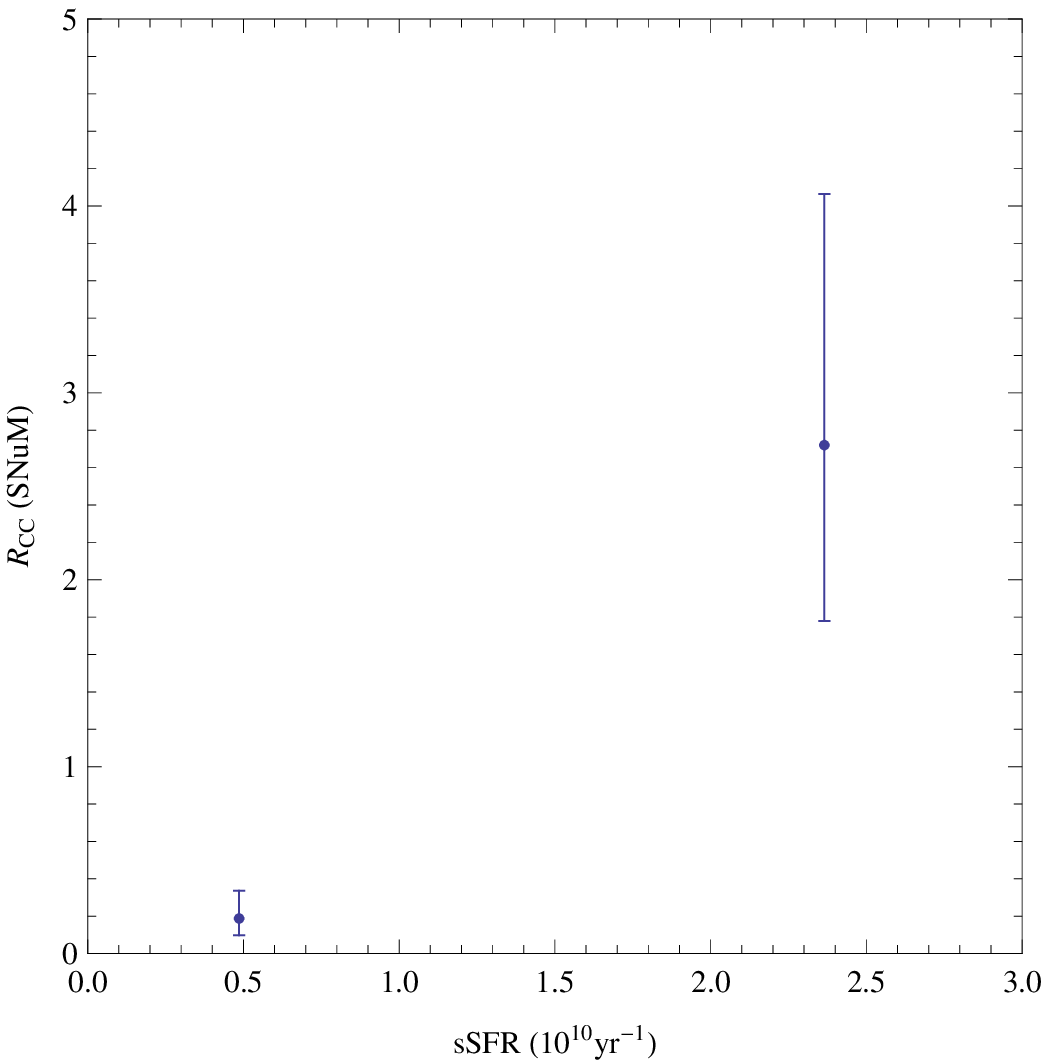}&
\\
\includegraphics[width=6cm]{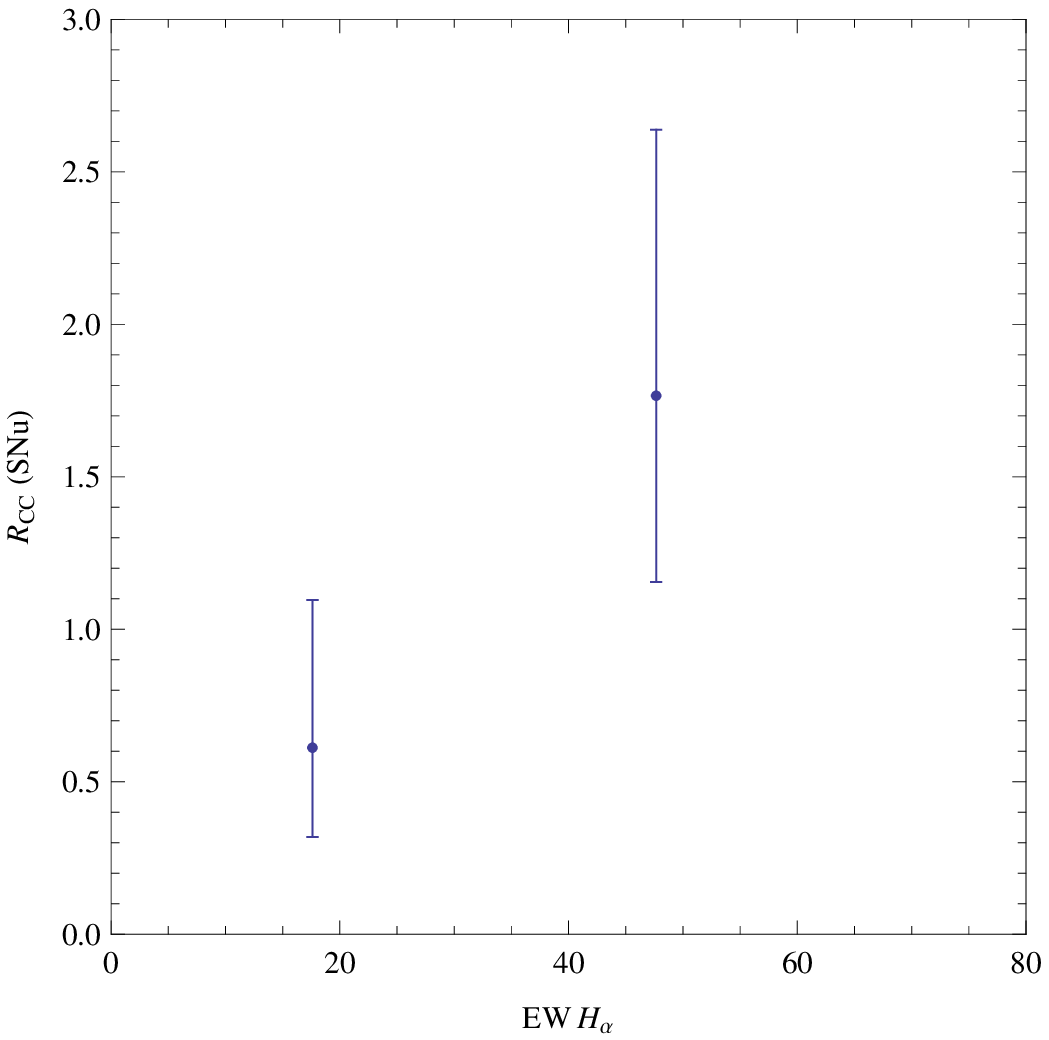} &
\includegraphics[width=6cm]{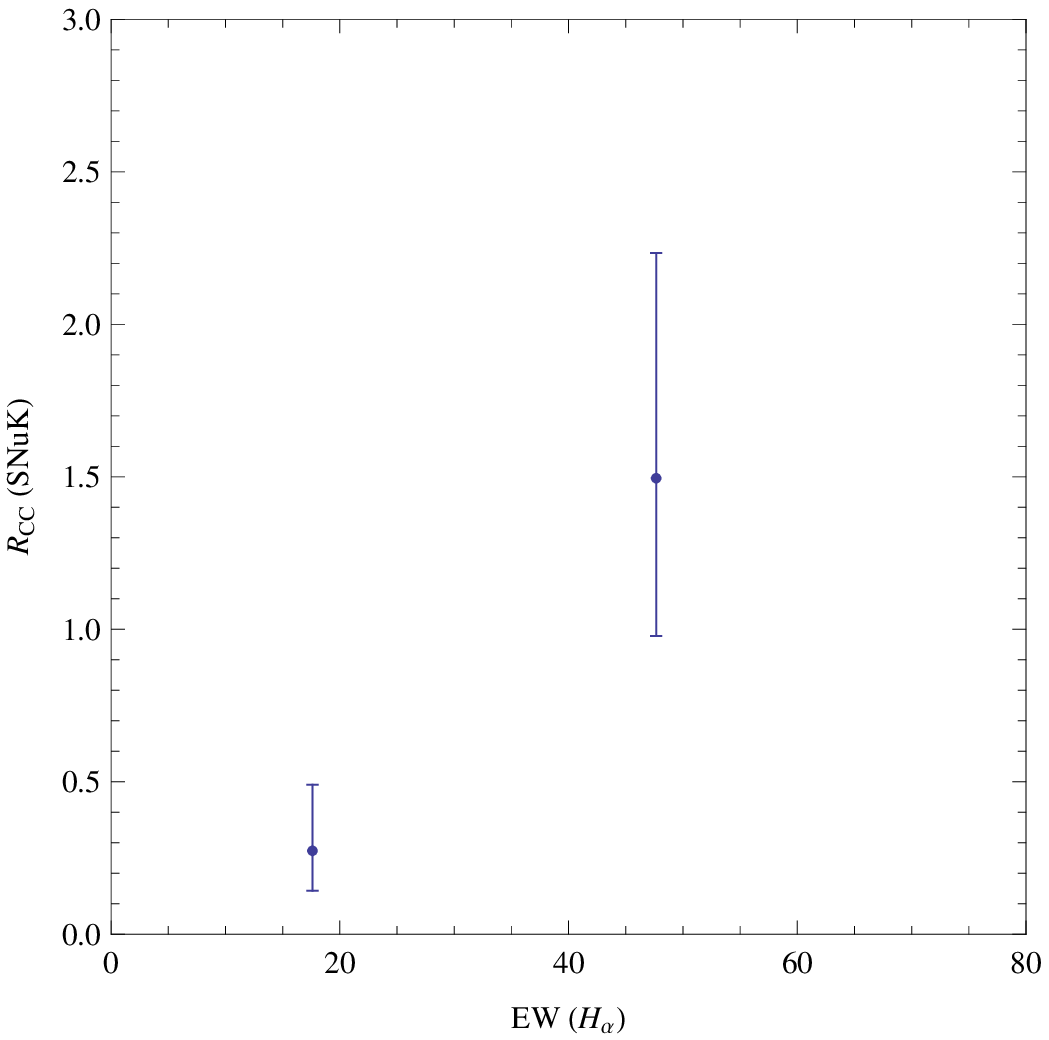} &
\includegraphics[width=6cm]{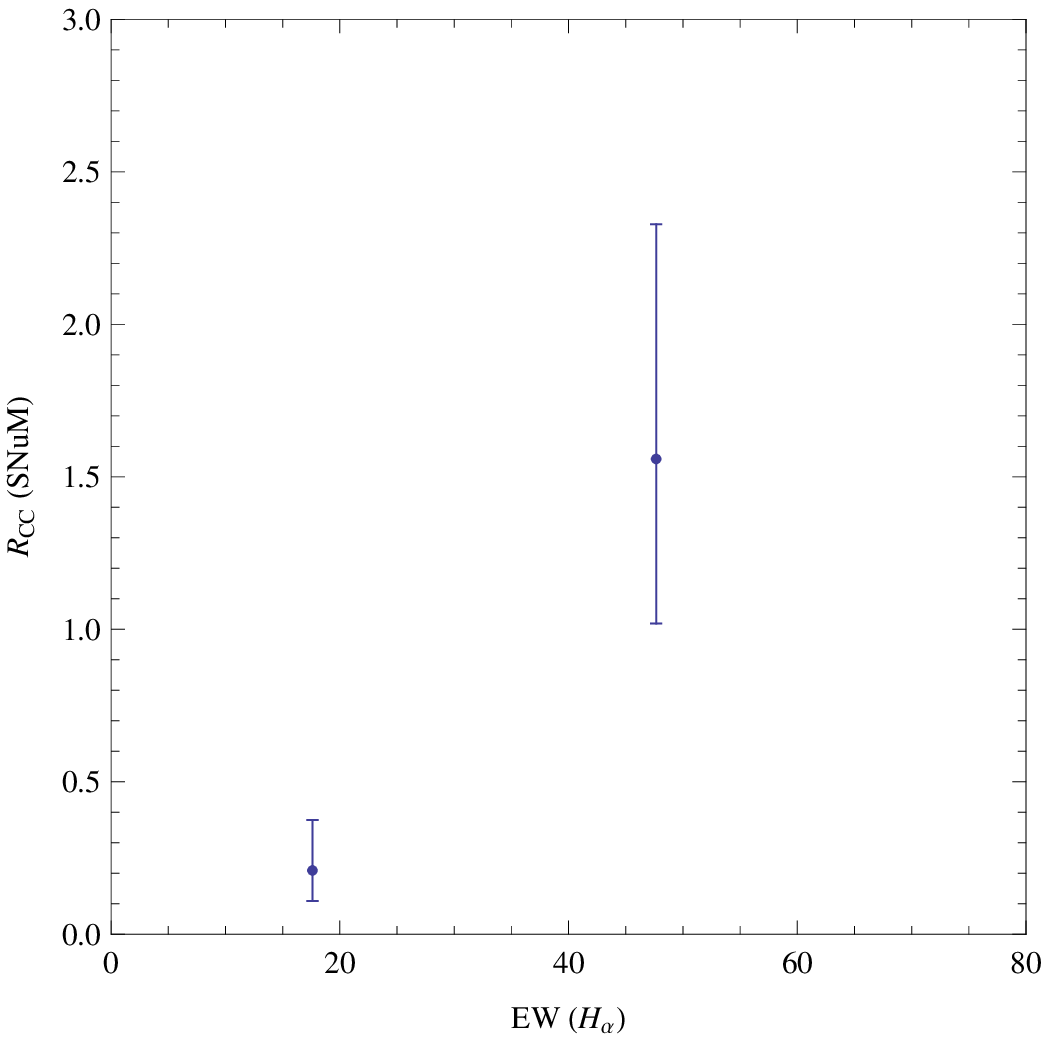}&
\end{array}
$
\end{center}
\caption{From the top to the bottom: the CC SN rates in SNu unit (left), SNuK unit (centre) and SNuM unit (right) as a function of $B-K$, mass, sSFR and \eha\, respectively.}\label{RCC_sampleC}
\end{figure*}

\end{appendix}

\end{document}